\documentclass[sigconf]{acmart}
\AtBeginDocument{%
  }

\usepackage{algorithm}
\usepackage[noend]{algorithmic}
\usepackage{multirow}
\usepackage{xurl}

\usepackage{graphicx}
\usepackage{subcaption}

\setcopyright{acmlicensed}
\copyrightyear{2018}
\acmYear{2018}
\acmDOI{XXXXXXX.XXXXXXX}
\acmConference[Conference acronym 'XX]{Make sure to enter the correct
  conference title from your rights confirmation email}{June 03--05,
  2018}{Woodstock, NY}
\acmISBN{978-1-4503-XXXX-X/2018/06}

\begin{document}

\title[3DPipe: GPU-Accelerated 3D Spatial Join]{3DPipe: A Pipelined GPU Framework for Scalable Generalized Spatial Join over Polyhedral Objects}

\author{Lyuheng Yuan, Da Yan, Saugat Adhikari, Akhlaque Ahmad}
\affiliation{%
  \institution{Indiana University Bloomington}
  \city{}
  \state{}
  \country{}  
}
\email{{lyyuan,yanda,adhiksa,akahmad}@iu.edu}

\author{Fusheng Wang}
\affiliation{%
  \institution{Stony Brook University}
  \city{}
  \state{}
  \country{}  
}
\email{fusheng.wang@stonybrook.edu}


\begin{abstract}
Spatial join is a fundamental operation in spatial databases. With the rapid growth of 3D data in applications such as LiDAR-based object detection and 3D digital pathology, there is an increasing need to support spatial join over 3D datasets. However, existing techniques are largely designed for 2D data, leaving 3D spatial join underexplored and computationally expensive.
We present {\bf 3DPipe}, a pipelined GPU framework for scalable spatial join over polyhedral objects. 3DPipe exploits GPU parallelism across both filtering and refinement stages, incorporates a multi-level pruning strategy for efficient candidate reduction, and employs chunked streaming to handle datasets exceeding GPU memory. Its pipelined execution overlaps CPU data preparation, host-device data transfer, and GPU computation to improve throughput.
Experiments show that 3DPipe achieves up to 9.0$\times$ speedup over the state-of-the-art GPU solution, TDBase, while maintaining excellent scalability. 
3DPipe is open-sourced at \url{https://github.com/lyuheng/3dpipe}.

\end{abstract}


\maketitle


%
%
%
%
%
%
%
%
%
%

\vspace{-1mm}
\section{Introduction}
\vspace{-1mm}
Spatial join is a fundamental operation that relates objects across two spatial datasets $R$ and $S$ based on their geometric relationships. Formally, for each object $r \in R$, spatial join retrieves objects $s \in S$ that satisfy a spatial predicate, including within-$\tau$ distance, intersection (i.e., special case with $\tau = 0$), or $k$-nearest neighbors ($k$-NN). 
These query types constitute a {\bf generalized} class of spatial join operations widely used in real spatial applications~\cite{geospark,DBLP:journals/geoinformatica/TengBPKW24,wang2011data,gnanakaran2003peptide,DBLP:journals/vldb/ZhangQLWW12}.

Nowadays, many applications require large-scale 3D spatial data processing, including digital pathology~\cite{farahani2017three}, human atlases~\cite{hubmap2019human}, GIS~\cite{esri}, mineral exploration~\cite{real2019large}, high-definition mapping~\cite{ZangLCT19}, and urban planning~\cite{Cesium}. Meanwhile, advances in AI and sensing technologies are rapidly accelerating the generation of such data. For instance, LiDAR-based autonomous driving systems produce massive point clouds~\cite{KITTI,nuScenes}, while modern 3D reconstruction~\cite{NeRF,Mega-NeRF}, generative~\cite{worldlabs_marble_labs}, and segmentation~\cite{chen2025sam3d,Point-SAM,SAM2Point} methods enable scalable transformation from raw inputs to structured 3D representations.

We adopt {\bf polyhedral representations} for 3D objects, which are widely adopted in spatial database systems such as PostGIS~\cite{postgis} and PolarDB~\cite{polardb} due to their ability to model complex geometries. Note that point clouds are typically converted to polyhedral representations via surface reconstruction methods~\cite{DBLP:conf/sgp/KazhdanBH06,DBLP:journals/tvcg/BernardiniMRST99,DBLP:conf/siggraph/LorensenC87,DBLP:conf/cvpr/ParkFSNL19}.

Most works on spatial join focus on 2D spatial objects (minimum bounding rectangles or polygons)~\cite{tspatial, pbsm, touch, DBLP:journals/geoinformatica/TengBPKW24,swiftspatial}. 
Compared to 2D geometries, 3D objects are substantially more complex, typically consisting of a large number of (triangular) facets, which leads to significant computational overhead. 
For example, a na\"{i}ve spatial join approach that directly evaluates object-pair relationships at the facet level incurs prohibitive cost. 
Such massively parallel facet-level computations naturally align with the SIMT execution model of GPUs, which is exploited by some recent studies~\cite{real2019large, 3dpro, tdbase}.

\begin{figure}[!t]
\centering
\includegraphics[width=0.78\columnwidth]{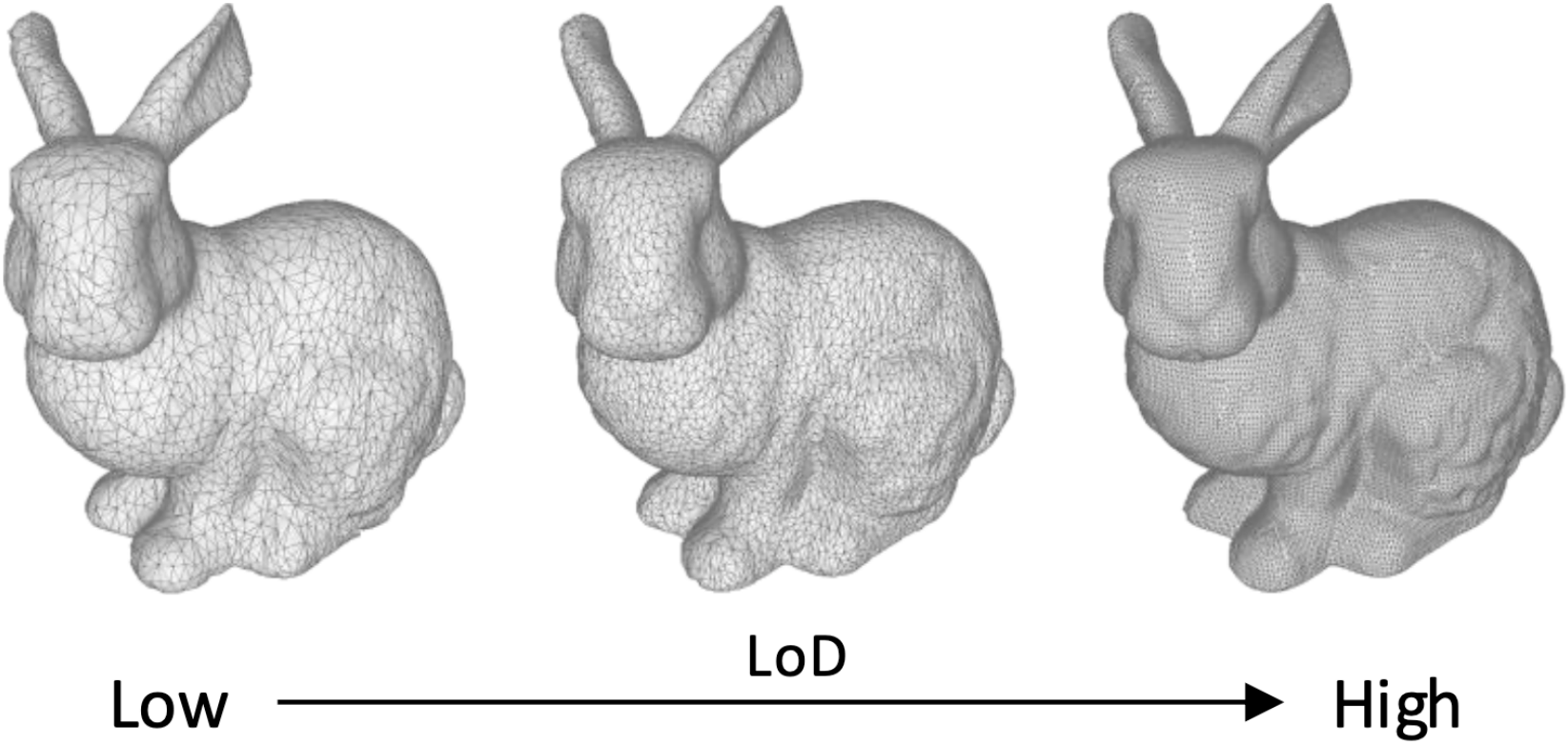}
\vspace{-3mm}
\caption{Multi-Resolution 3D Object Compression}\label{fig:lod}
\vspace{-6mm}
\end{figure}

We are only aware of three works that accelerate spatial join over polyhedral objects: iSPEED~\cite{DBLP:journals/tsas/TengLVKW22} which is built with MapReduce, and 3DPro~\cite{3dpro} and TDBase~\cite{tdbase} which support GPU execution. These works all follow a {\em Filter-and-Refine} paradigm, where coarse-grained filtering is first performed using lightweight object-level approximations to remove invalid object pairs, followed by a refinement stage that progressively evaluates candidate pairs using multi-resolution representations derived from geometric simplification (i.e., different levels of detail, or LODs, see Figure~\ref{fig:lod}), enabling early termination whenever the spatial relationship can be determined at lower resolutions, and only falling back to the original high-resolution geometries when necessary. To mitigate the high cost of geometric computations during refinement, 3DPro~\cite{3dpro} and TDBase~\cite{tdbase} leverage GPU to exploit massive parallelism.

While 3DPro~\cite{3dpro} and TDBase~\cite{tdbase} only use CPU for filtering, we observe that the filtering stage can take up to 82\% of the total running time even with OpenMP enabled for multithreaded execution. This is because, besides using object-level minimum bounding box (MBB) for filtering, they also utilize a more precise approximation of object geometry called {\em skeleton-based partitioning}, which decomposes each object by grouping nearby facets into intermediate geometric units (i.e., facet clusters) to enable more effective pruning than object-level MBBs. For convenience, we refer to such facet clusters as {\em voxels}, where each voxel is defined as the MBB enclosing the corresponding cluster of facets.

\begin{figure}[!t]
\centering
\includegraphics[width=0.8\columnwidth]{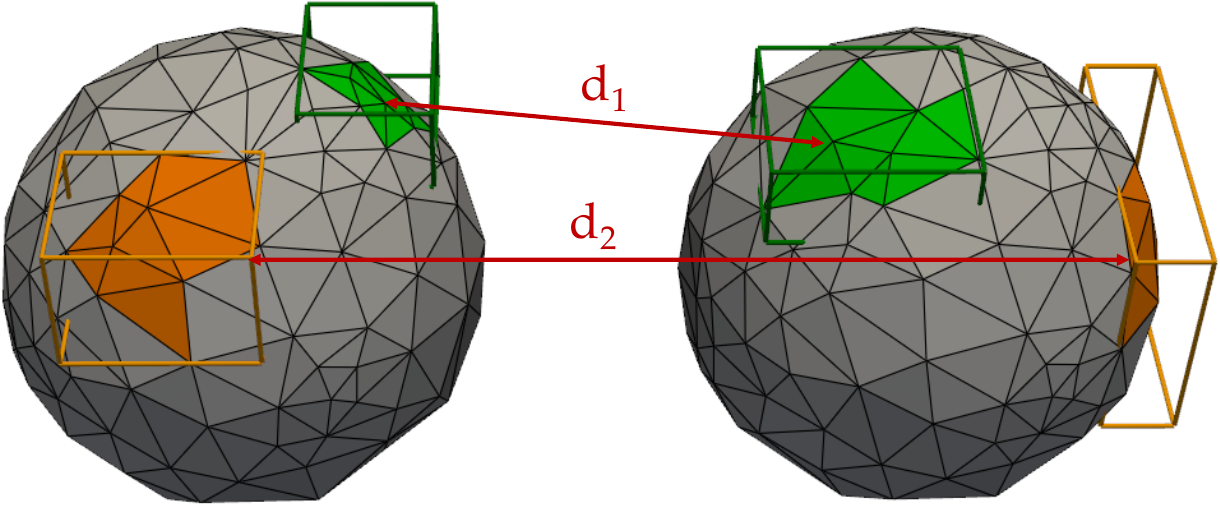}
\vspace{-3mm}
\caption{Illustration of Voxels and Voxel Pairs}\label{fig:voxel_def}
\vspace{-5mm}
\end{figure}

We illustrate why skeleton-based partitioning is essential for reducing facet-level computations during refinement using Figure~\ref{fig:voxel_def}, which shows a pair of objects $(r, s)$, each with two voxels highlighted. 
To compute the distance between $r$ and $s$ (i.e., the closest pair of points on their surfaces), denoted by $d(r,s)$, consider a distance $d_1$ between a vertex in $r$'s green voxel and a vertex in $s$'s green voxel. Clearly, $d_1$ serves as an upper bound of $d(r,s)$. 
For the orange voxel pair, let $d_2$ denote the minimum distance (MINDIST) between their MBBs (efficiently computed by Definition~2 of~\cite{DBLP:conf/sigmod/RoussopoulosKV95}). Since $d_2 > d_1$, this voxel pair can be safely pruned, as it cannot contain the closest pair of points that determines $d(r,s)$. 
By applying this principle across all voxel pairs, a large fraction of voxel pairs can be pruned early, significantly reducing the number of facet-level comparisons required during refinement. Since voxel-pair filtering introduces much computing overhead, it often becomes a performance bottleneck in TDBase, requiring GPU acceleration.

Even in the refinement stage, 3DPro~\cite{3dpro} and TDBase~\cite{tdbase} do not fully exploit advanced GPU features. 
For example, they perform distance-bound aggregation in global memory rather than the faster shared memory, 
and do not leverage CUDA streams to overlap data transfer (of facets) and CPU-GPU computation (of facet-pair distances). 
In addition, they use a poor implementation that incurs excessive kernel launches, causing substantial scheduling overhead.

In this paper, we propose a pipelined GPU framework, called {\bf 3DPipe}, for scalable generalized spatial join over polyhedral objects. 
{\color {red}Besides extensive performance evaluation, we also demonstrate the practical utility of 3DPipe through a real-world case study on post-diaster forest assessment (in Section~\ref{ssec:case_study}), showing how generalized 3D spatial join can delineate the tornado corridor from downed trees.} 
The main contributions are summarized as follows:
\begin{itemize}
 \item 3DPipe exploits GPU parallelism across both voxel-pair filtering and facet-level refinement stages, with (i)~chunked streaming to handle datasets exceeding GPU memory, and (ii)~a pipelined execution strategy to overlap data transfer, CPU and GPU computations to improve GPU utilization.
 \item 3DPipe adopts a block-centric design, where each GPU (thread) block serves as a basic processing unit to process object pairs (resp.\ voxel pairs) during voxel-pair filtering (resp.\ facet-level refinement). Distance-bound aggregations (used for candidate pruning and early termination) are conducted in the fast shared memory using block-wise Hillis-Steele scan~\cite{DBLP:journals/cacm/HillisS86}.
 \item 3DPipe utilizes a workload flattening strategy to evenly distribute the all voxel-pair (resp.\ facet-pair) computations for each object pair (resp.\ voxel pair) among the threads of a block, effectively minimizing the number of kernel launches. 
 \item Extensive experiments demonstrate the efficiency and scalability of 3DPipe, achieving up to 9.0$\times$ end-to-end speedup over TDBase, the current state-of-the-art GPU solution.
\end{itemize}

The rest of this paper is organized as follows. Section~\ref{sec:preliminaries} presents the preliminaries, including our offline preprocessing of 3D objects and the used GPU algorithms. 
Then, Section~\ref{sec:method} introduces the detailed design of our 3DPipe framework. 
Finally, Section~\ref{sec:experiment} reports our experiments, Section~\ref{sec:related_work} reviews the related work, and Section~\ref{sec:conclusion} concludes this paper and discusses the future work.

\vspace{-1mm}
\section{Preliminaries}\label{sec:preliminaries}
We explain the preliminaries that are used by our algorithms to be presented in Section~\ref{sec:method}, including an offline preprocessing stage for object indexing, a brief review of GPU basics and advanced features, and some GPU algorithms used by 3DPipe as basic modules.

\vspace{-2mm}
\subsection{Offline Data Preprocessing}\label{ssec:offline}
We assume each 3D object is represented as a polyhedron composed of multiple polygonal faces. We also assume each face is a triangle with 3 vertices (called as a facet), since we can always divide a polygon into multiple triangles and regard them as different facets. 

Given two polyhedra $P_1$ and $P_2$, we define their distance as $d(P_1, P_2) = \min_{p_1 \in P_1,\, p_2 \in P_2} \|p_1 - p_2\|_2$, i.e., the minimum Euclidean distance between any pair of points on their surfaces.

\vspace{1mm}
\noindent{\bf Object Voxelization.} 
We preprocess each object by clustering its facets into a set of voxels. 
Given a target number of voxels $k$, we represent each facet by its centroid and apply $k$-means clustering over these centroids to group facets into $k$ clusters. 
Each cluster corresponds to a voxel that aggregates spatially proximate facets. 
For efficient preprocessing, we perform only two iterations of $k$-means, i.e., the cluster centroids are updated twice. 
The initial $k$ centroids are uniformly sampled from the vertices of the polyhedron.

\vspace{1mm}
\noindent{\bf Object and Voxel Centers.} 
We leverage a simple yet effective upper bound for object- and voxel-level distance-based pruning (recall Figure~\ref{fig:voxel_def}). 
Specifically, selecting one vertex from each side (either two objects or two voxels) and computing their Euclidean distance yields an upper bound of the minimum distance between the corresponding pair. 
This bound enables early result confirmation and candidate pruning for object pairs and voxel pairs.

To facilitate such computation, we associate each object and each voxel with a representative point, referred to as its {\bf anchor} (point). 
For an object, if its MBB center lies inside the polyhedron, the center is directly used as the anchor; otherwise, the anchor is selected as the closest polyhedron vertex to the MBB center. 
For a voxel, we similarly compute the center of its bounding box, and then select its closest vertex among all facets assigned to the voxel.

Given a pair of objects (or voxels), the distance between their anchors provides an efficient upper bound of their minimum distance, which can be computed with negligible overhead (since anchors are precomputed offline) and enables effective pruning.

\begin{figure}[!t]
\centering
\includegraphics[width=0.9\columnwidth]{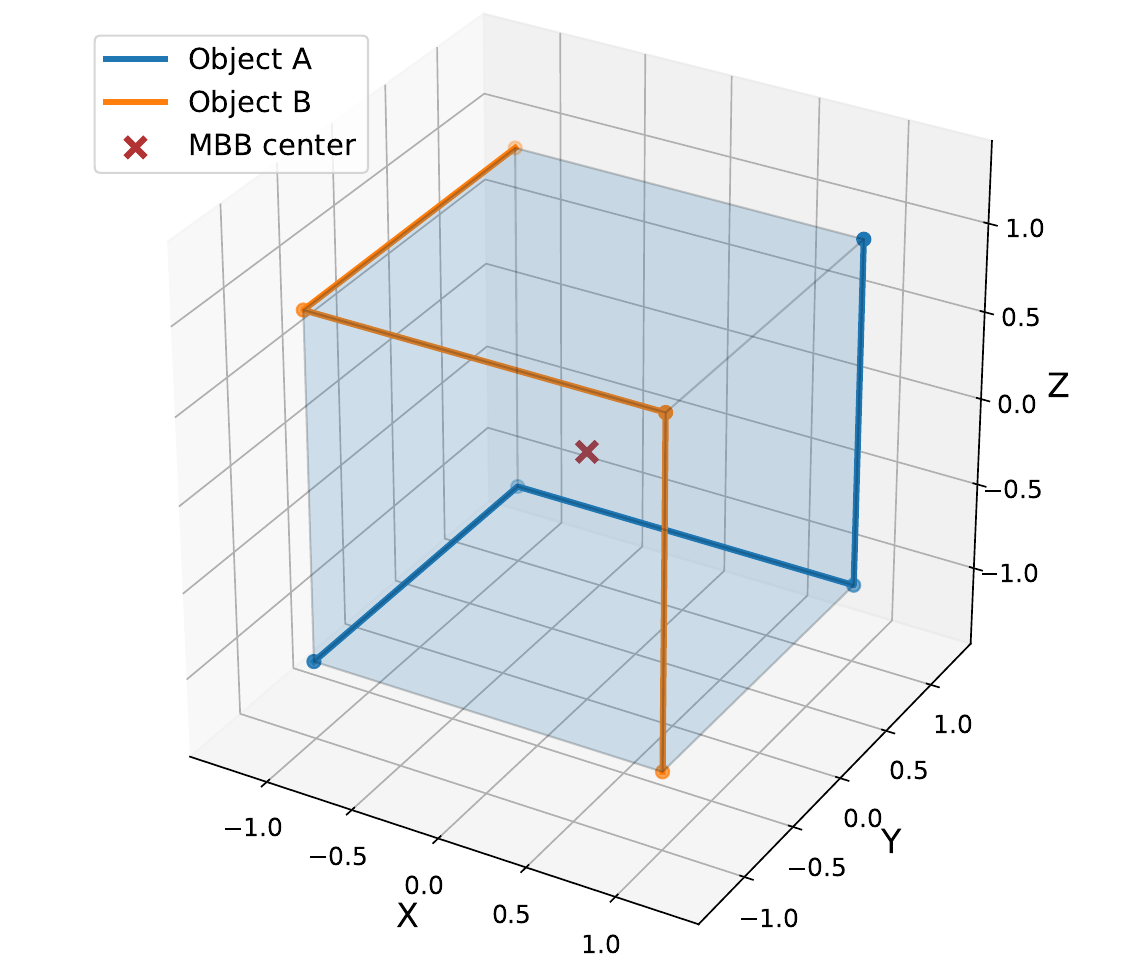}
\vspace{-3mm}
\caption{Failure Case of Upper-Bounding with MBB Centers}\label{fig:failure}
\vspace{-7mm}
\end{figure}

Note that TDBase computes the distance upper bound directly as the distance between bounding box centers, which may lead to  incorrect pruning decisions.
Figure~\ref{fig:failure} shows two objects (e.g., blood vessels) with coincident MBB centers while they remain separated, resulting in a non-zero minimum distance but a zero center-to-center distance. Our anchor selection avoids this issue by ensuring that an anchor always lies on (or within) the object geometry.

\vspace{1mm}
\noindent{\bf Level of Detail (LoD).} 
Following 3DPro and TDBase, we adopt a multi-resolution representation by constructing progressively simplified polyhedra via iterative mesh simplification~\cite{DBLP:journals/csur/MagloLDH15}, where vertices are removed to minimize approximation error and the resulting holes are filled with new facets to preserve connectivity. 
We materialize a sequence of LoDs along this process, from coarse to fine, with the finest level being the original polyhedron.

Simplification strategies differ in geometric guarantees. 
3DPro~\cite{3dpro} removes only protruding vertices, ensuring each LoD is contained within its higher-resolution counterpart. 
This monotonic containment yields progressively tighter lower bounds for pruning, but may degrade shape fidelity. 
In contrast, PPMC~\cite{DBLP:journals/cg/MagloCAH12} minimizes approximation error without enforcing containment, producing higher-quality approximations but lacking valid distance bounds.
We follow TDBase~\cite{tdbase} and decouple simplification from distance bounding using facet-level Hausdorff bounds (introduced next), enabling progressive tightening of both lower and upper bounds across LoDs while supporting arbitrary simplification algorithms. 
As in TDBase, we adopt PPMC due to its strong approximation quality.

\vspace{1mm}
\noindent{\bf Consistent Voxelization across LoDs.} We require voxel partition to remain consistent across all LoDs, since voxel-pair filtering relies on distance bounds computed over voxel pairs (see Figure~\ref{fig:voxel_def}), and its pruning decisions are only valid if the underlying voxelization is invariant throughout the multi-resolution refinement process. We enforce a fixed voxel partition by performing $k$-means clustering on the coarsest LoD to determine voxel boundaries, which are then shared across all LoDs. We track the correspondence between facets across LoDs induced by the facet-splitting mesh simplification process, allowing each facet at any LoD to be consistently mapped to its associated voxel. Finally, we compute anchor points for voxels on the original object resolution (highest LoD).

\vspace{1mm}
\noindent{\bf Facet-Level Hausdorff Bounds.} 
To enable valid distance bounding under arbitrary LoD simplification, we quantify the geometric deviation between a simplified polyhedron and its original counterpart using Hausdorff distances. 
Intuitively, given two objects $P_1$ and $P_2$ and their simplified versions $P'_1$ and $P'_2$, the true distance $d(P_1, P_2)$ can be bounded by the distance between $P'_1$ and $P'_2$, adjusted by their respective approximation errors.

A straightforward approach is to use polyhedron-level Hausdorff distances. However, such bounds are often too loose, as the approximation error may vary significantly across different regions of an object (e.g., concave vs. convex areas), leading to ineffective pruning~\cite{tdbase}. 
To obtain tighter bounds, we follow TDBase~\cite{tdbase} and refine the error quantification to the facet level. 
Specifically, for each facet $f'$ in a low-LoD polyhedron $P'$, we precompute two directional distances: (i)~Hausdorff distance $hd(f', P)$, measuring how far the facet deviates from the original object, and (ii)~proxy Hausdorff distance $ph(P, f')$, which helps derive a tighter lower bound than using $hd(P, f')$. See~\cite{tdbase} for the their detailed definitions. 

These bidirectional distances allow us to derive tighter lower and upper bounds by considering all facet pairs between two objects. Specifically, \cite{tdbase} proves the following distance bounds:
\begin{equation}\label{eq:e1}
d(P_1, P_2) \le \min_{f_1' \in P_1', f_2' \in P_2'} \{ d(f_1', f_2') + hd(f_1', P_1) + hd(f_2', P_2) \},
\end{equation}
\begin{equation}\label{eq:e2}
d(P_1, P_2) \ge \min_{f_1' \in P_1', f_2' \in P_2'} \{ d(f_1', f_2') - ph(P_1, f_1') - ph(P_2, f_2') \}.
\end{equation}
As the resolution (LoD) of $P'_1$ and $P'_2$ increases, $hd(\cdot)$ and $ph(\cdot)$ decrease, yielding progressively tighter bounds on $d(P_1, P_2)$.

In summary, for each facet $f'$ in each simplified polyhedron $P'$ (i.e., each LoD), we precompute and store two values offline: $f'.hd = hd(f', P)$ and $f'.ph = ph(P, f')$. 
These values are later used to compute tight distance bounds during progressive refinement.

\vspace{-2mm}
\subsection{CUDA Concepts and GPU Algorithms}\label{ssec:cuda}

\begin{figure}[!t]
\centering
\includegraphics[width=0.88\columnwidth]{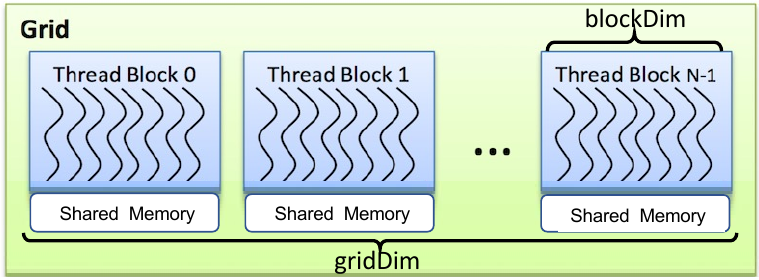}
\vspace{-3mm}
\caption{GPU and CUDA Concepts}\label{fig:threadblock}
\vspace{-6mm}
\end{figure}

\noindent{\bf CUDA Basics.} 
The host (CPU) launches a {\bf kernel} to execute massively parallel computation on the GPU, using the syntax \texttt{<<<gridDim, blockDim>>>}. 
As Figure~\ref{fig:threadblock} shows, the kernel invocation creates a grid of \texttt{gridDim} thread blocks, where each block contains \texttt{blockDim} threads.
Threads are the basic execution units that execute the same kernel function, identified by their intra-block indices \texttt{threadIdx} (starting from 0), while blocks are identified by their indices \texttt{blockIdx} within the grid (also starting from 0). 
A global thread ID in the grid can be computed as \texttt{blockIdx\,*\,blockDim + threadIdx}. 

On the hardware side, a GPU consists of multiple streaming multiprocessors (SMs), each containing multiple CUDA cores. 
Thread blocks are scheduled onto SMs for execution. Threads within a block execute on the same SM and can efficiently cooperate via fast on-chip {\bf shared memory} and barrier synchronization.
Execution is further organized in units of \textbf{warps}, each consisting of 32 threads. 
Threads in a warp execute in an SIMT fashion, i.e., the same instruction is issued to all threads in the warp. 
A block is thus decomposed into multiple warps, which are scheduled independently by the SM.

\setcounter{figure}{5}
\begin{figure*}[!t]
\centering
\includegraphics[width=1.5\columnwidth]{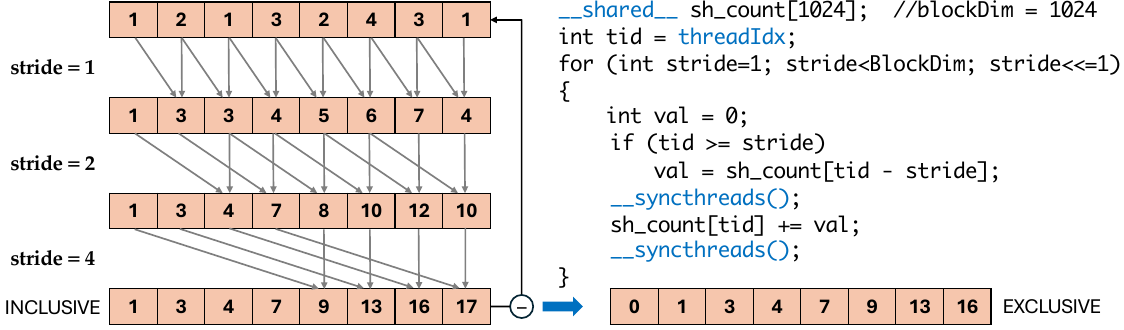}
\vspace{-2mm}
\caption{An Illustration of the Hillis-Steele Scan}\label{fig:prefix}
\vspace{-5mm}
\end{figure*}

\vspace{1mm}
\noindent{\bf CUDA Streams.} 
In a typical CUDA program, memory copy between host and GPU (\texttt{cudaMemcpy}) are blocking, resulting in a strictly sequential execution between host-device data transfer and kernel execution.
To enable concurrency, CUDA introduces {\bf streams}, where each stream represents an ordered sequence of operations (memory copies or kernels) executed on the GPU.
Operations within the same stream are executed in issue order, while operations from different streams may overlap. 
CUDA provides asynchronous primitives such as \texttt{cudaMemcpyAsync} and stream-based kernel launches, allowing data transfer and computation to proceed without blocking the host (which issues the operations to streams).
To ensure true asynchrony for memory transfers, host buffers are typically allocated as {\bf pinned memory} (\texttt{cudaMallocHost}).

Dependencies across streams are managed using {\bf events}.
An event can be recorded at a specific point in a stream via \texttt{cudaEvent-\\Record}, 
which marks the completion of all preceding operations in that stream.
Another stream can invoke \texttt{cudaStreamWaitEvent} on this event, ensuring that its subsequent operations will not start until the recorded point is reached. 
This mechanism enables fine-grained cross-stream synchronization without global barriers.

\setcounter{figure}{4}
\begin{figure}[!t]
\centering
\vspace{1mm}
\includegraphics[width=0.8\columnwidth]{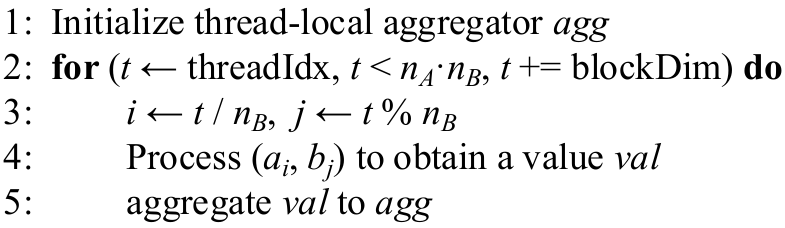}
\vspace{-1mm}
\caption{Workload Flattening \& Thread-Local Aggregation}\label{fig:agg}
\vspace{-6mm}
\end{figure}

\setcounter{figure}{6}

\vspace{1mm}
\noindent{\bf Pairwise Workload Flattening.} In both filtering and refinement, a thread block is assigned two local sets and needs to process all cross-set pairs between them.
This occurs when enumerating all voxel pairs of an object pair in filtering, and all facet pairs of a voxel pair in refinement. 
Given two sets $A=\{a_0,\dots,$ $a_{n_A-1}\}$ and $B=\{b_0,\dots,b_{n_B-1}\}$ assigned to a thread block, we need to process all the $n_An_B$ pairs of their elements $(a_i,b_j)$.
To evenly distribute this work among the $\texttt{blockDim}$ threads, we flatten the 2D pair space into a 1D index range in row-major order.
Specifically, the pair $(a_i,b_j)$ is mapped to the linear index $t = i \cdot n_B + j$. 

As Lines 2--4 in Figure~\ref{fig:agg} illustrates, the linear indices are assigned to all threads in a round-robin manner to achieve a balanced workload within the block:
thread $\texttt{threadIdx}$ processes $t=\texttt{threadIdx}, \texttt{threadIdx}+\texttt{blockDim}, \texttt{threadIdx}+2\cdot\texttt{blockDim}, \dots$,
until all $n_A n_B$ pairs are covered. For each assigned index $t$, the thread recovers the corresponding pair indices by computing $i = \lfloor t / n_B \rfloor$ and $j = t \bmod n_B$,
and then reads $a_i$ and $b_j$ for computation.

\vspace{1mm}
\noindent{\bf Fast Blockwise Aggregation in Shared Memory.} Our algorithm frequently needs to aggregate the thread-local results of pairwise computations by taking the minimum or summation. This can be achieved in two steps.
In the {\bf first} step, as shown in Figure~\ref{fig:agg}, we let each thread of a block aggregate all the results of its pairwise computations onto its thread-local aggregator variable {\em agg}. 

In the {\bf second} step, the \texttt{blockDim} threads of the block then write their aggregated values from Step~1 onto an array (e.g., \texttt{sh\_count} in Figure~\ref{fig:prefix}) in the block's shared memory for fast aggregation.

This is achieved by the Hillis-Steele scan as illustrated in Figure~\ref{fig:prefix} for summation aggregation, which completes in $\log(\texttt{blockDim})$ rounds. At round $i$, the stride equals $2^i$, and each thread $\texttt{tid}$ (when $\texttt{tid} \ge \texttt{stride}$) updates its value by adding the element located $\texttt{stride}$ positions before it, thereby maintaining the invariant that, after round $i$, each position $\texttt{tid}$ stores the sum over the range $[\max(0, \texttt{tid} - 2^i + 1),\ \texttt{tid}]$ in the initial array. Block-level synchronization barriers are inserted between rounds to ensure that all threads observe a consistent view of shared memory, preventing read-after-write hazards due to independently scheduled warps.

After all rounds, each position contains the sum over all preceding elements in the initial array, yielding the final {\bf inclusive} prefix sum. 
In particular, the rightmost element stores the final block-level aggregation result. 
While Figure~\ref{fig:prefix} illustrates the summation case, the same procedure can be directly applied to other associative operations such as minimum aggregation, which is useful for aggregating facet-pair distances (or their bounds) for an object pair to determine its object-pair distance (or its bounds). Notably, TDBase performs such aggregation in global memory which is much slower than shared memory, and multiple threads update a shared result using \texttt{atomicMin}, which leads to significant contention overhead.

When the aggregation operator is summation, the inclusive prefix sums produced by the Hillis-Steele scan can be converted to {\bf exclusive} prefix sums by subtracting the original thread-local value at each position.
This variant is particularly useful for computing per-thread output offsets within a block, so that each thread can independently determine its write position in a shared output buffer and avoid write conflicts when emitting
multiple results.

\begin{figure*}[!t]
\centering
\includegraphics[width=1.7\columnwidth]{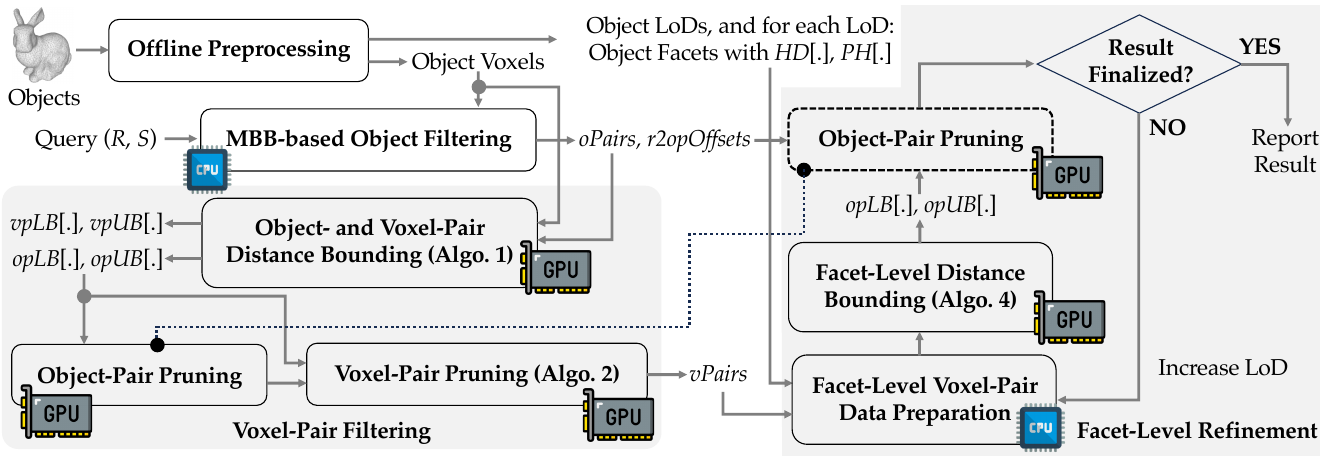}
\vspace{-3mm}
\caption{Overview of the Execution Pipeline of 3DPipe}\label{fig:overview}
\vspace{-4mm}
\end{figure*}

\vspace{1mm}
\noindent{\bf Triangle-Triangle Distance Computation.} 
We adopt the M\"{o}ller triangle-triangle distance algorithm~\cite{moller97} to compute the minimum distance between two triangles.
The algorithm decomposes the problem into two types of distance computations:
(i)~point-to-triangle distance from each vertex of one triangle to the other triangle, and (ii)~segment-to-segment distance for each pair of edges from the two triangles.
Concretely, for two triangles, this amounts to six vertex-to-triangle checks (three vertices from each triangle) and nine edge-to-edge checks ($3\times 3$ edge pairs).
The global minimum triangle-triangle distance is guaranteed to be found among these 15 candidate distances~\cite{moller97} by taking their minimum.

In the refinement stage, we assign one thread to each facet pair within a voxel pair. Since every thread executes the same M\"{o}ller routine with the same fixed sequence of 15 candidate distance checks, the computation is highly regular and thus embarrassingly parallel on the GPU. This thread-per-facet-pair design is also adopted by prior GPU-based methods such as 3DPro and TDBase.




\vspace{-2mm}
\section{The 3DPipe Framework}
\label{sec:method}

Our 3DPipe framework supports GPU-accelerated 3D spatial join between $R$ and $S$ for three types of queries: 
{\bf (1)~intersection:} which for each object $r \in R$, retrieves all objects $s \in S$ that intersects with $r$, i.e., $d(r, s) = 0$; 
{\bf (2)~within-$\tau$ distance:} which for each $r \in R$, retrieves all $s \in S$ with $d(r, s) \le \tau$; 
{\bf (3)~$k$-NN:} which for each $r \in R$, retrieves $k$ objects in $S$ that are closest to $r$.

In this section, we first overview the execution pipeline of 3DPipe, and then present the details of its voxel-pair filtering, facet-level refinement stages, and additional techniques for $k$-NN queries.

\vspace{-2mm}
\subsection{Overview}
Figure~\ref{fig:overview} presents the overall execution pipeline of 3DPipe, where operations involving GPU processing are marked with a GPU icon.

\vspace{1mm}
\noindent{\bf $\bullet$ Offline Processing.} Each polyhedral object is first preprocessed as described in Section~\ref{ssec:offline}. This includes (1)~voxelizing the object, (2)~computing anchor points for the object MBB and its voxels, and (3)~progressively simplifying the polyhedron $P$ via iterative mesh simplification~\cite{DBLP:journals/cg/MagloCAH12} to obtain a sequence of representations with increasing levels of detail (LoDs), and for each simplified polyhedron (i.e., LoD)  $P'$, we compute the Hausdorff distance $hd(f, P)$ and proxy Hausdorff distance $ph(P, f)$ for each facet $f\in P'$.

\vspace{1mm}
\noindent{\bf $\bullet$ MBB-based Object Filtering.} 
Given a join query over object sets $R$ and $S$, we first perform lightweight MBB-based filtering on the CPU using an R-tree built on $S$, denoted by $T_S$.

For a within-$\tau$ distance query, for each object $r \in R$, we traverse $T_S$ from the root and recursively visit a child node $N$ only if $\mathrm{MINDIST}(r, N) \le \tau$. 
If a leaf node corresponding to an object $s$ is reached, we obtain lightweight distance bounds of $d(r, s)$ using their MBBs, denoted by $[lb, ub]$. This yields three possible cases: 
(1)~if $ub \le \tau$, $(r, s)$ is directly added to the result set; 
(2)~if $lb > \tau$, $(r, s)$ is safely pruned; 
(3)~otherwise, the pair $(r, s)$ remains undecided. All undecided object pairs form a candidate set \texttt{oPairs} to be subsequently processed by the voxel-pair filtering module.

For a $k$-NN query, we compute the $k$ nearest neighbors for each $r \in R$ over $T_S$ using a variant of the best-first search algorithm~\cite{DBLP:conf/sigmod/RoussopoulosKV95}. Specifically, nodes are expanded in ascending order of $\mathrm{MINDIST}$ maintained in a priority queue. The search terminates when the smallest $\mathrm{MINDIST}$ in the queue exceeds a threshold $\theta$. Since only MBBs are considered for objects $s$, the exact distance $d(r,s)$ is bounded by an interval $[lb, ub]$. Therefore, we set $\theta$ to the $k$\textsuperscript{th} smallest upper bound among the current candidate objects. All candidate object pairs retained in the queue at termination form the set \texttt{oPairs} to be processed by the voxel-pair filtering module.

\vspace{1mm}
\noindent{\bf $\bullet$ Voxel-Pair Filtering.} Given candidates \texttt{oPairs}, we next perform a finer-grained voxel-level filtering. This stage enables additional object-pair filtering and, for the remaining pairs, applies voxel-pair pruning to reduce the cost of subsequent facet-level refinement.

The voxel-pair filtering module has three stage. First, the {\bf object- and voxel-pair distance bounding} stage utilizes GPU to compute pairwise distance bounds for all voxel pairs of each object pair in
\texttt{oPairs}. Specifically, for each object pair, we enumerate all cross-object voxel pairs, compute a lower bound and an upper bound for each voxel pair, and store them in \texttt{vpLB} and
\texttt{vpUB}, respectively. At the same time, these voxel-pair bounds are (minimum-)aggregated into object-pair bounds, stored in \texttt{opLB} and \texttt{opUB}. 

Next, the {\bf object-pair pruning} stage uses the object-pair bounds \texttt{opLB} and \texttt{opUB} (which are tighter than MBB-based bounds) to prune pairs that can be safely discarded and to directly move those that can be determined to the final result set. 

Finally, the {\bf voxel-pair pruning} stage utilizes GPU to prune cross-object voxel pairs whose lower bounds exceed the current object-pair upper bound, 
since such voxel pairs cannot contribute to the final object-pair minimum distance. 
The remaining voxel pairs form a reduced set \texttt{vPairs} for subsequent facet-level refinement.

\vspace{1mm}
\noindent{\bf $\bullet$ Facet-Level Refinement.} Given the reduced voxel-pair set \texttt{vPairs}, we proceed to the final refinement stage to compute exact object distances. Refinement proceeds progressively over LoDs from coarse to fine. {\bf At each LoD}, the CPU prepares the facet data associated with the current \texttt{vPairs} and transfers them to the GPU. 
The {\bf facet-level distance bounding} stage then computes facet-pair distance bounds (using Eqs.~\eqref{eq:e1}--\eqref{eq:e2}) for all facet pairs within each voxel pair, and aggregates them to update object-pair bounds \texttt{opLB} and \texttt{opUB}. 

For $k$-NN queries, {\bf object-pair pruning} is further invoked on GPU to use the tightened object-pair bounds \texttt{opLB} and \texttt{opUB} to prune object pairs that can be safely discarded and to directly move those that can be determined to the final result set.

\begin{figure*}[!t]
\centering
\includegraphics[width=2.1\columnwidth]{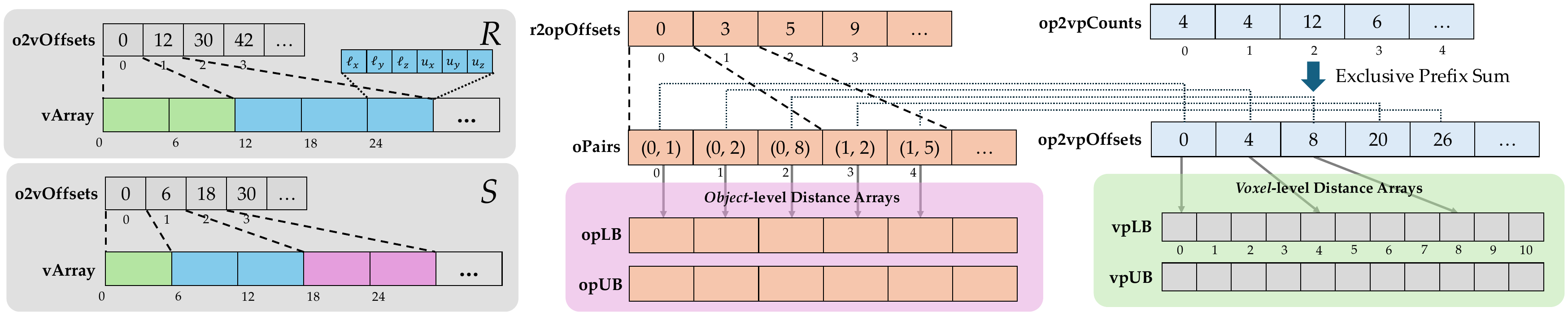}
\vspace{-6mm}
\caption{Data Structures for Object Voxels (Left), Object-Level (Middle) and Voxel-Level (Right) Distance Bounds}\label{fig:filter_data}
\vspace{-3mm}
\end{figure*}

After each LoD, we check whether every object pair can be resolved based on the current object-pair bounds; if so, the results are reported immediately. Otherwise, we advance to the next finer LoD and repeat the refinement process until the original-resolution geometry is reached, at which point the bounds become exact object-pair distances and all remaining object pairs can be resolved.

\vspace{-4mm}
\subsection{GPU Algorithms for Voxel-Pair Filtering}\label{sec:3.2}
Recall that given a query $(R, S)$, MBB-based object filtering searches R-tree $T_S$ to identify object pairs $(r,s)$ that can be directly included in the results. The undecided candidate pairs are then transferred to GPU global memory, where their objects and associated voxels are organized using the data structures shown in Figure~\ref{fig:filter_data} (left). 

Specifically, undecided object pairs are stored in \texttt{oPairs} shown in Figure~\ref{fig:filter_data} (middle), with \texttt{r2opOffsets} indexing the candidates associated with each object $r \in R$. The voxels of objects in $R$ and $S$ are stored in contiguous arrays \texttt{vArray}, with \texttt{o2vOffsets} indicating the starting location for reading voxel coordinates of each object. Each voxel is represented by six coordinates, corresponding to the minimum and maximum values along the $x$, $y$, and $z$ axes. 

Note that for each object $r\in R$, the number of voxels can be obtained by computing $n_r = (\texttt{o2vOffsets}[r+1] - \texttt{o2vOffsets}[r])/6$. We can similarly compute the number of voxels, $n_s$, for object $s\in S$.

\begin{algorithm}[t]
\caption{Voxel-Pair Distance Bounding Kernel}
\label{alg:vp_kernel}
\begin{algorithmic}[1]
\STATE $(r,s) \gets \texttt{oPairs}[\texttt{blockIdx}]$
\STATE $n_r, n_s \gets$ voxel counts of $r$ and $s$ from \texttt{o2vOffsets}
\STATE $\texttt{offset} \gets \texttt{op2vpOffsets}[\texttt{blockIdx}]$
\STATE $N \gets \texttt{op2vpCounts}[\texttt{blockIdx}]$ \quad \textcolor{blue}{//$N=n_r\cdot n_s$}
\FOR{$t \gets \texttt{threadIdx}$; $t < N$; $t$ $+$$=$ $\texttt{blockDim}$}
    \STATE $i \gets t / n_s$, \quad $j \gets t \bmod n_s$
    \STATE Load the $i$-th voxel of $r$ and $j$-th voxel of $s$ from \texttt{vArray}
    \STATE Compute $(lb, ub)$ for the voxel pair using their coordinates
    \STATE \texttt{vpLB}[\texttt{offset} $+$ $t$] $\gets lb$, \quad \texttt{vpUB}[\texttt{offset} $+$ $t$] $\gets ub$
\ENDFOR
\STATE \texttt{opLB}[\texttt{blockIdx}] $\gets$ blockwise minimum of \texttt{vpLB}
\STATE \texttt{opUB}[\texttt{blockIdx}] $\gets$ blockwise minimum of \texttt{vpUB}
\end{algorithmic}
\end{algorithm}
\setlength{\textfloatsep}{5pt}

\vspace{1mm}
\noindent\textbf{Object- and Voxel-Pair Distance Bounding.}
For each object pair $(r,s)$, this stage computes lower and upper distance bounds 
for all voxel pairs between $r$ and $s$. The results are stored in 
\texttt{vpLB} and \texttt{vpUB}, shown in Figure~\ref{fig:filter_data} (right). 
Specifically, the host first computes an array \texttt{op2vpCounts}, which is aligned one-to-one with \texttt{oPairs}. 
For each $(r,s)$, the corresponding entry records the number of voxel pairs, i.e., $n_r \cdot n_s$. 
The host then transfers \texttt{op2vpCounts} to the GPU, where a parallel exclusive prefix sum is performed 
using an optimized GPU scan primitive (i.e., CUB DeviceScan) to produce \texttt{op2vpOffsets}, which defines, for each object pair $(r,s)$, 
the starting location in \texttt{vpLB} and \texttt{vpUB} for writing the voxel-pair bounds.

Given \texttt{op2vpOffsets}, we compute voxel-pair distance bounds on the GPU with $\texttt{gridDim}=|\texttt{oPairs}|$, as shown in Algorithm~\ref{alg:vp_kernel}. 
Each thread block is assigned to an object pair $(r,s)$ and enumerates all cross-object voxel pairs via workload flattening. 
For each voxel pair, we compute its lower and upper distance bounds, stored in \texttt{vpLB} and \texttt{vpUB}, respectively. 
Meanwhile, these bounds are aggregated within each thread block to obtain object-pair bounds \texttt{opLB} and \texttt{opUB}, 
using the fast blockwise aggregation technique in shared memory described in Section~\ref{ssec:cuda}.

\vspace{1mm}
\noindent\textbf{Object-Pair Pruning.} Recall that Lines~11--12 have computed tightened object-pair distance bounds $[lb, ub]$. 
For \emph{within-$\tau$ distance} queries, we use CPU to classify the object pairs into three cases: (i)~if $ub \le \tau$, the pair can be directly reported as a result; 
(ii)~if $lb > \tau$, the pair is safely pruned; and (iii)~otherwise, the pair remains undecided and is forwarded to the subsequent voxel-pair pruning stage for finer-grained processing. 
For $k$-NN queries, object-pair filtering follows a different paradigm since no fixed threshold $\tau$ is given; instead, candidates are maintained and progressively pruned based on relative distance bounds. We defer the  discussion to Section~\ref{sec:3.4}.

\vspace{1mm}
\noindent\textbf{Voxel-Pair Pruning.} 
Despite prior filtering, the number of voxel pairs within the remaining object pairs can still be substantial.
This stage leverages the voxel-pair distance bounds \texttt{vpLB} and \texttt{vpUB} computed by Algorithm~\ref{alg:vp_kernel} to prune voxel pairs that cannot influence the final object-pair distances (recall Figure~\ref{fig:voxel_def}).
Specifically, given a voxel pair with distance bound $[lb_v, ub_v]$ (obtained from \texttt{vpLB} and \texttt{vpUB}) and the aggregated bound $[lb_o, ub_o]$ of its corresponding object-pair (obtained from \texttt{opLB} and \texttt{opUB}), we discard the voxel~pair 
if $lb_v > ub_o$. This voxel-pair pruning stage significantly reduces the workload of the subsequent facet-level refinement, and it is im-plemented 
using three GPU kernels which are invoked sequentially.


\begin{figure}[!t]
\centering
\includegraphics[width=0.84\columnwidth]{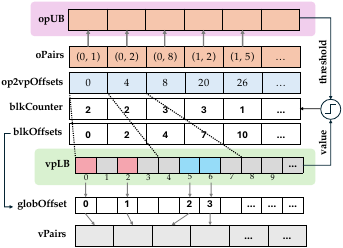}
\vspace{-3mm}
\caption{Voxel-Pair Pruning by Object-Pair Upper Bound}\label{fig:algo2}
\vspace{-1mm}
\end{figure}

Algorithm~\ref{algo:voxel_prune} shows this three-step procedure for voxel-pair pruning, which is illustrated in Figure~\ref{fig:algo2}. 
Object pairs that have already been pruned in the preceding stage are skipped in this step; their corresponding thread blocks simply return without performing any computation. For simplicity, we omit this detail in Algorithm~\ref{algo:voxel_prune}.

In the first kernel, each block processes one object pair and scans its voxel pairs to count the number of valid pairs satisfying $lb_v \le ub_o$ by workload flattening (Lines~3--5). The per-thread counts are aggregated within the block via shared-memory Hillis-Steele scan to obtain a block-level counter, stored in \texttt{blkCounter} (Line~6).

Next, the second kernel performs an exclusive prefix sum over \texttt{blkCounter} to compute \texttt{blkOffsets} using CUB DeviceScan, which determines the starting write position of each thread block in the global output array \texttt{vPairs}. 
Finally, in the third kernel, each thread block re-scans its voxel pairs and writes the surviving pairs to \texttt{vPairs}. 
Within each block, the valid voxel pairs are first counted by per-thread \texttt{threadCounter} values (Line~10), and the resulting \texttt{localOffsets} are computed via a shared-memory Hillis-Steele scan (Line~11). 
Each thread then uses its \texttt{localOffset} to determine the write position, and the final global offset is computed as
$\texttt{globOffset}=\texttt{blkOffsets}[\texttt{blockIdx}]+\texttt{localOffset}$. 
This design ensures that all valid voxel pairs are written into a contiguous array without conflicts, enabling coalesced memory access.

\begin{algorithm}[t]
\caption{Voxel-Pair Pruning}
\label{algo:voxel_prune}
\begin{algorithmic}[1]
\STATE \textcolor{blue}{// Kernel 1}
\STATE Use \texttt{blockIdx} to obtain $(r,s)$, $n_r$, $n_s$, \texttt{offset} and $N$, the same as in Algorithm~\ref{alg:vp_kernel} Lines 1--4; \quad $ub_o \leftarrow \texttt{opUB}[\texttt{blockIdx}]$
\STATE \texttt{threadCounter} $\leftarrow 0$
\FOR{$t \leftarrow \texttt{threadIdx}; \ t < N; \ t \mathrel{+}= \texttt{blockDim}$}
    \STATE \textbf{if} \texttt{vpLB}[$\texttt{offset}+t$] $\le ub_o$ \textbf{then} \quad $\texttt{threadCounter} \mathrel{++}$
\ENDFOR
\STATE \texttt{blkCounter}[\texttt{blockIdx}] $\gets$ blockwise sum of \texttt{threadCounter}
\STATE \textcolor{blue}{// Kernel 2}
\STATE \texttt{blkOffsets} $\leftarrow$ exclusive prefix sum of \texttt{blkCounter}
\STATE \textcolor{blue}{// Kernel 3}
\STATE Perform the same operations as in Lines~2--5
\STATE \texttt{localOffsets} $\leftarrow$ exclusive prefix sum of \texttt{threadCounter}
\STATE \texttt{localOffset} $\leftarrow \texttt{localOffsets}[threadIdx]$
\FOR{$t \leftarrow \texttt{threadIdx}; \ t < N; \ t \mathrel{+}= \texttt{blockDim}$}
    \STATE $lb_v \leftarrow \texttt{vpLB}[\texttt{offset}+t]$
    \STATE \textbf{if} $lb_v \le ub_o$ \textbf{then}
        \STATE \hspace{1em}$i \gets t / n_s$, \quad $j \gets t \bmod n_s$ \quad \textcolor{blue}{// Get Object IDs}
        \STATE \hspace{1em}\texttt{globOffset} $\leftarrow$ \texttt{blkOffsets}[\texttt{blockIdx}] $+$ \texttt{localOffset}
        \STATE \hspace{1em}\texttt{vPairs}[\texttt{globOffset}] $\leftarrow (i, j)$
        \STATE \hspace{1em}$\texttt{localOffset} \mathrel{++}$
\ENDFOR
\end{algorithmic}
\end{algorithm}
\setlength{\textfloatsep}{5pt}

\vspace{1mm}
\noindent\textbf{Chunked Streaming.} The major memory cost of voxel-pair processing comes from the voxel-pair distance-bound arrays \texttt{vpLB} and \texttt{vpUB}, which can be prohibitively large. For example, our experiments show that the total number of voxel pairs can reach 15 billion (see Figure~\ref{exp:stream}), requiring about $15 \times 2 \times 4 = 120$~GB to store \texttt{vpLB} and \texttt{vpUB}, exceeding the memory capacity of commodity GPUs.
To address this issue, we process object pairs in a {\bf chunked} manner. 
Specifically, we maintain fixed-size buffers on GPU for \texttt{vpLB} and \texttt{vpUB}, and iteratively load a subset of object pairs whose total number of voxel pairs fits within the buffers. 
For each chunk, we invoke Algorithms~\ref{alg:vp_kernel} and~\ref{algo:voxel_prune} to compute and prune voxel pairs, and send the surviving pairs back to the host to be appended to \texttt{vPairs}. 

A na\"ive alternative is to rely on {\em unified memory} (allocated via \texttt{cudaMallocManaged}) to transparently migrates data (i.e., \texttt{vpLB}, \texttt{vpUB}, \texttt{vPairs}) between host and device memory. 
However, unified memory operates at the granularity of pages, whereas voxel-pair processing exhibits large, sequential access over GB-scale arrays. 
As a result, data is migrated in a page-by-page manner, leading to a massive number of page faults that incur severe overhead.

\vspace{1mm}
\noindent\textbf{Double-Stream Pipelining.} While chunked streaming effectively bounds GPU memory usage, a na\"ive implementation suffers due to synchronous device-to-host (D2H). In particular, after processing each chunk, the valid voxel pairs must be copied back to host memory before the next chunk can be processed, leaving the GPU idle during the transfer. To further improve throughput, we overlap D2H transfer with GPU computation using two CUDA streams: a \texttt{compute} stream for kernel execution and a \texttt{memcpy} stream for asynchronous output transfer. Algorithm~\ref{algo:chunk_stream_pipe} provides the details.

\begin{algorithm}[t]
\caption{Double-Stream Chunked Streaming}
\label{algo:chunk_stream_pipe}
\begin{algorithmic}[1]
\STATE Initialize two CUDA streams: \texttt{compute} and \texttt{memcpy}
\STATE $begin \leftarrow 0$, $iter \leftarrow 0$
\STATE Allocate \texttt{vpLB}, \texttt{vpUB} buffers of size \texttt{CHUNK\_SIZE}
\STATE Allocate two output buffers \texttt{vBuf}[0], \texttt{vBuf}[1]
\WHILE{$begin < |\texttt{oPairs}|$}
    \STATE $end \leftarrow begin$, \quad $voxCnt \leftarrow 0$
    \STATE $curr \leftarrow iter \bmod 2$, \quad $prev \leftarrow 1 - curr$
    \WHILE{$end < |\texttt{oPairs}|$ \AND $voxCnt + \texttt{op2vpCounts}[end] \le \texttt{CHUNK\_SIZE}$}
        \STATE $voxCnt \leftarrow voxCnt + \texttt{op2vpCounts}[end]$
        \STATE $end \leftarrow end + 1$
    \ENDWHILE
    \IF{$iter > 0$}
        \STATE (Async) D2H copy of \texttt{vBuf}[$prev$] via \texttt{memcpy} stream
    \ENDIF
    \STATE (Async) launch Algorithms~\ref{alg:vp_kernel} and~\ref{algo:voxel_prune} on \texttt{compute} stream, store results in \texttt{vBuf}[$curr$]
    \STATE $begin \leftarrow end$, \quad $iter \leftarrow iter + 1$
\ENDWHILE
\end{algorithmic}
\end{algorithm}

Specifically, Line~3 allocates a single pair of \texttt{vpLB} and \texttt{vpUB} buffers for use by the \texttt{compute} stream in Line~13, which is sufficient since the \texttt{compute} stream processes one chunk at a time. 
Note that Lines~6 \& 8--10 construct the next chunk of object pairs from \texttt{oPairs} so that the total number of their voxel pairs does not exceed \texttt{CHUNK\_SIZE}, hence \texttt{vpLB} and \texttt{vpUB} buffers will not overflow. 

Line~4 allocates two output buffers \texttt{vBuf}[0] and \texttt{vBuf}[1] to store the valid voxel pairs from consecutive chunking iterations, so that when one buffer is used for computation, the other is being transferred. 
Specifically, at each iteration, we alternate between the two buffers, where \texttt{curr} stores the output of the current chunk and \texttt{prev} holds the output from the previous chunk. 
When $\texttt{iter} > 0$, the results in \texttt{vBuf}[\texttt{prev}] are asynchronously copied back to host memory via the \texttt{memcpy} stream (Lines~11--12), while the \texttt{compute} stream simultaneously processes the current chunk and writes its results to \texttt{vBuf}[\texttt{curr}] (Line~13). 
Since the host issues both operations without blocking, and CUDA allows concurrent execution across streams, the D2H transfer of one chunk overlaps with the computation of the next chunk. This design improves GPU utilization.

The double-buffering scheme is implemented using two events to coordinate the two CUDA streams. 
First, after the \texttt{compute} stream finishes producing \texttt{vBuf}[\texttt{curr}] for the current chunk in Line~13, it records a \texttt{computeDone} event. 
Before the \texttt{memcpy} stream issues the D2H transfer for that buffer in Line~12, it waits on this event to ensure that the output has been fully generated. 
Second, after the D2H copy of \texttt{vBuf}[\texttt{prev}] completes in Line~12, the \texttt{memcpy} stream records a \texttt{copyDone} event. 
When the same buffer is reused by a later iteration, the \texttt{compute} stream waits on the corresponding \texttt{copyDone} event before writing new results into it in Line~13, thereby avoiding overwrite of data that is still in transit. 

\begin{figure}[!t]
\centering
\includegraphics[width=\columnwidth]{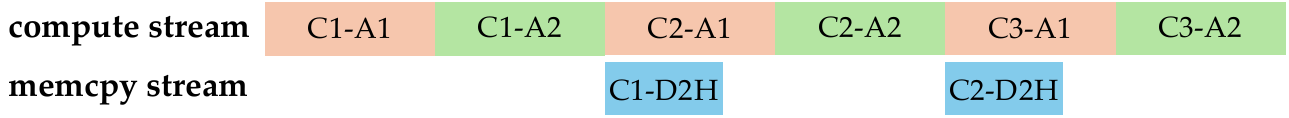}
\vspace{-5mm}
\caption{Pipelined Computation and Memcpy}\label{fig:pipeline}
\end{figure}

Figure~\ref{fig:pipeline} illustrates the pipelined execution of computation and D2H memory transfer. 
Here, ``C1'' denotes chunk~1, and ``A1'' (resp.\ ``A2'') corresponds to Algorithm~1 (resp.\ Algorithm~2). 

\vspace{1mm}
\noindent\textbf{Summary.} Although the filtering stage accounts for only a small fraction (e.g., 15\%) of the total runtime after GPU acceleration, it is critical for enabling low-latency coarse search. 
In practice, it produces the majority of final results (up to 90\%) within a few seconds, while refining the remaining undecided object pairs is significantly more expensive. 
This makes our GPU-based filtering particularly valuable in scenarios requiring partial results ASAP.

\vspace{-2mm}
\subsection{Facet-Level Refinement with GPU}\label{sec:3.3}
The filtering stage in Section~\ref{sec:3.2} produces valid voxel pairs which are further evaluated in the refinement stage. Recall from Figure~\ref{fig:voxel_def} that each voxel contains multiple triangle facets in close proximity. The refinement stage computes pairwise facet-level distances for all valid voxel pairs to obtain precise object distances. 

Recent GPU solutions 3DPro~\cite{3dpro} and TDBase~\citep{tdbase} adopt the M\"{o}ller algorithm~\cite{moller97} for triangle-triangle distance computation as described in Section~\ref{ssec:cuda}, to exploit its regular structure and massive parallelism. Specifically, they assign each triangle pair to a thread, and every thread executes the same fixed sequence of 15 candidate distance checks. This design leads to an SIMT-friendly computation.

Despite this parallel formulation, we identify two performance bottlenecks that significantly limit their efficiency. 

The first bottleneck is {\bf excessive kernel launches}. 
Given two voxels $M$ and $N$, existing approaches such as TDBase launch one kernel per triangle in $M$, where each launch computes the distances between a single triangle in $M$ and all triangles in $N$. 
As a result, computing all triangle-level distances between $M$ and $N$ requires $|M|$ kernel launches. 
This design is inefficient, as frequent kernel launches incur non-trivial scheduling overhead on the GPU.

The other bottleneck is the inefficient use of memory during aggregation. 
After computing all pairwise facet-level distances within a voxel pair, the results must be aggregated to obtain the minimum distance for the a voxel pair. 
This operation is memory-intensive and ideally suited for GPU shared memory due to its low latency and high bandwidth. 
However, existing approaches such as TDBase perform aggregation directly in the slower global memory by letting threads update current aggregated value by \texttt{atomicMin}, 
which incurs higher access latency and contention that forces serialization.

\begin{algorithm}[t]
\caption{Facet-Level Refinement} \label{algo:refine}
\textbf{Input:} \raggedright \texttt{vPairs}, facet-level Hausdorff distance arrays \texttt{HD\_r}, \texttt{HD\_s}, and facet-level proxy Hausdorff distance arrays \texttt{PH\_r}, \texttt{PH\_s}. \\
\textbf{Output:} Voxel-pair distance bounds \texttt{vpLB}, \texttt{vpUB} \\
\begin{algorithmic}[1]
\STATE $(v_r, v_s) \leftarrow \texttt{vPairs}[\texttt{blockIdx}]$ \label{refine_valid}
\STATE $n_r \leftarrow$ number of triangles in voxel $v_r$ \label{refine_s1}
\STATE $n_s \leftarrow$ number of triangles in voxel $v_s$ \label{refine_s2}
\STATE $\texttt{localLB} \leftarrow \texttt{FLT\_MAX}$,\ \ $\texttt{localUB} \leftarrow \texttt{FLT\_MAX}$
\FOR{$t \leftarrow \texttt{threadIdx};\ t < n_r \cdot n_s;\ t \mathrel{+}= \texttt{blockDim}$} \label{refine_for}
\STATE $i \leftarrow t / n_s$, \hspace{0.6em}$j \leftarrow t \bmod n_s$ \label{refine_flat}
\STATE Read coordinates of the $i$-th facet in $v_r$ and $j$-th facet in $v_s$ \label{refine_coord}
\STATE Compute their triangle-triangle distance $dist$ \label{refine_comp}
\STATE $\texttt{lb} \leftarrow dist - \texttt{PH\_r}[v_r][i] - \texttt{PH\_s}[v_s][j]$ \quad \textcolor{blue}{// Eq.~\eqref{eq:e2}}
\STATE $\texttt{ub} \leftarrow dist + \texttt{HD\_r}[v_r][i] + \texttt{HD\_s}[v_s][j]$ \quad \textcolor{blue}{// Eq.~\eqref{eq:e1}}
\STATE $\texttt{localLB} \leftarrow \min(\texttt{localLB}, \texttt{lb})$
\STATE $\texttt{localUB} \leftarrow \min(\texttt{localUB}, \texttt{ub})$
\ENDFOR
\STATE Aggregate thread-local bounds in shared memory to obtain block-wise bounds $(\texttt{vPairLB}, \texttt{vPairUB})$ \label{refine_aggregate}
\STATE $\texttt{vpLB}[\texttt{blockIdx}] \leftarrow \texttt{vPairLB}$,\ \ $\texttt{vpUB}[\texttt{blockIdx}] \leftarrow \texttt{vPairUB}$
\end{algorithmic}
\end{algorithm}
\setlength{\textfloatsep}{5pt}

\vspace{1mm}
\noindent\textbf{Facet-Level Refinement.}
We design an algorithm that launches only a single kernel for efficient facet-pair distance computation and aggregation at the voxel-pair level. Algorithm~\ref{algo:refine} shows this kernel where each thread block processes one voxel pair and computes its distance bounds based on all its facet pairs. The resulting voxel-pair bounds (\texttt{vpLB} and \texttt{vpUB}) are then transferred back to the host, which aggregates them into the object-pair bounds (\texttt{opLB} and \texttt{opUB}) for determining whether each object pair can be pruned or finalized. This refinement is then repeated for the next LoD if needed.

In Lines~\ref{refine_valid}--\ref{refine_s2} of Algorithm~\ref{algo:refine}, each thread block is assigned a voxel pair $(v_r, v_s)$ and retrieves the numbers of triangles in the two voxels. The loop at Line~\ref{refine_for} enumerates all facet pairs between $v_r$ and $v_s$. Specifically, Line~\ref{refine_flat} flattens the 2D facet-pair space into a 1D index, so that each thread processes facet pairs in a round-robin manner. For each assigned pair, we read the triangle coordinates (Line~\ref{refine_coord}) and compute their triangle-triangle distance (Line~\ref{refine_comp}). Using the precomputed facet-level Hausdorff and proxy Hausdorff distances, Lines~9 and~10 derive the lower and upper bounds according to Equations~\eqref{eq:e2} and~\eqref{eq:e1}. Each thread maintains its local bounds during the iteration (see Lines~11--12). After all facet pairs are processed, thread-local results are aggregated in shared memory to obtain blockwise voxel-pair bounds, which are written to \texttt{vpLB} and \texttt{vpUB}.

\vspace{1mm}
\noindent\textbf{Facet-Level Voxel-Pair Data Preparation.}
Algorithm~\ref{algo:refine} requires accessing several pieces of facet-level data for each valid voxel pair. 
Specifically, in Line~\ref{refine_valid}, each thread block retrieves a voxel pair $(v_r, v_s)$ from \texttt{vPairs}, which encodes the offsets and lengths of the corresponding facet arrays. 
Lines~\ref{refine_s1}--\ref{refine_s2} obtain the numbers of triangles in the two voxels, determining the total number of facet pairs to be processed. 
Inside the main loop from Line~\ref{refine_for}, each thread reads the triangle coordinates of its assigned facet pair (Line~7) for distance computation. 
In addition, Lines~9--10 access the precomputed facet-level Hausdorff and proxy Hausdorff values (\texttt{HD} and \texttt{PH}) associated with each facet to derive tight lower and upper bounds. 

\begin{figure}[!t]
\centering
\includegraphics[width=0.8\columnwidth]{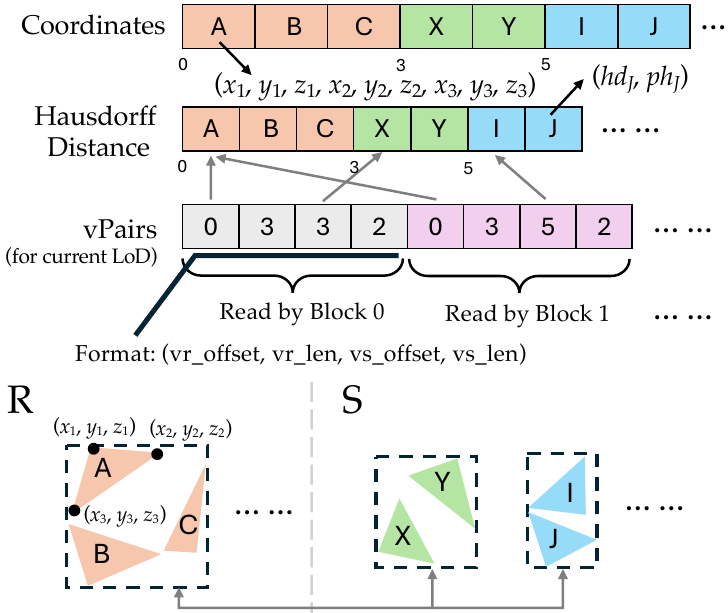}
\vspace{-3mm}
\caption{Voxel and Voxel-Pair Data Layout}\label{fig:refine_data}
\vspace{-4mm}
\end{figure}

To support efficient GPU execution, these data for the current LoD are first prepared on the CPU and organized into contiguous arrays following the layout illustrated in Figure~\ref{fig:refine_data}, and then transferred to GPU global memory. 
Specifically, the CPU first collects all valid voxel pairs produced by the filtering stage, {\bf deduplicates} their voxels to get a list of {\bf unique} voxels, and obtains their facets which are then organized into two compact facet arrays, one for facet coordinates and the other for facet-level Hausdorff distances. 
Each voxel pair $(v_r, v_s)$ is then represented by a tuple $(v_r^{\text{off}}, v_r^{\text{len}}, v_s^{\text{off}}, v_s^{\text{len}})$, which records the offsets and lengths of their corresponding facet segments in the two facet arrays.

This organization allows each thread block of Algorithm~\ref{algo:refine} to directly access the required data via simple pointer arithmetic.

\vspace{1mm}
\noindent\textbf{Chunked Pipelining by CUDA Streams.} 
We observe that the facet-level data preparation phase incurs a non-negligible time cost (nearly 59\% of the total refinement time) when data preparation and GPU-accelerated facet-level refinement are executed in sequence. 

\begin{figure}[!t]
\centering
\includegraphics[width=\columnwidth]{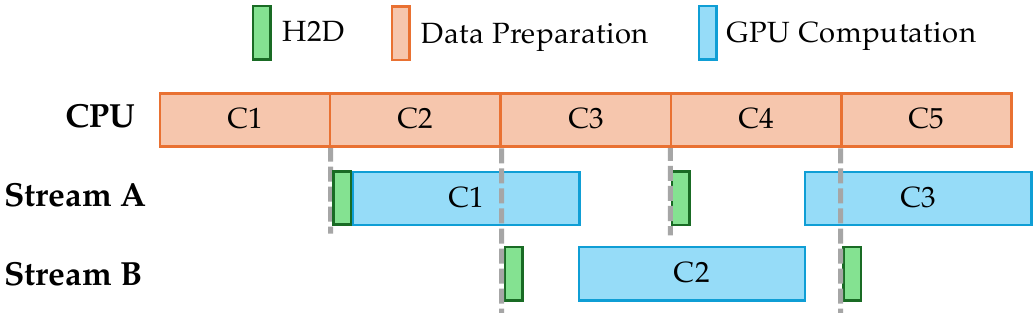}
\vspace{-5mm}
\caption{CPU-GPU Pipelining in Refinement Stage}\label{fig:refine_pipe}
\vspace{-1mm}
\end{figure}

To overlap data preparation with GPU computation, we partition the valid voxel pairs into multiple chunks and process them in a pipelined manner. 
We implement this pipelining via two independent CUDA streams. Given a sequence of chunks $[c_0, c_1, c_2, \cdots]$, the host assigns even-indexed chunks to one stream and odd-indexed chunks to the other stream. The benefit is when one stream uses GPU for computation, another stream may copy prepared data from CPU to GPU, while CPU simultaneously prepares data for a subsequent chunk. As a result, GPU-based computation, CPU-based data preparation, and host-to-device (H2D) data transfer are effectively overlapped to minimize resource waste, as Figure~\ref{fig:refine_pipe} illustrates.


\begin{algorithm}[t]
\caption{CPU-GPU Pipelining in Refinement Stage} \label{algo:gpu_cpu_pipe}
\begin{algorithmic}[1] 
\STATE Launch a dedicated CPU thread for data preparation \label{pipe_thread}
\STATE Initialize two CUDA streams \label{pipe_stream}
\STATE {\bf for} $c_i$ {\bf in} chunks $[c_0, c_1, c_2, \cdots, c_n]$ {\bf do}
\STATE \ \ \ \ $curr \leftarrow i\bmod 2$, \quad $prev \leftarrow 1-curr$
\STATE \ \ \ \ Wait until the facet-level data of $c_i$ is prepared
\STATE \ \ \ \ (Async) perform H2D facet-level data transfer for $c_i$, and launch Algorithm~\ref{algo:refine} for $c_i$, both on stream $curr$ \label{pipe_async}
\STATE \ \ \ \ {\bf if} $i > 0$ {\bf then}
\STATE \ \ \ \ \ \ \ Wait for stream $prev$ to complete the computation on $c_{i-1}$
\STATE \ \ \ \ \ \ \ Perform D2H transfer of \texttt{vpLB} and \texttt{vpUB} for $c_{i-1}$
\STATE \ \ \ \ \ \ \ Aggregate them to object-pair bounds (\texttt{opLB}, \texttt{opUB}) for $c_{i-1}$\!\!
\STATE Perform D2H bounds transfer and aggregation for $c_n$ \label{pipe_host2}
\end{algorithmic} 
\end{algorithm}
\setlength{\textfloatsep}{5pt}

Algorithm~\ref{algo:gpu_cpu_pipe} describes the implementation of the pipelined execution. First, the main thread launches a CPU thread dedicated for preparing data (Line~1), and initializes two CUDA streams (Line~2). Then, we iterate every chunk $c_i$, identify which stream $c_i$ should use (Line~4). Once the data of $c_i$ is prepared by the CPU thread (notified via a condition variable, Line~5), the main thread launches H2D facet-level data transfer and then the refinement kernel for $c_i$ on the current GPU stream (Line~6). Meanwhile, we proceed to receive \texttt{vpLB} and \texttt{vpUB} for $c_{i-1}$ (Line~9) after being computed by the other stream (Line~8, waiting via \texttt{cudaStreamSynchronize}) and aggregate them to object-pair bounds (\texttt{opLB}, \texttt{opUB}) (Line~10). Finally, Line~11 completes this host-side processing for the last chunk. 

Chunked pipelining also reduces GPU memory consumption: instead of materializing facet-level data for all valid voxel pairs at once, the GPU maintains only the data for two consecutive chunks at any time, one for each stream under a double-buffering scheme.

\subsection{GPU Algorithms for Object-Pair Pruning}\label{sec:3.4}

\begin{figure}[!t]
\centering
\vspace{-3mm}
\includegraphics[width=0.8\columnwidth]{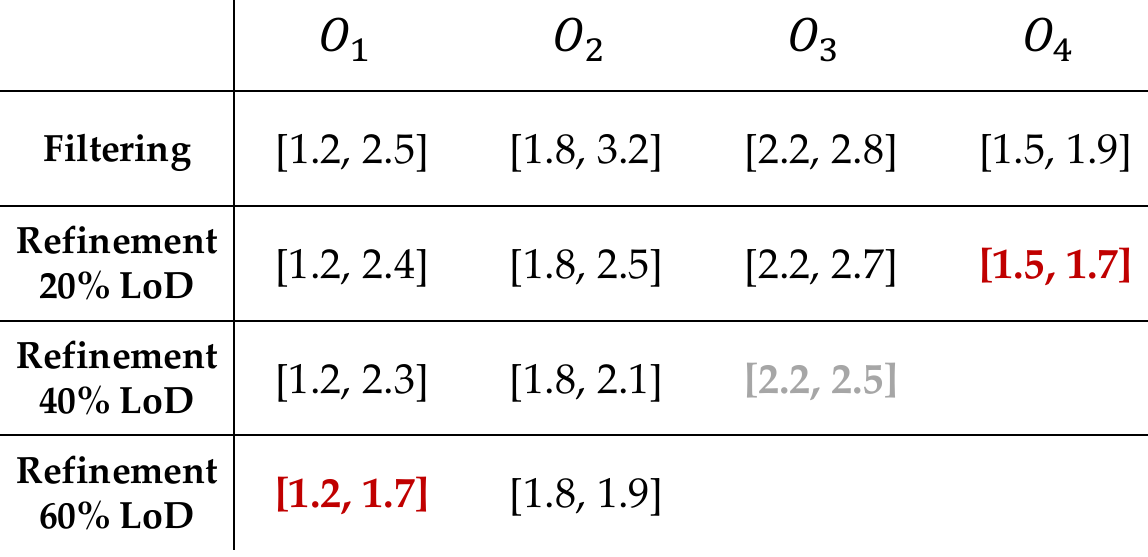}
\vspace{-2mm}
\caption{Progressive Candidate Pruning for a $k$-NN Query}\label{fig:knn}
\vspace{-1mm}
\end{figure}

Recall from Figure~\ref{fig:overview} that {\em Object-Pair Pruning} is used in both stages of {\em Voxel-Pair Filtering} and {\em Facet-Level Refinement}, which classifies the object pairs into three status \texttt{CONFIRMED}, \texttt{REMOVED} or \texttt{UNDECIDED} based on their distance bounds in \texttt{opLB} and \texttt{opUB}.
In Section~\ref{sec:3.2}, we have seen how to perform {\em Object-Pair Pruning} using CPU for \emph{within-$\tau$ distance} queries based on the fixed threshold $\tau$. 
This section designs a GPU kernel for {\em Object-Pair Pruning} in $k$-NN queries, which determines the top-$k$ results progressively as object-pair bounds become tighter across refinement levels, as Figure~\ref{fig:knn} illustrates.

Specifically, consider four candidates $O_1,O_2,O_3,O_4$ for a top-2 query. After filtering, the bounds are still too loose to finalize any result: although $O_4$ is already known to be closer than $O_3$, its relative order with respect to $O_1$ and $O_2$ remains unresolved.

At 20\% LoD, refinement tightens the bounds enough to prove that $O_4$ is closer than both $O_2$ and $O_3$. Hence, $O_4$ must belong to the top-2 result set and is marked as \texttt{CONFIRMED}. At 40\% LoD, the updated bounds show that $O_3$ is necessarily farther than $O_2$, so $O_3$ is marked as \texttt{REMOVED}. At 60\% LoD, $O_1$ is further shown to be closer than $O_2$, which makes $O_1$ the remaining top-1 candidate and therefore \texttt{CONFIRMED}. This example shows that $k$-NN evaluation can progressively confirm results and discard impossible candidates without waiting for exact distances of all object pairs.

While Figure~\ref{fig:knn} illustrates the process of {\em Object-Pair Pruning} for four candidate objects, in large datasets CPU-based top-$k$ selection becomes a bottleneck (e.g., tens of seconds); we therefore offload it to GPU, reducing the per-round cost to subseconds. 

\begin{algorithm}[t]
\caption{$k$-NN Object-Pair Pruning}
\label{algo:knn}
\begin{algorithmic}[1]
\STATE $r \leftarrow \texttt{blockIdx}$
\STATE $\textit{kLeft} \leftarrow k - \texttt{numConfirmed}[r]$
\STATE $\textit{offset} \leftarrow \texttt{r2opOffsets}[r]$, \ $N \leftarrow \texttt{r2opOffsets}[r+1] - \textit{offset}$
\FOR{$t \leftarrow \texttt{threadIdx}$; $t < N$; $t += \texttt{blockDim}$}
    \STATE $\_\,, m \leftarrow \texttt{oPairs}[\textit{offset}+t]$
    \STATE \textbf{if} \texttt{status}[$m$] $\neq$ \texttt{UNDECIDED} \textbf{then continue}
    \STATE $closer \leftarrow 0$, \ $\textit{farther} \leftarrow 0$
    \FOR{$j \leftarrow 0$; $j < N$; $j++$}
        \STATE $\_\,, n \leftarrow \texttt{oPairs}[\textit{offset}+j]$
        \STATE \textbf{if} $n=m$ \textbf{or} \texttt{status}[$n$] $\neq$ \texttt{UNDECIDED} \textbf{then continue}
        \STATE \textbf{if} \texttt{opLB}[$m$] $\ge$ \texttt{opUB}[$n$] \textbf{then} $closer{+}{+}$
        \STATE \textbf{if} \texttt{opUB}[$m$] $\le$ \texttt{opLB}[$n$] \textbf{then} $\textit{farther}{+}{+}$
    \ENDFOR
    \STATE \textbf{if} $N-\textit{farther} < \textit{kLeft}$\ \ \textbf{then} \texttt{status}[$m$] $\leftarrow$ \texttt{CONFIRMED}
    \STATE \textbf{else if} $closer \ge \textit{kLeft}$\ \ \textbf{then} \texttt{status}[$m$] $\leftarrow$ \texttt{REMOVED}
    \STATE \textbf{else} \texttt{status}[$m$] $\leftarrow$ \texttt{UNDECIDED}
\ENDFOR
\end{algorithmic}
\end{algorithm}
\setlength{\textfloatsep}{15pt}

Algorithm~\ref{algo:knn} shows our GPU kernel. We assign one query object $r \in R$ to each thread block, and let the threads in the block evaluate the candidate object pairs associated with $r$. Specifically, Line~1 identifies the query object $r$ by \texttt{blockIdx}. Line~2 computes \textit{kLeft}, the number of result objects that still need to be determined for $r$, by subtracting \texttt{numConfirmed}[$r$] from $k$. Here, \texttt{numConfirmed}[$r$] records how many candidates of $r$ have already been confirmed as top-$k$ results in previous LoD pruning rounds (initialized as 0). Line~3 then obtains the contiguous subarray of candidate object pairs of $r$ in \texttt{oPairs} using \texttt{r2opOffsets} (recall Figure~\ref{fig:filter_data}).

The for-loop from Line~4 lets the threads examine the candidates of $r$ in parallel. Each thread takes one candidate object pair $m$ at a time (Line~5), skipping those whose status has already been determined (Line~6). For each remaining candidate $m$, we compare its current distance bounds with every other undecided candidate $n$ of the same query object $r$ (Lines~8--12), where we count how many candidates are guaranteed to be closer (resp.\ farther) than $m$, which is recorded by local variable $closer$ (resp.\ {\em farther}). Lines~13--15 then determine the status of $m$ based on these two counters.

This pruning procedure is invoked after each update of object-pair bounds, so more candidates can be progressively confirmed or removed as refinement proceeds from coarse to fine LoDs.





\section{Experiments}
\label{sec:experiment}
Besides our 3DPipe framework, only 3DPro~\cite{3dpro} and TDBase~\citep{tdbase} support 3D spatial join on GPU. However, 3DPro is not open-sourced, while TDBase is reported to outperform 3DPro by up to 4$\times$ in time~\cite{tdbase} and is open-sourced~\cite{tdbase_code}. 
In this section, we compare 3DPipe with TDBase as the state-of-the-art baseline to evaluate efficiency and scalability. We also conduct ablation studies to assess the effectiveness of our optimizations for filtering and refinement. 
3DPipe is open-sourced at \url{https://github.com/lyuheng/3dpipe}.

\subsection{Experiment Setup}

\begin{table}
\centering
\caption{Dataset Parameters} \label{table:parameter}
\vspace{-3mm}
\begin{tabular}{c | c c c c} 
	\hline 
  Dataset & Join Task & Alias &  $k$  &  $\tau$ \\
  \hline
  \multirow{2}{*}{Digital Pathology} & Nuclei, Vessel & NV & [1, 10] & [0, 400] \\
  	& Nuclei, Nuclei & NN & [1, 10] & [0, 200] \\
  \hline
   \multirow{2}{*}{ModelNet40} & Train, Test & TI & [1, 10] & [0, 20] \\
  	& Train, Train & TT & [5, 50] & [0, 10] \\
  \hline
\end{tabular}
\end{table}

\noindent \textbf{Datasets.} Following existing works~\cite{3dpro, tdbase, DBLP:journals/tsas/TengLVKW22}, we use two real-world 3D digital pathology objects which are generated by reconstructing 2D segmented objects from brain tissues~\cite{ispeed, liang2017development}, represented in Object File Format (OFF) format~\cite{off}. Specifically, one object is a blood vessel containing 30,000 facets and 5 bifurcations; another object is a nucleus cell which has 300 facets. We follow the procedure of~\cite{3dpro, tdbase} to replicate the vessel into multiple objects and shift them to different locations (without overlap of their bounding boxes) to simulate blood vessels in human body. We denote the resulting vessel object dataset as \texttt{V}. We also replicate the nucleus cell and uniformly distribute the cells in the space of \texttt{V} to form a nuclei cell dataset, denoted as \texttt{N}. 
Unless otherwise specified, we assume $|\texttt{V}|=5000$ and $|\texttt{N}|=$ 1 million. We can change spatial join workload by varying the numbers of replicated objects in \texttt{V} and \texttt{N}.

We also use ModelNet40~\cite{modelnet_dataset,modelnet}, which contains 12,311 pre-aligned 3D CAD models in OFF format from 40 categories such as beds and chairs. It is split into 9,843 training objects and 2,468 test objects. By replicating each object 100 times and shifting the replicas to different locations, we obtain a training dataset \texttt{T} of 984,300 objects and a test dataset \texttt{I} of 246,800 objects.

\vspace{1mm} 
\noindent \textbf{Queries.} We evaluate all three query types: intersection, within-$\tau$ distance, and $k$-NN, with parameter settings summarized in Table~\ref{table:parameter}. We choose the ranges of $k$ and $\tau$ according to the characteristics of each dataset so as to generate representative join workloads without making the running time excessively long; note that $\tau=0$ is a special case corresponding to intersection queries.

\vspace{1mm} 
\noindent \textbf{Configurations.} All experiments were conducted on a Polaris node at Argonne National Laboratory with 4 AMD EPYC 7543P CPUs (32 cores), 512~GB DRAM, and one NVIDIA A100 GPU with 40~GB memory connected via PCIe Gen4. We use CUDA 12.8 and compile all implementations with \texttt{-O3}.

CPU-side processing in both 3DPipe and TDBase uses OpenMP with 32 threads whenever applicable, e.g., to parallelize MBB-based Object Filtering over all objects $r\in R$ against the static R-tree $T_S$.

For voxelization, the number of voxels is set to 2\% of the number of facets per object. For pipelining, the chunk sizes in Algorithms~\ref{algo:chunk_stream_pipe} and~\ref{algo:gpu_cpu_pipe} are set to 5~GB and 500,000 voxel pairs, respectively.

\begin{figure}[!t]
\centering
\includegraphics[width=0.4\linewidth]{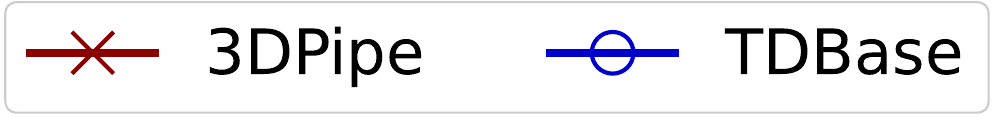}  \\
\begin{subfigure}{0.48\linewidth}
	\centering
	\includegraphics[width=\linewidth]{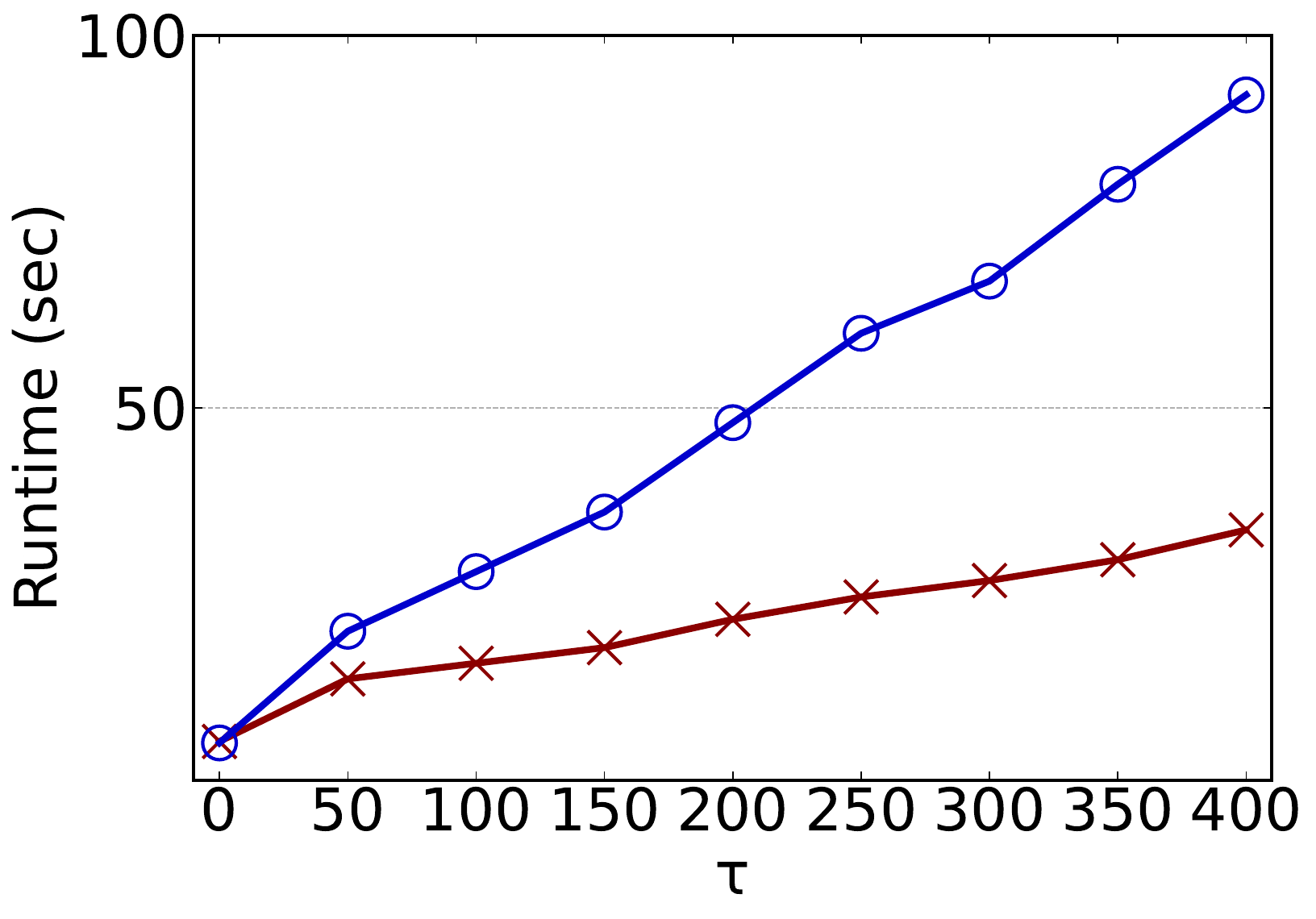} 
	\vspace{-6mm}
	\caption{Within-$\tau$ NV (varying $\tau$)}
	\label{exp:nv_within}
	\vspace{1mm}
\end{subfigure}
\hfill
\begin{subfigure}{0.48\linewidth}
	\centering
	\includegraphics[width=\columnwidth]{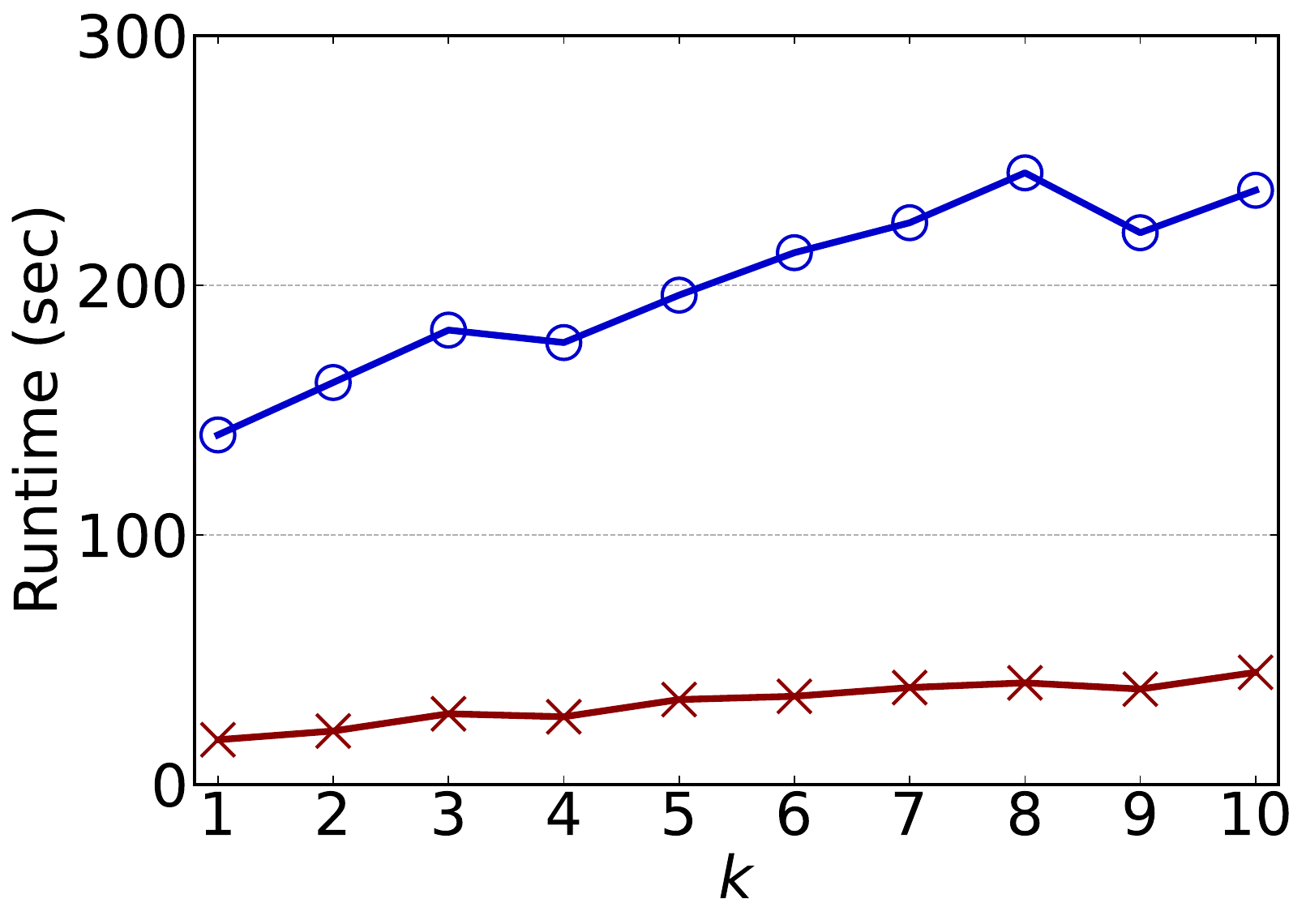}
	\vspace{-6mm}
	\caption{$k$-NN NV (varying $k$)}
	\label{exp:nv_knn}
	\vspace{1mm}
\end{subfigure}
\begin{subfigure}{0.48\linewidth}
	\centering
	\includegraphics[width=\linewidth]{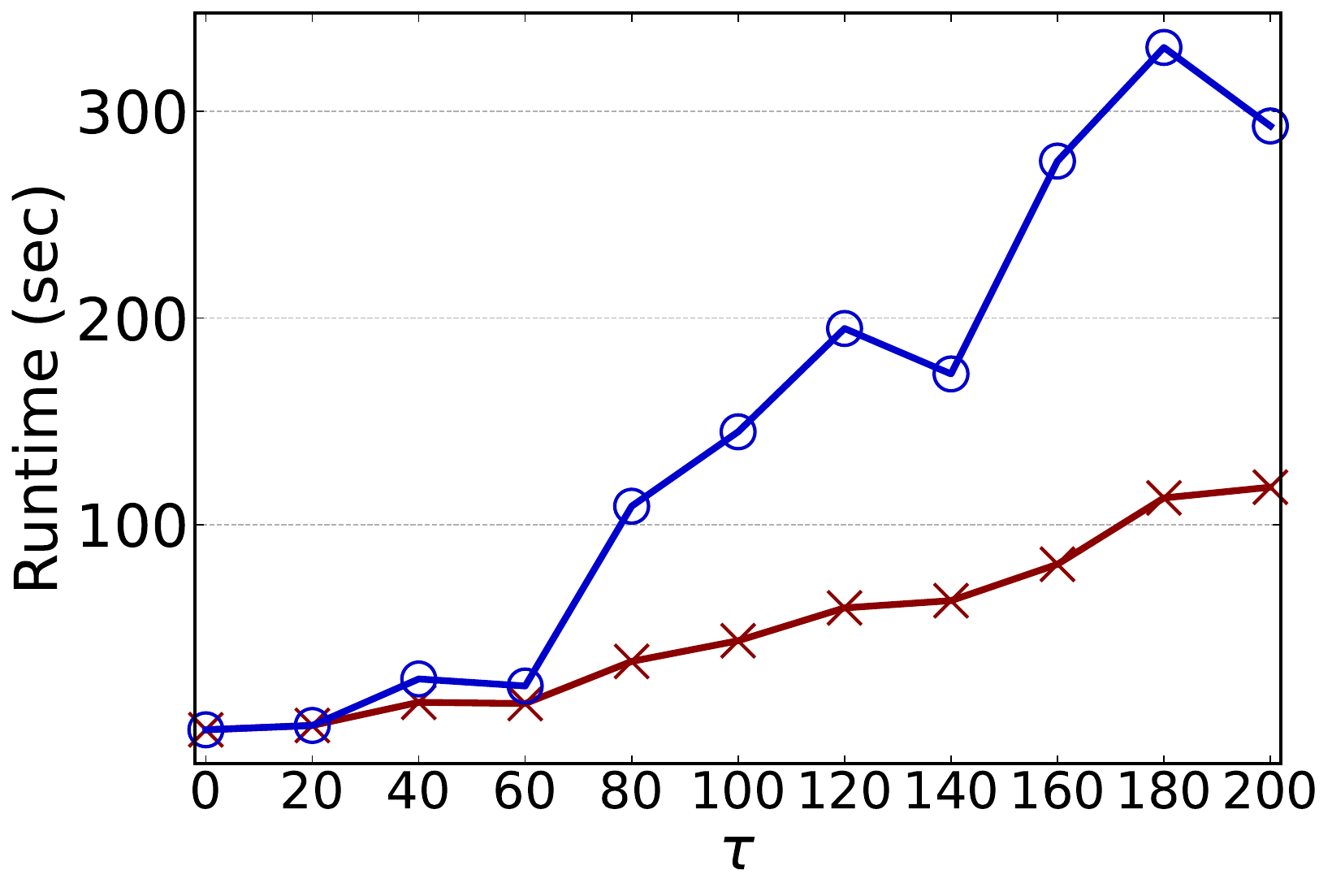} 
	\vspace{-6mm}
	\caption{Within-$\tau$ NN (varying $\tau$)}
	\label{exp:nn_within}
	\vspace{1mm}
\end{subfigure}
\hfill
\begin{subfigure}{0.48\linewidth}
	\centering
	\includegraphics[width=\columnwidth]{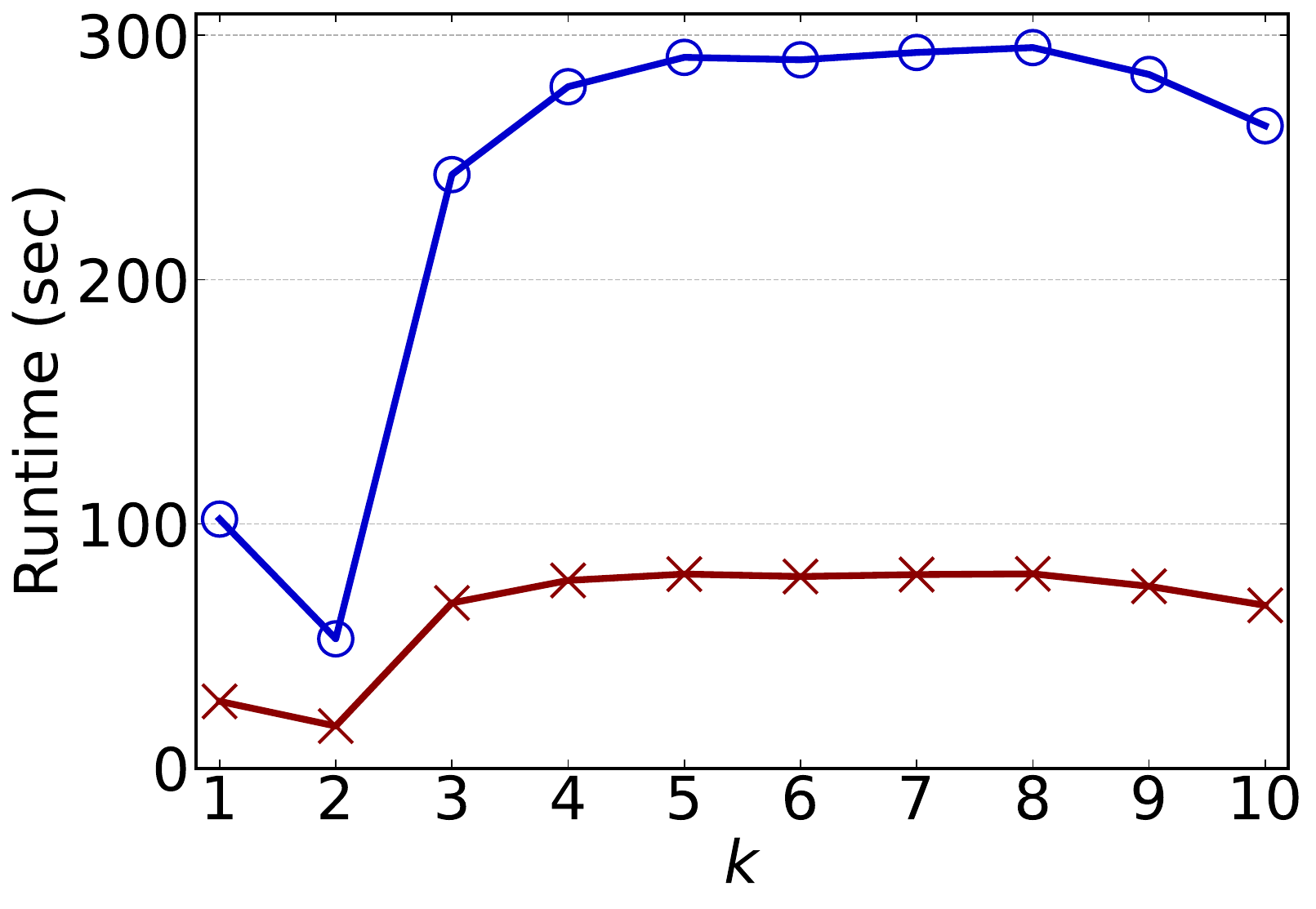}
	\vspace{-6mm}
	\caption{$k$-NN NN (varying $k$)}
	\label{exp:nn_knn}
	\vspace{1mm}
\end{subfigure}

\begin{subfigure}{0.48\linewidth}
	\centering
	\includegraphics[width=\linewidth]{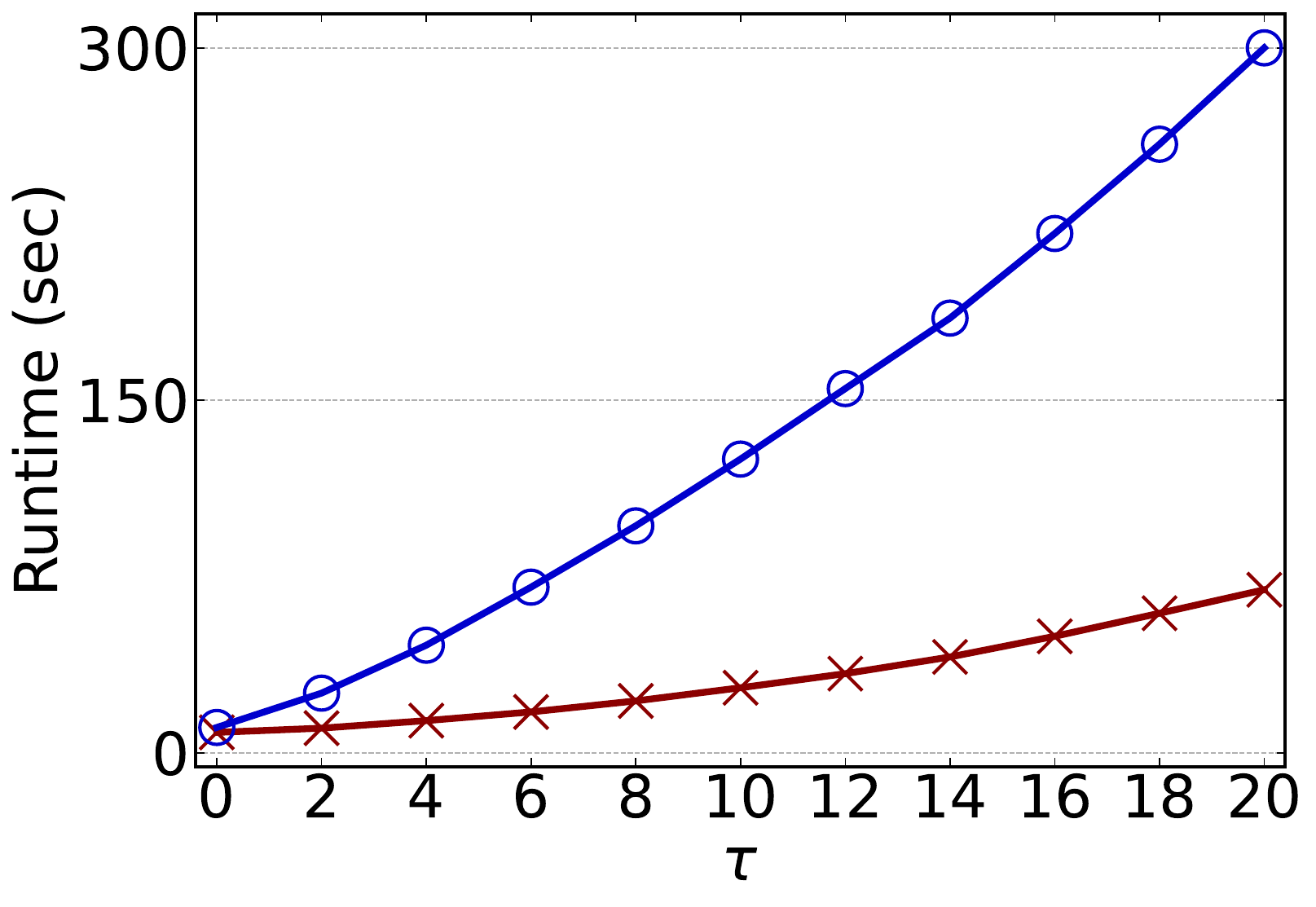} 
	\vspace{-6mm}
	\caption{Within-$\tau$ TI (varying $\tau$)}
	\label{exp:mt_within}
	\vspace{1mm}
\end{subfigure}
\hfill
\begin{subfigure}{0.48\linewidth}
	\centering
	\includegraphics[width=\columnwidth]{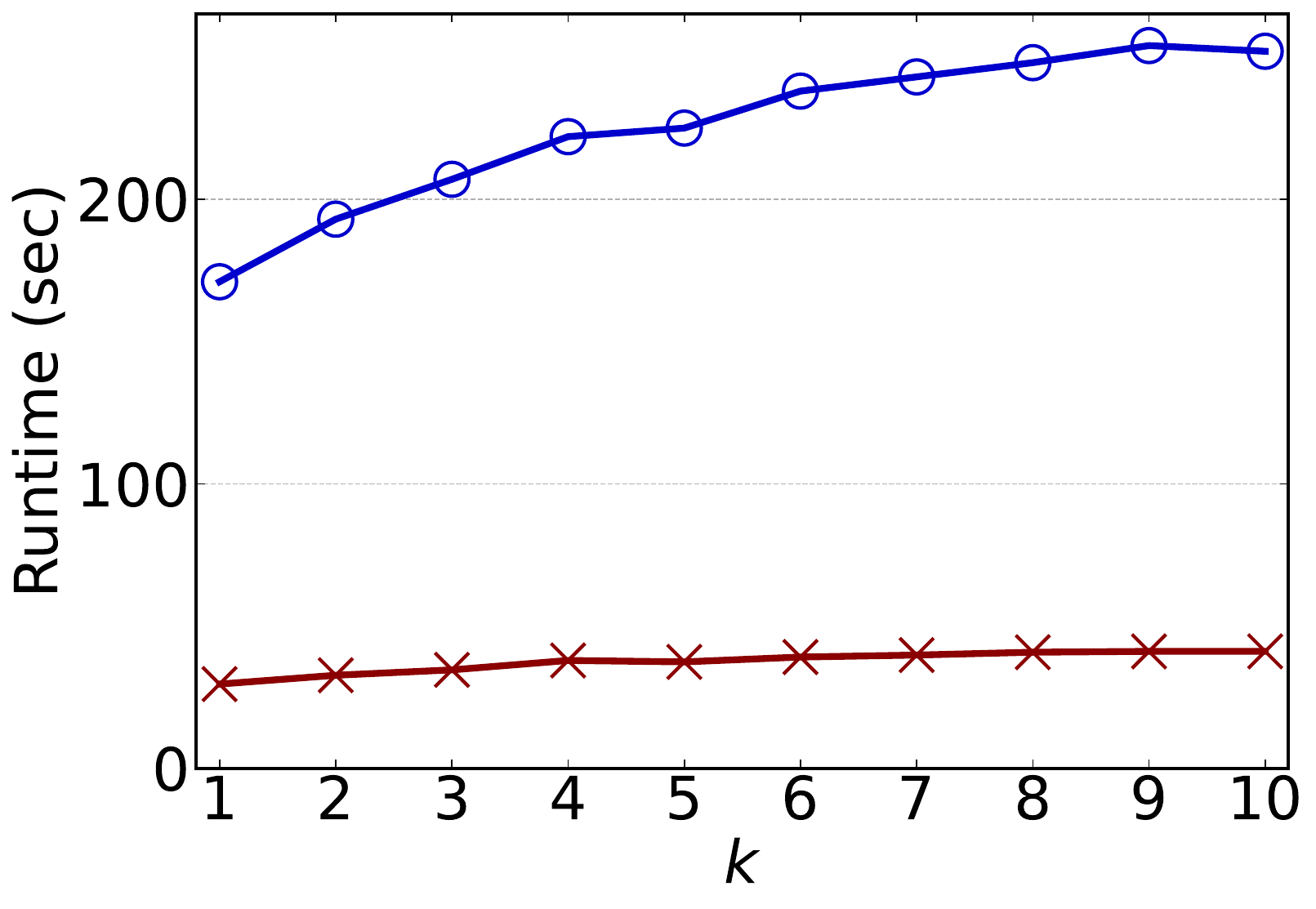}
	\vspace{-6mm}
	\caption{$k$-NN TI (varying $k$)}
	\label{exp:mt_knn}
	\vspace{1mm}
\end{subfigure}
\begin{subfigure}{0.48\linewidth}
	\centering
	\includegraphics[width=\linewidth]{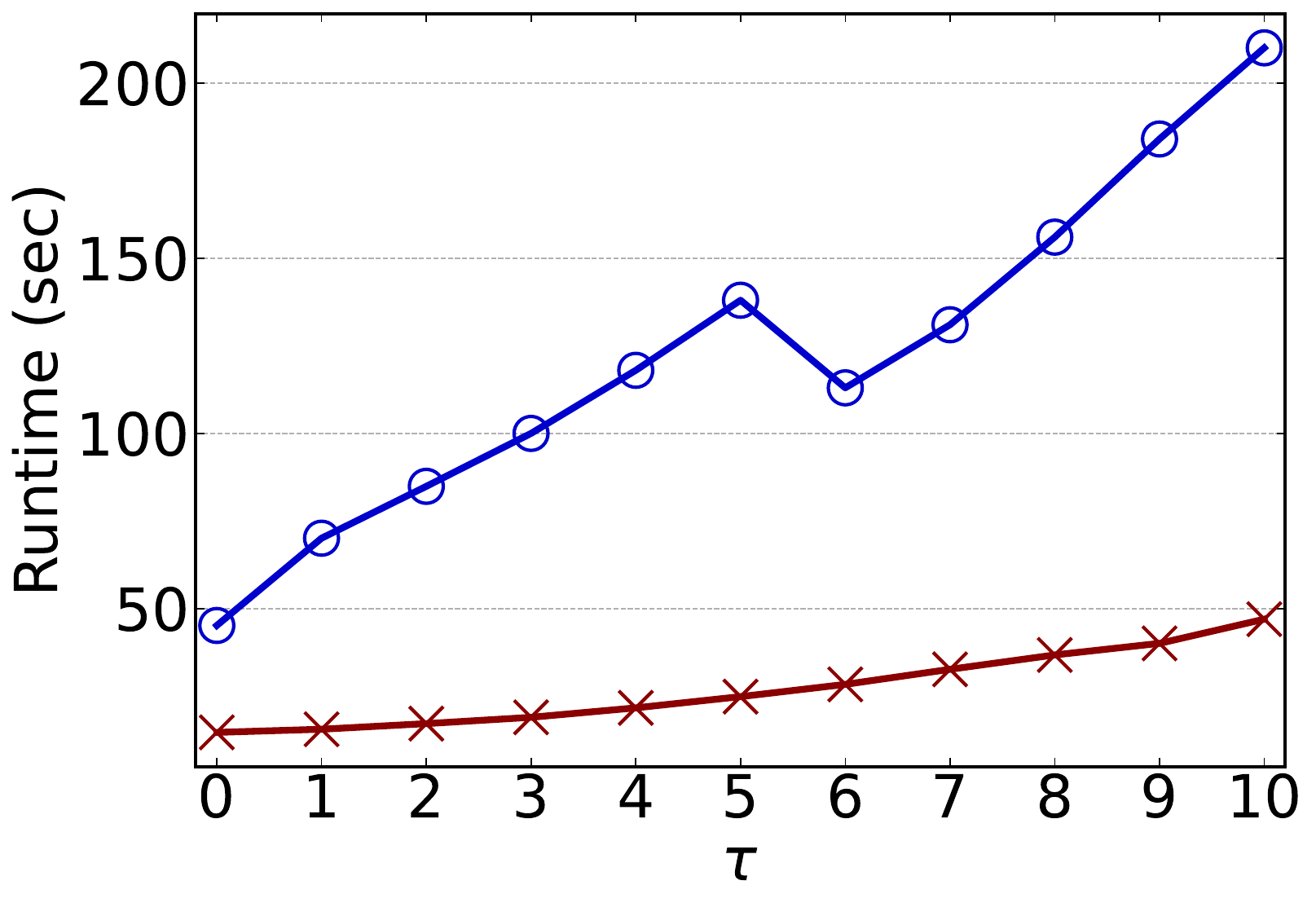} 
	\vspace{-6mm}
	\caption{Within-$\tau$ TT (varying $\tau$)}
	\label{exp:tt_within}
\end{subfigure}
\hfill
\begin{subfigure}{0.48\linewidth}
	\centering
	\includegraphics[width=\columnwidth]{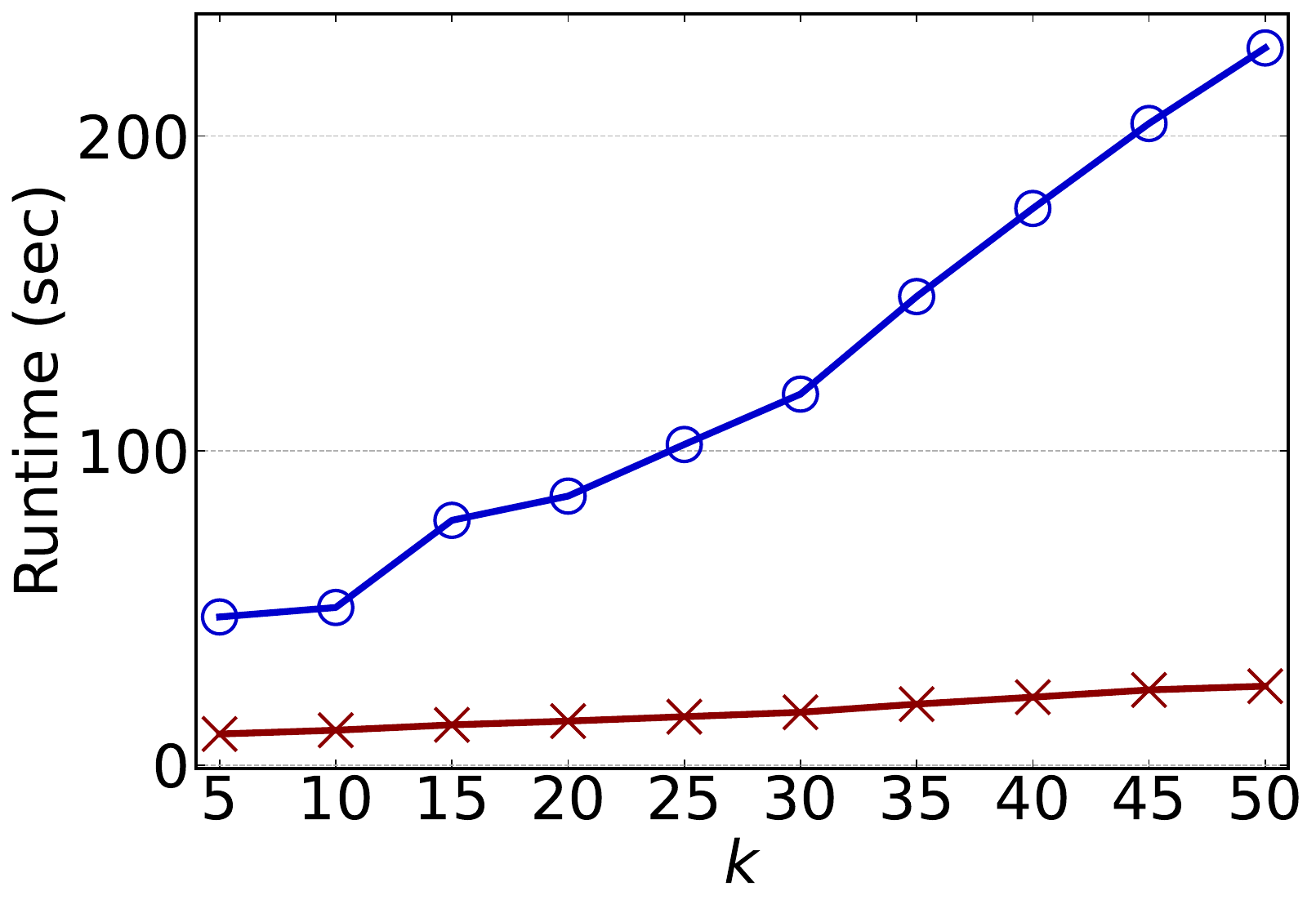}
	\vspace{-6mm}
	\caption{$k$-NN TT (varying $k$)}
	\label{exp:tt_knn}
\end{subfigure}
\vspace{-3mm}
\caption{Performance Comparison with TDBase}
\label{exp:tdbase}
\end{figure}

\vspace{-2mm}
\subsection{Comparison with TDBase}

Figure~\ref{exp:tdbase} compares 3DPipe with TDBase on all three query types, including within-$\tau$ distance, intersection ($\tau=0$), and $k$-NN. Overall, 3DPipe consistently outperforms TDBase across all datasets and query settings, with especially large gains on $k$-NN queries.

On NV, Figures~\ref{exp:nv_within} and \ref{exp:nv_knn} show that 3DPipe achieves up to 2.7$\times$ speedup for within-$\tau$ queries and 5.2$\times$--7.8$\times$ speedup for $k$-NN queries. The improvement comes from accelerating both filtering and refinement on GPU, rather than leaving key bottlenecks on CPU (TDBase uses only CPU for filtering). The advantage is particularly pronounced for $k$-NN, where TDBase spends substantial CPU time on object-pair filtering, while our GPU-based design in Section~\ref{sec:3.4} keeps each round of $k$-NN processing at the sub-second level.

A similar trend is observed on NN in Figures~\ref{exp:nn_within} and \ref{exp:nn_knn}. 3DPipe reaches up to 3.4$\times$ speedup for within-$\tau$ queries and 3.9$\times$ for $k$-NN queries. In addition, our runtime curves are noticeably more stable than those of TDBase. For example, in Figure~\ref{exp:nn_within}, when $\tau$ increases to 80, TDBase rises from 22\,s to 109\,s, whereas 3DPipe increases only from 13\,s to 33\,s. A similar jump appears in Figure~\ref{exp:nn_knn} when $k$ increases from 2 to 3: TDBase grows from 53\,s to 243\,s, while 3DPipe increases from 17\,s to 67\,s. This indicates that 3DPipe scales more robustly as the number of voxel or facet comparisons grows.

The same pattern holds on ModelNet40. As shown in Figures~\ref{exp:mt_within} and \ref{exp:mt_knn}, 3DPipe achieves up to 4.6$\times$ and 6.1$\times$ speedup, respectively. On TT, shown in Figures~\ref{exp:tt_within} and \ref{exp:tt_knn}, the speedups further increase to 5.4$\times$ for within-$\tau$ and 9.0$\times$ for $k$-NN. These results show that 3DPipe consistently outperforms TDBase across datasets and query types, with larger gains under heavier workloads, confirming the importance of 3DPipe's fully GPU-oriented optimization.

\vspace{-2mm} 
\subsection{Performance Breakdown}

\begin{figure}[!t]
\centering
\includegraphics[width=0.4\linewidth]{new_exp/ours_tdbase_header}  \\
\begin{subfigure}{0.48\linewidth}
	\centering
	\includegraphics[width=\linewidth]{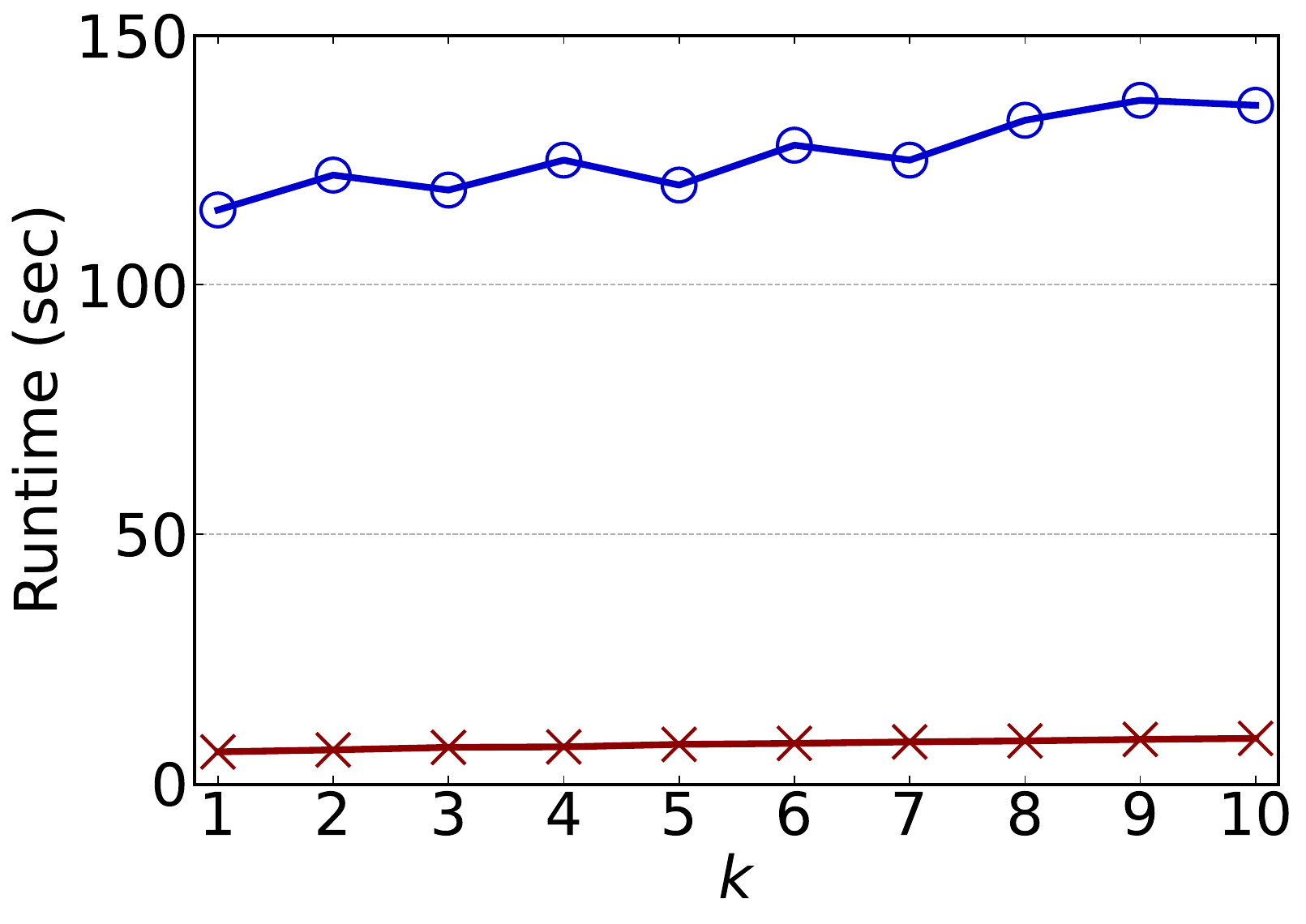} 
	\vspace{-6mm}
	\caption{$k$-NN NV (varying $k$)}
	\label{exp:nv_knn_filter}
	\vspace{0.5mm}
\end{subfigure}
\hfill
\begin{subfigure}{0.48\linewidth}
	\centering
	\includegraphics[width=\columnwidth]{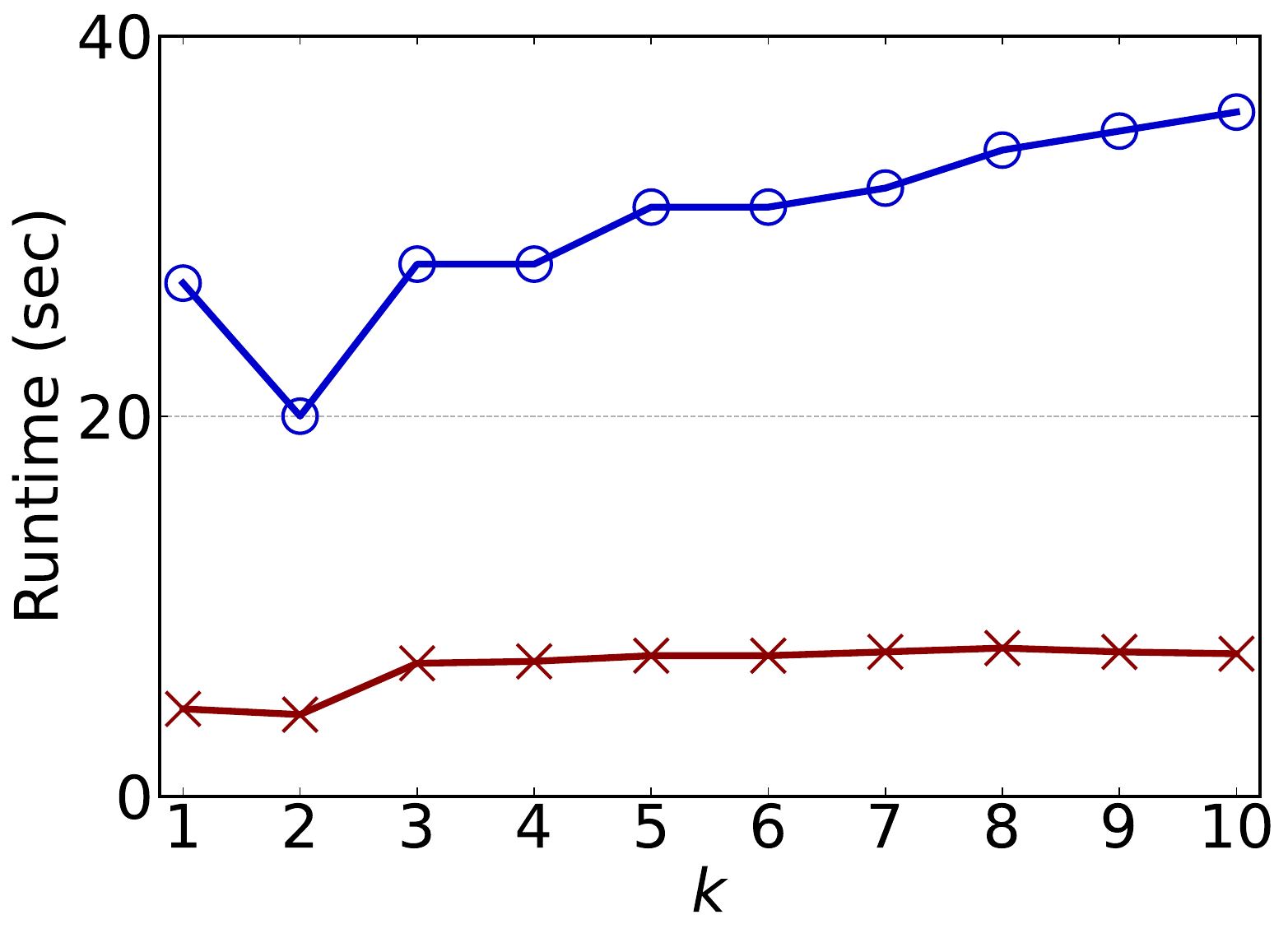}
	\vspace{-6mm}
	\caption{$k$-NN NN (varying $k$)}
	\label{exp:nn_knn_filter}
	\vspace{0.5mm}
\end{subfigure}

\begin{subfigure}{0.48\linewidth}
	\centering
	\includegraphics[width=\linewidth]{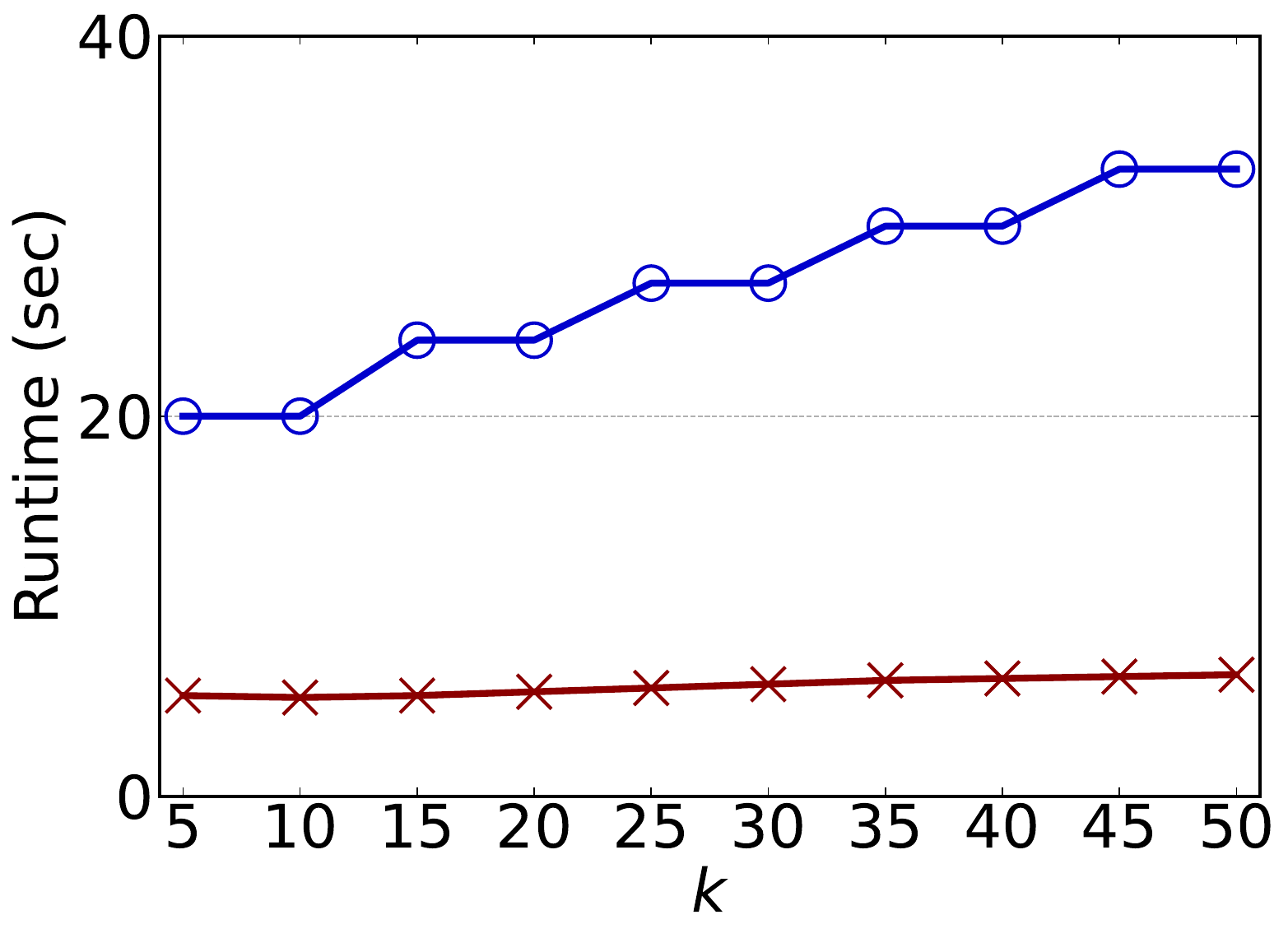} 
	\vspace{-6mm}
	\caption{$k$-NN TI (varying $k$)}
	\label{exp:mt_knn_filter}
\end{subfigure}
\hfill
\begin{subfigure}{0.48\linewidth}
	\centering
	\includegraphics[width=\columnwidth]{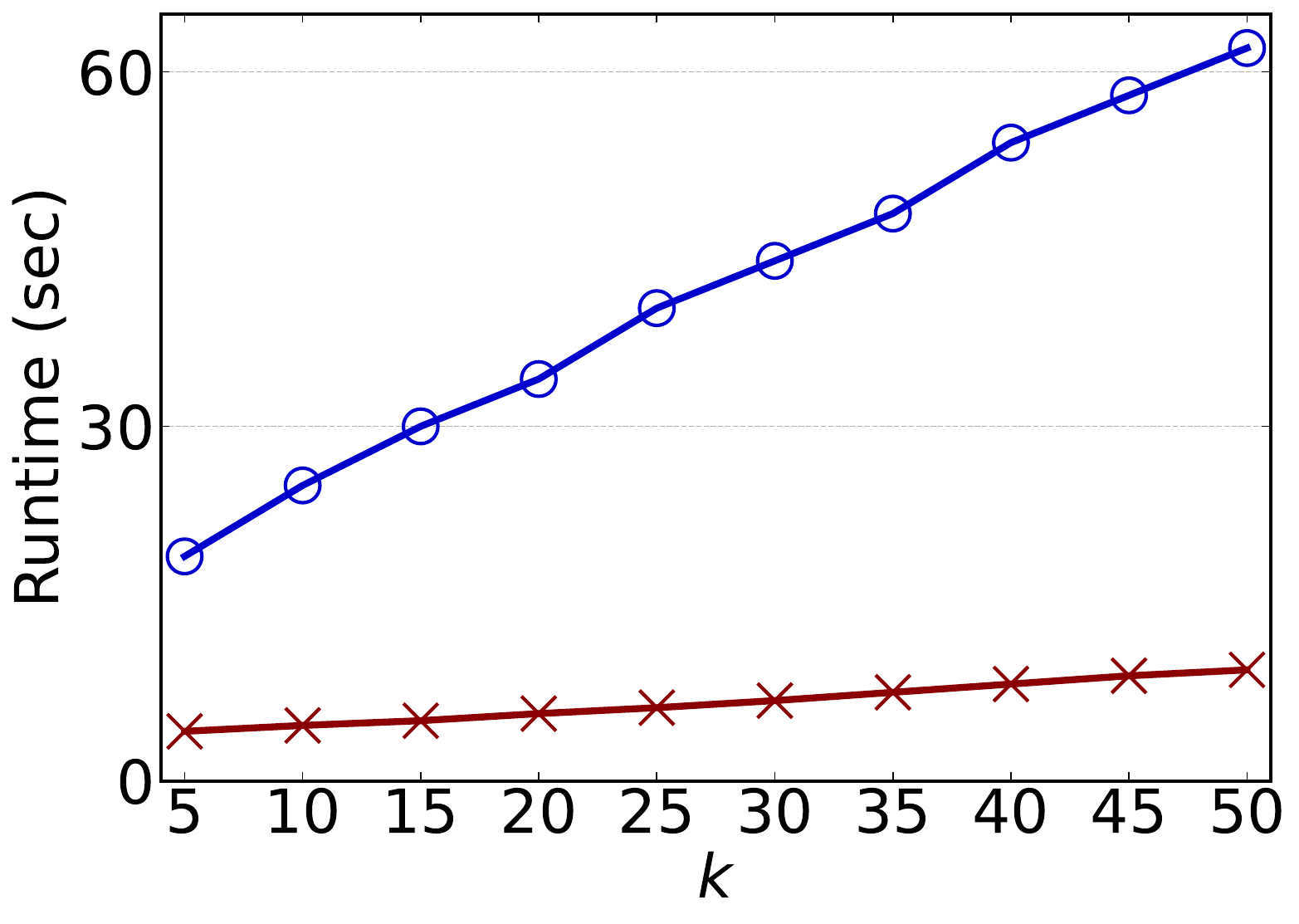}
	\vspace{-6mm}
	\caption{$k$-NN TT (varying $k$)}
	\label{exp:tt_knn_filter}
\end{subfigure}
\vspace{-3.5mm}
\caption{Runtime of Filtering Stage}
\label{exp:filter_stage}
\vspace{-3mm}
\end{figure}

\noindent \textbf{Filtering Stage.} 
Figure~\ref{exp:filter_stage} reports the runtime of the filtering stage for $k$-NN queries on NV, NN, TI, and TT datasets. We focus on $k$-NN queries, as filtering is a major performance bottleneck in this setting, whereas for within-$\tau$ queries the filtering stage completes only within a few seconds and does not dominate the overall runtime.

Our filtering stage consists of two phases: (i) MBB-based object filtering on CPU, and (ii) voxel-pair filtering with GPU. Overall, 3DPipe significantly outperforms TDBase, achieving 15$\times$--18$\times$ speedup on NV, 4$\times$--6$\times$ on NN, and up to 5.3$\times$--6.6$\times$ on ModelNet40. 

The sources of improvement differ across datasets. On NV, MBB-based filtering is lightweight, and the runtime is dominated by voxel-pair filtering. In this case, our GPU-based voxel-pair distance computation and pruning (Algorithms~\ref{alg:vp_kernel} and \ref{algo:voxel_prune}) effectively exploit massive parallelism, while TDBase relies on CPU parallelism (OpenMP), leading to a substantial performance gap. 
In contrast, on NN, TI, and TT, MBB-based filtering becomes the bottleneck, while voxel-pair filtering completes quickly. Here, the advantage comes from our best-first search strategy, which prioritizes candidates with smaller distances and prunes unpromising pairs earlier, whereas TDBase adopts a less efficient depth-first traversal.

\begin{figure}[!t]
\centering
\begin{subfigure}{0.45\linewidth}
	\centering
	\includegraphics[width=\linewidth]{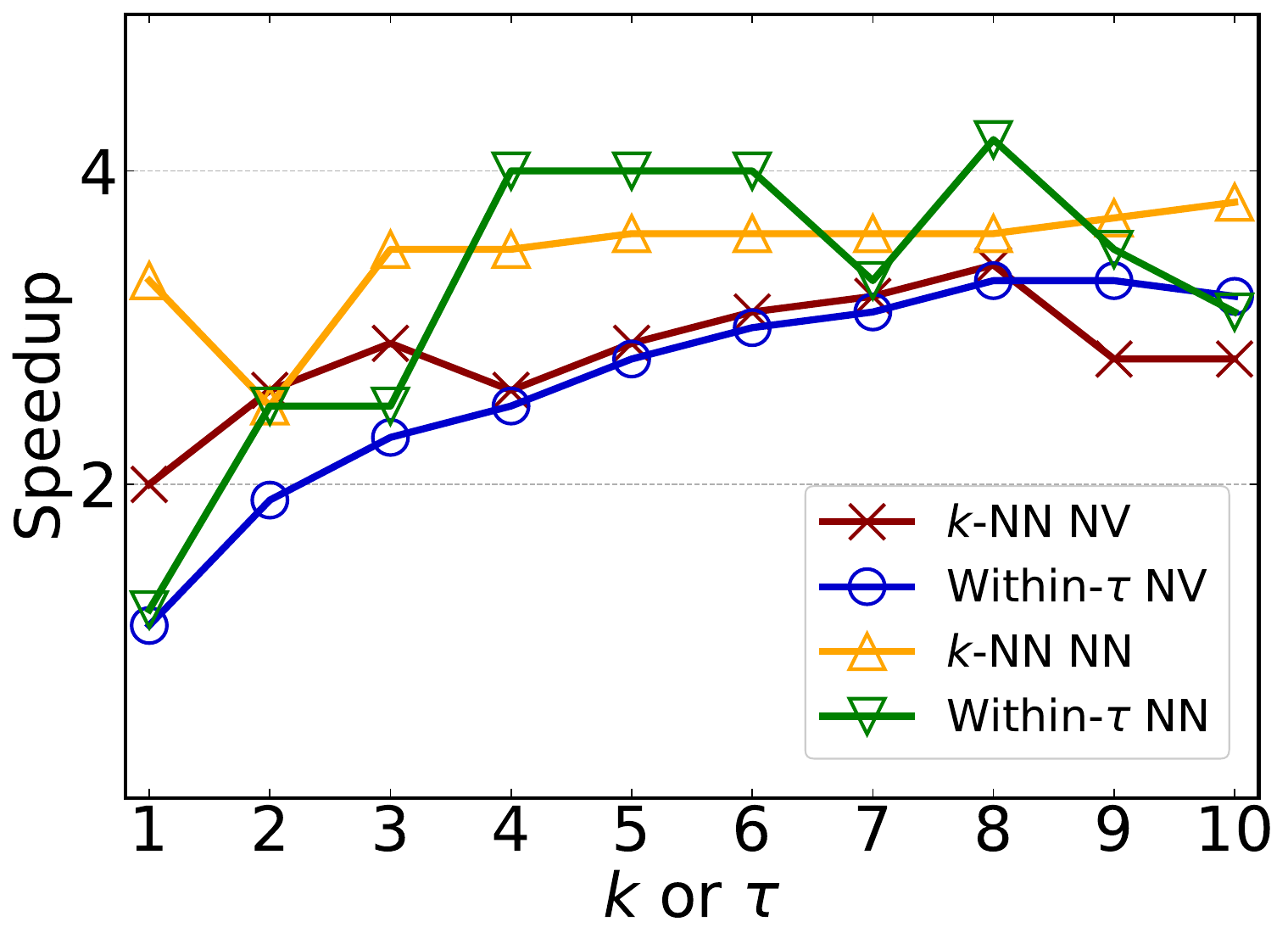} 
	\vspace{-6mm}
	\caption{Refinement on NV and NN}
	\label{exp:nv_nn_refine}
\end{subfigure}
\hfill
\begin{subfigure}{0.45\linewidth}
	\centering
	\includegraphics[width=\columnwidth]{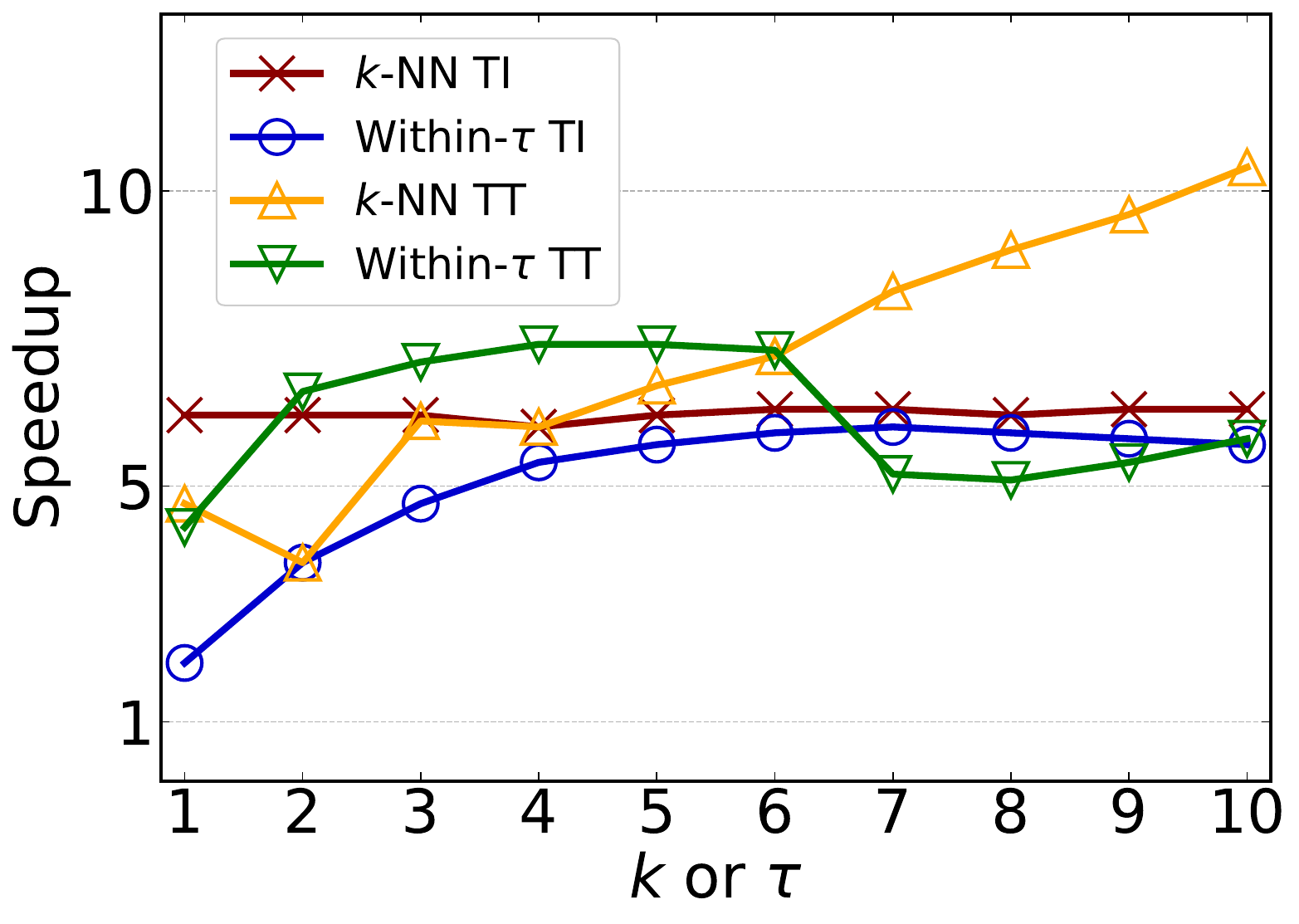}
	\vspace{-6mm}
	\caption{Refinement on TI and TT}
	\label{exp:mt_tt_refine}
\end{subfigure}
\vspace{-3mm}
\caption{Runtime Speedup of Refinement Stage}
\label{exp:refine}
\vspace{-1mm}
\end{figure}

\vspace{1mm} 
\noindent \textbf{Refinement Stage.} 
Figure~\ref{exp:refine} shows the refinement-stage speedup of 3DPipe over TDBase for both within-$\tau$ and $k$-NN queries across all datasets, where the x-axis varies $k$ or $\tau$. 

Overall, 3DPipe achieves up to $4\times$ speedup on NV/NN (Figure~\ref{exp:nv_nn_refine}) and up to $10\times$ on TI/TT (Figure~\ref{exp:mt_tt_refine}). Notably, although TDBase already leverages GPU acceleration, its performance remains suboptimal due to inefficient hardware utilization.

In contrast, 3DPipe consistently delivers higher performance by fully exploiting GPU capabilities. In particular, workload flattening and shared-memory aggregation significantly improve parallel efficiency, while CPU-GPU pipelining overlaps data preparation with computation to further enhance utilization.

\vspace{-2mm}
\subsection{Effectiveness of Chunked Streaming}

\begin{figure}[!t]
\includegraphics[width=0.7\linewidth]{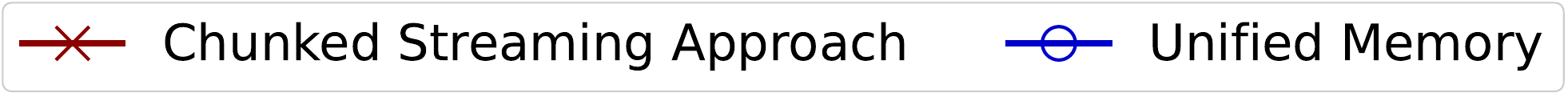}  \\
\centering
\begin{subfigure}{0.49\linewidth}
	\centering
	\includegraphics[width=\linewidth]{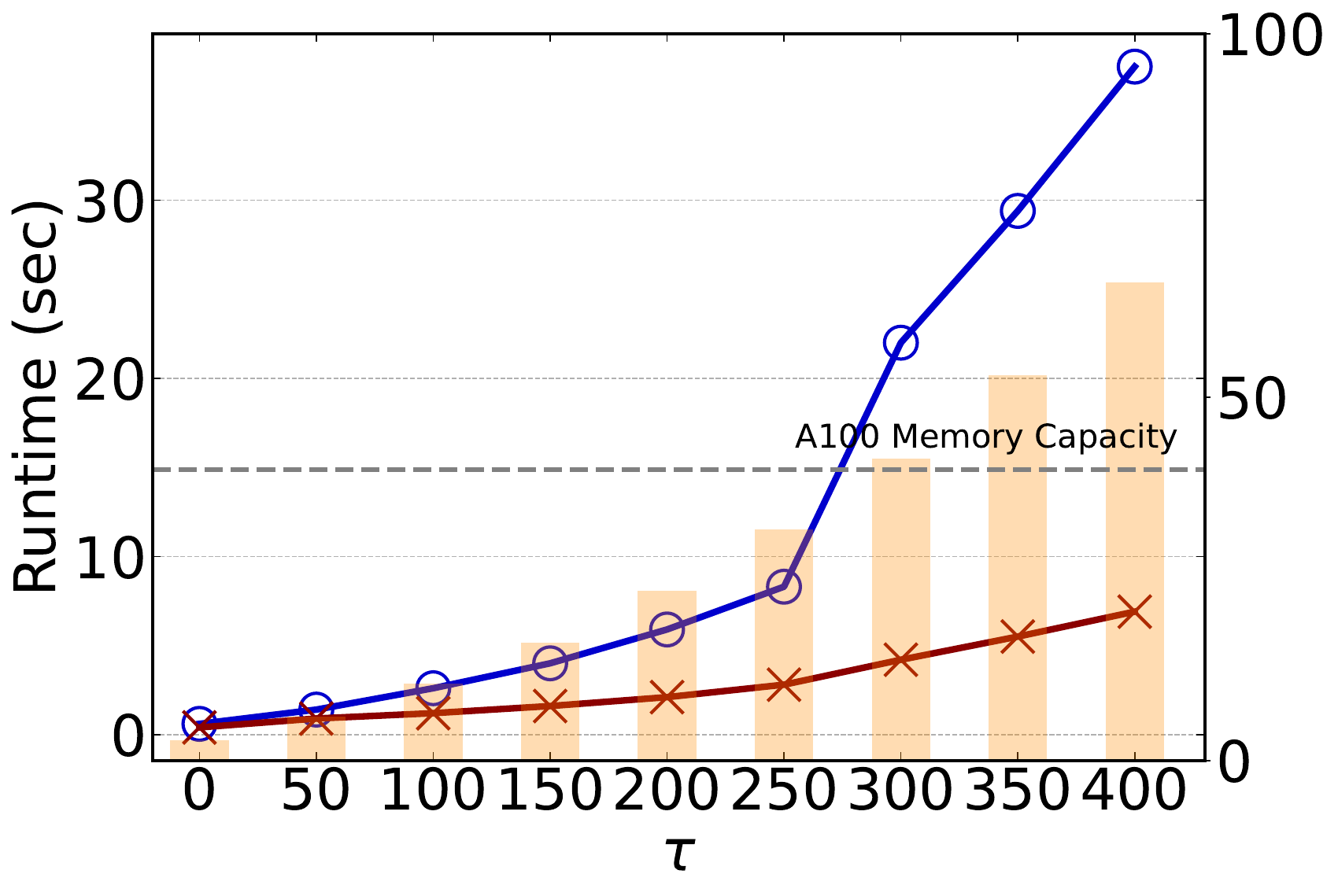} 
	\vspace{-6mm}
	\caption{Within-$\tau$ NV (varying $\tau$)}
	\label{exp:nv_wd_stream}
	\vspace{1mm}
\end{subfigure}
\begin{subfigure}{0.49\linewidth}
	\centering
	\includegraphics[width=\columnwidth]{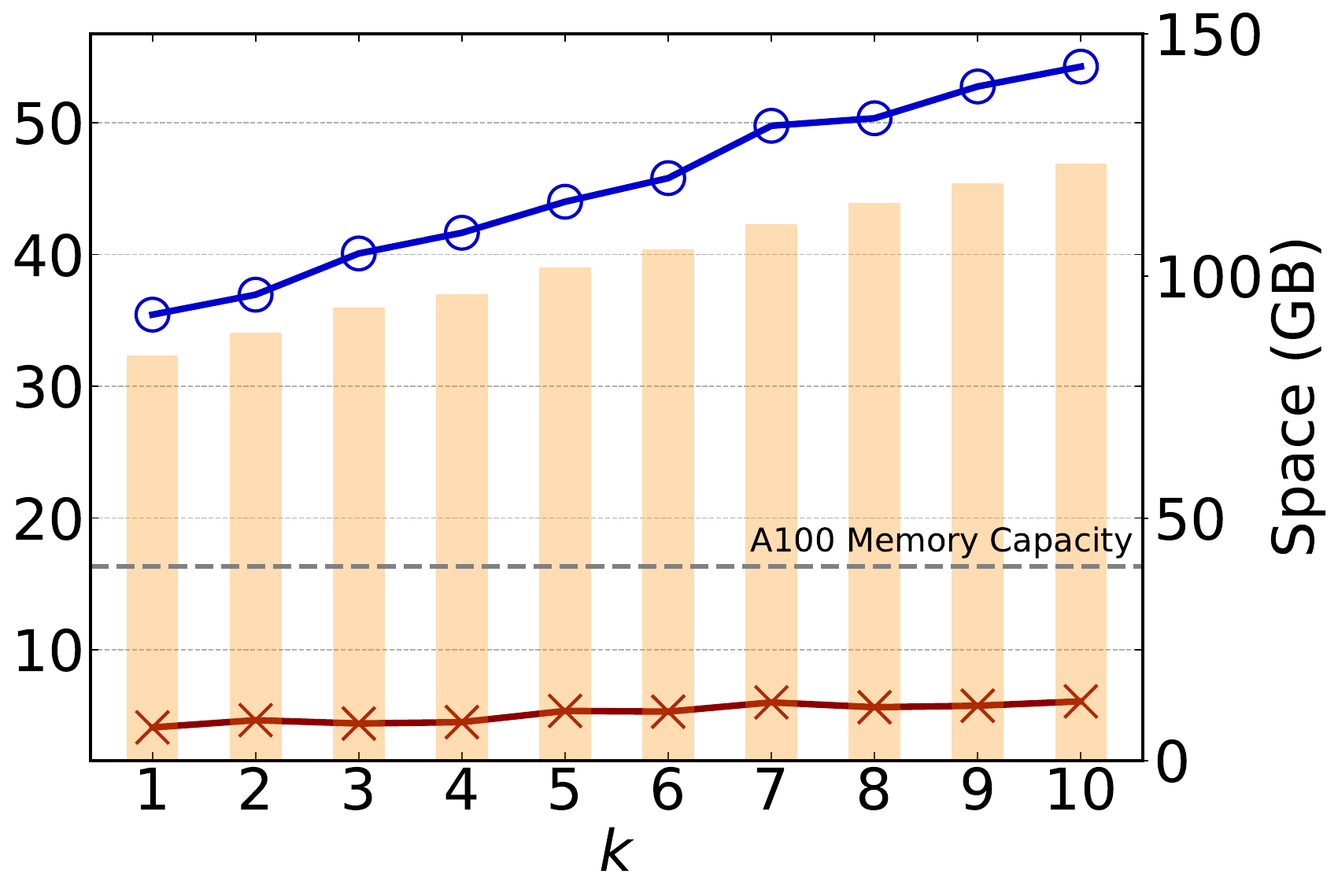}
	\vspace{-6mm}
	\caption{$k$-NN NV (varying $k$)}
	\label{exp:nv_knn_stream}
	\vspace{1mm}
\end{subfigure}
\begin{subfigure}{0.45\linewidth}
	\centering
	\includegraphics[width=\linewidth]{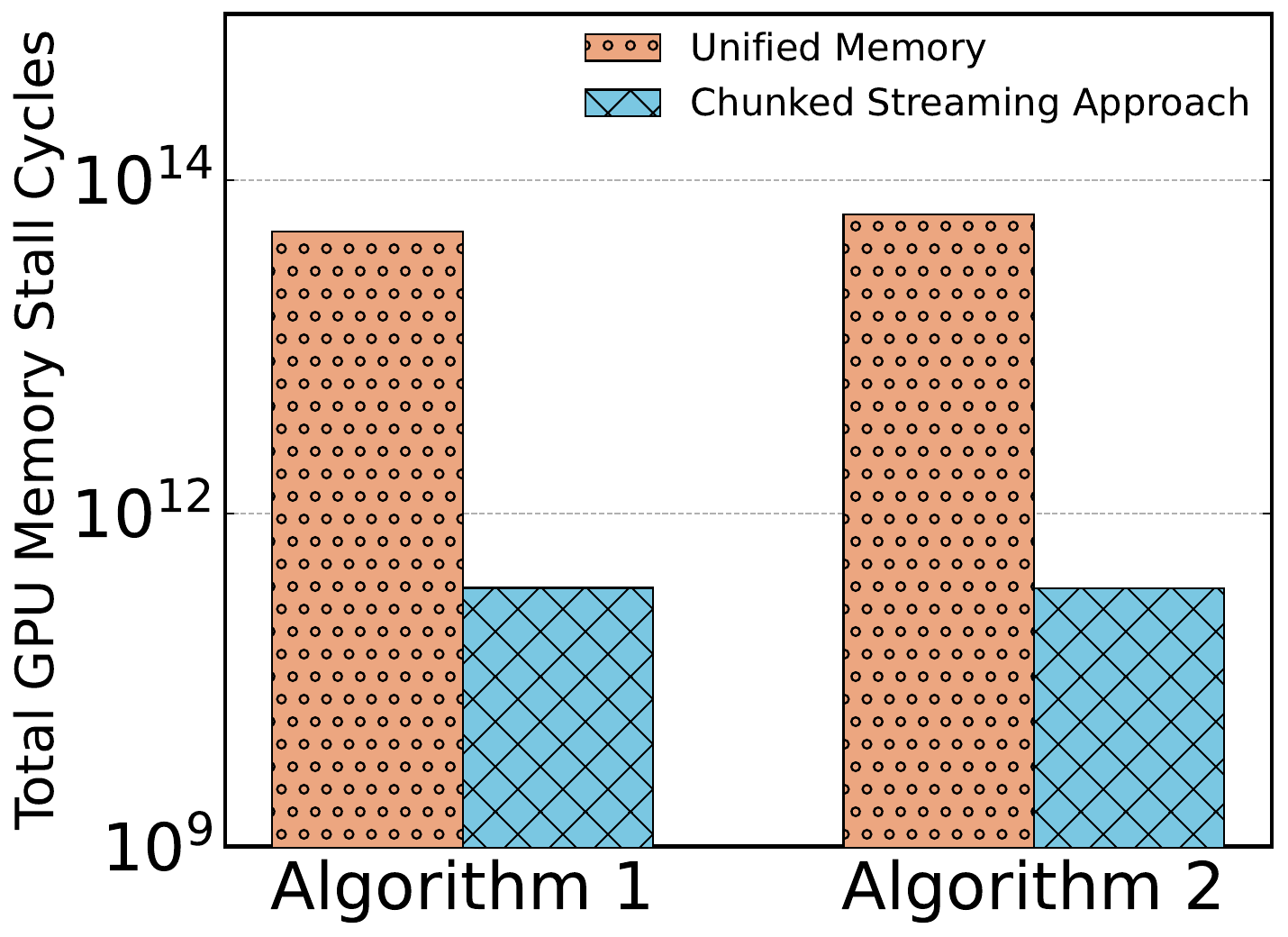} 
	\vspace{-6mm}
	\caption{Within-400 NV}
	\label{exp:wd_memory_install}
\end{subfigure}
\begin{subfigure}{0.45\linewidth}
	\centering
	\includegraphics[width=\columnwidth]{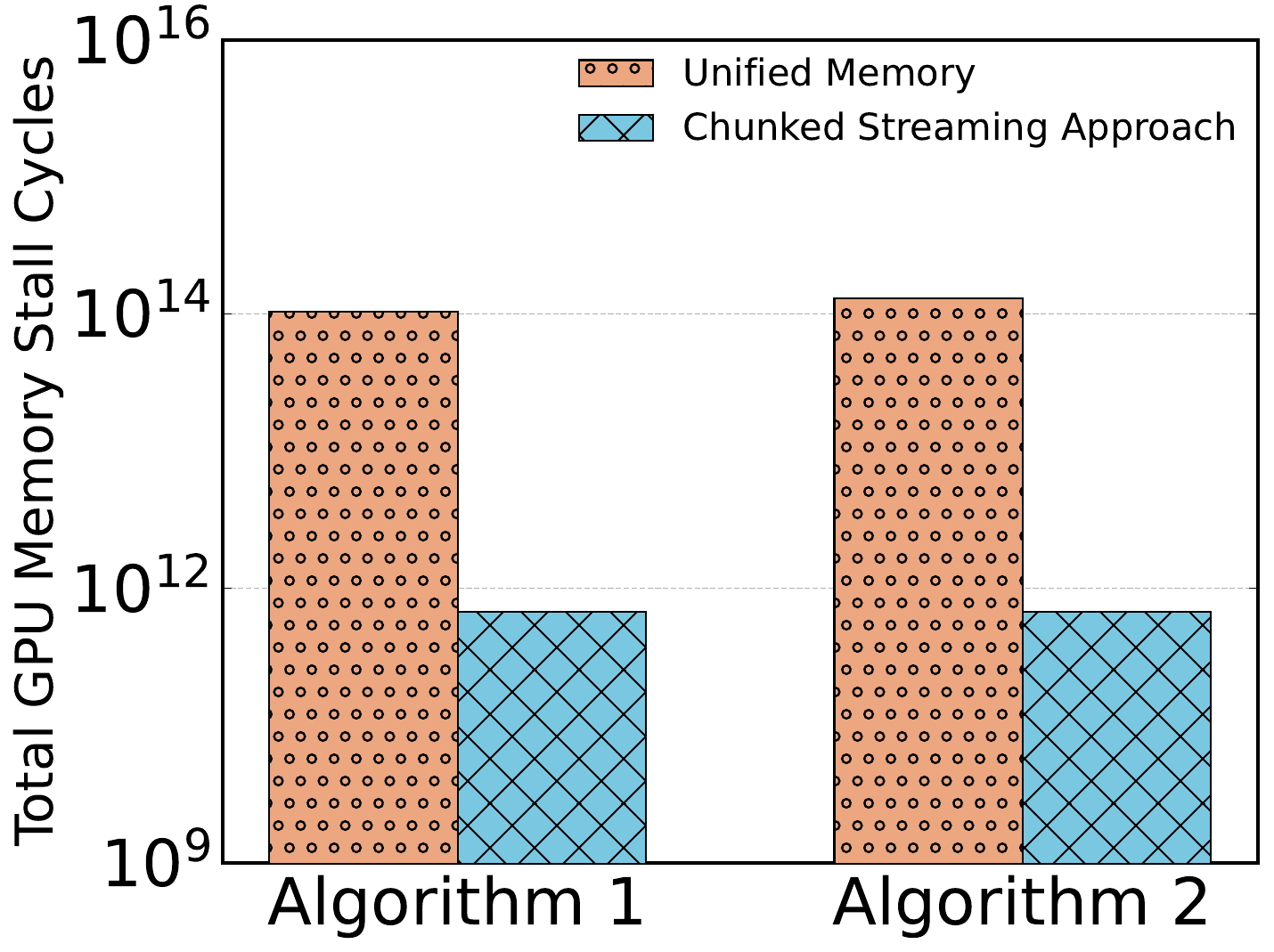}
	\vspace{-6mm}
	\caption{10-NN NV}
	\label{exp:knn_memory_install}
\end{subfigure}
\vspace{-3mm}
\caption{Unified Memory v.s. Chunked Streaming}
\label{exp:stream}
\vspace{-1mm}
\end{figure}

Recall that our chunked streaming design (Algorithms~\ref{algo:chunk_stream_pipe} and~\ref{algo:gpu_cpu_pipe}) enables processing voxel pairs beyond GPU memory capacity. We compare it with unified memory (\texttt{cudaMallocManaged}), which relies on on-demand page migration so can exceed the 40GB A100 memory capacity. 
Figure~\ref{exp:stream} reports the performance on NV for both within-$\tau$ and $k$-NN queries, where the curves show runtime (left y-axis), and the bars indicate memory usage (right y-axis).

In Figure~\ref{exp:nv_wd_stream}, when $\tau \le 250$, the working set fits within A100 memory, and our approach already outperforms unified memory with a modest margin. Once $\tau \ge 300$, memory demand exceeds capacity, and unified memory exhibits a sharp slowdown, while our approach remains stable, leading to a much larger performance gap.
This effect is more pronounced in Figure~\ref{exp:nv_knn_stream}, since memory demand exceeds capacity even at $k=1$.

To further understand this gap, we profile two representative queries (within-400 and 10-NN) using NVIDIA Nsight Compute. Figures~\ref{exp:wd_memory_install} and \ref{exp:knn_memory_install} report GPU memory stall cycles, which indicate stalled warps due to memory access delays. Unified memory incurs 146$\times$ and 184$\times$ more stall cycles than our approach for Algorithms~\ref{alg:vp_kernel} and \ref{algo:voxel_prune}, respectively. 
These results show that unified memory suffers from significant overhead once page migration is triggered, whereas chunked streaming avoids this issue, achieving up to 5.4$\times$ and 8.9$\times$ speedup for within-400 and 10-NN queries, respectively.

\vspace{-2mm}
\subsection{Effectiveness of CPU-GPU Pipelining}

\begin{figure}[!t]
\centering
\begin{subfigure}{0.9\linewidth}
	\centering
	\includegraphics[width=\linewidth]{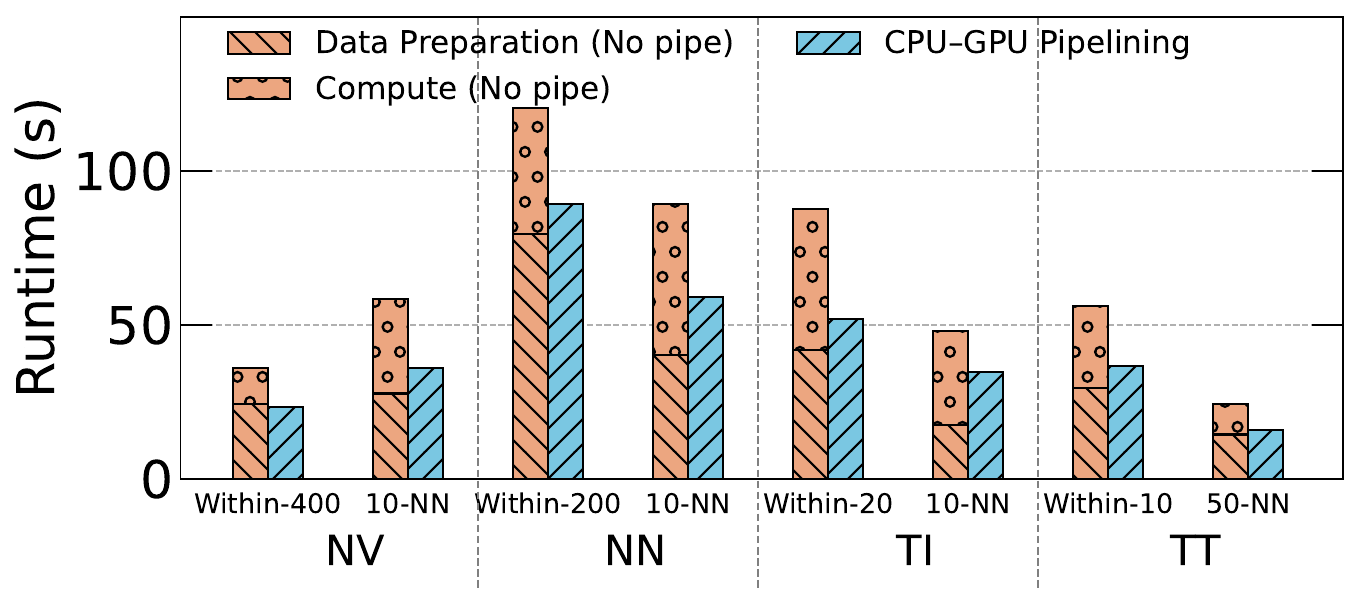} 
	\vspace{-6mm}
	\caption{Running Time of Refinement Stage}
	\label{exp:cpu_gpu_pipe_time}
	\vspace{2mm}
\end{subfigure}
\hfill
\begin{subfigure}{0.9\linewidth}
	\centering
	\includegraphics[width=\columnwidth]{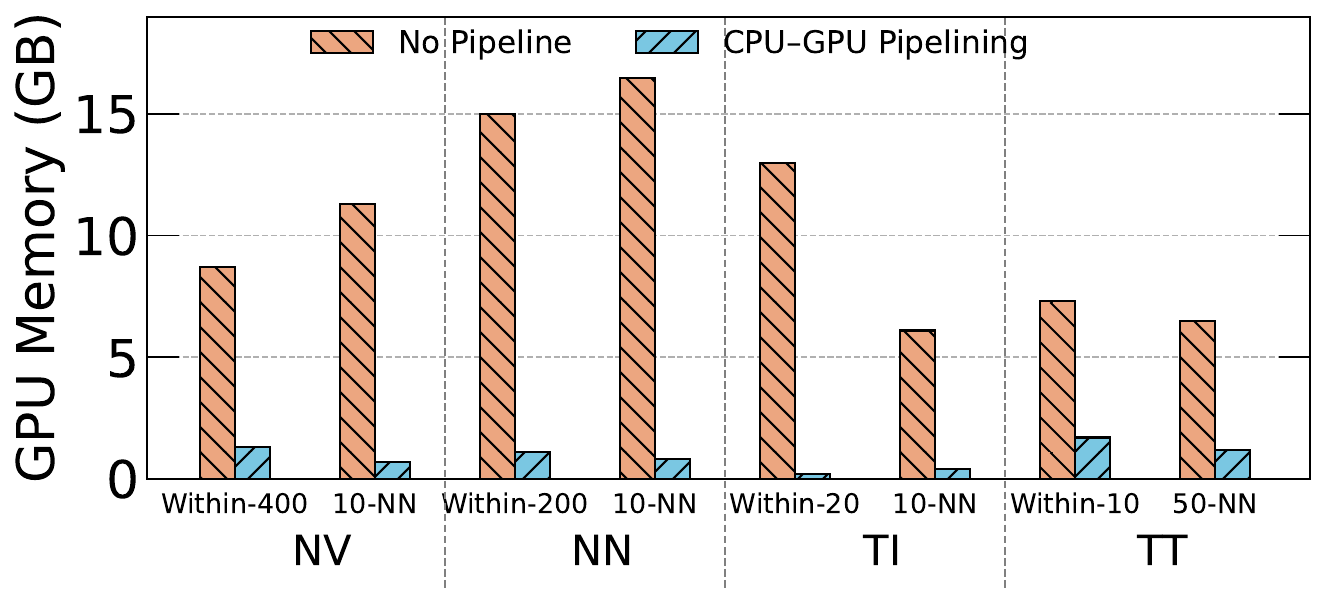}
	\vspace{-6mm}
	\caption{GPU Memory Consumption of Refinement Stage}
	\label{exp:cpu_gpu_pipe_memory}
\end{subfigure}
\vspace{-1mm}
\caption{CPU-GPU Pipelining v.s. No Pipelining}
\label{exp:cpu_gpu_pipe}
\vspace{-2mm}
\end{figure}

Recall that refinement involves both CPU data preparation and GPU computation. We compare our CPU-GPU pipelining (Algorithm~\ref{algo:gpu_cpu_pipe}) with a non-pipelined baseline, where CPU data preparation and GPU computation are executed sequentially.

Figure~\ref{exp:cpu_gpu_pipe} shows the impact on runtime and GPU memory usage. As shown in Figure~\ref{exp:cpu_gpu_pipe_time}, pipelining reduces the refinement time by up to 69\% by overlapping CPU and GPU execution. It also significantly reduces memory usage: since at most two chunks reside on the GPU simultaneously, the peak memory consumption is only 1.7~GB, compared to 16.5~GB for the baseline (Figure~\ref{exp:cpu_gpu_pipe_memory}), while eliminating GPU idle time during data preparation.

\vspace{-2mm}
\subsection{Evaluation of Object-Pair Pruning on GPU}

\begin{figure}[!t]
\includegraphics[width=0.3\linewidth]{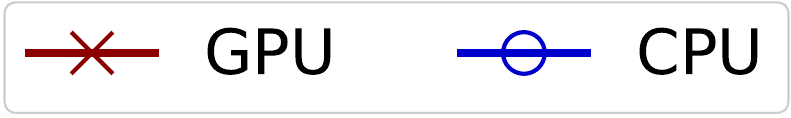}  \\
\centering
\begin{subfigure}{0.45\linewidth}
	\centering
	\includegraphics[width=\linewidth]{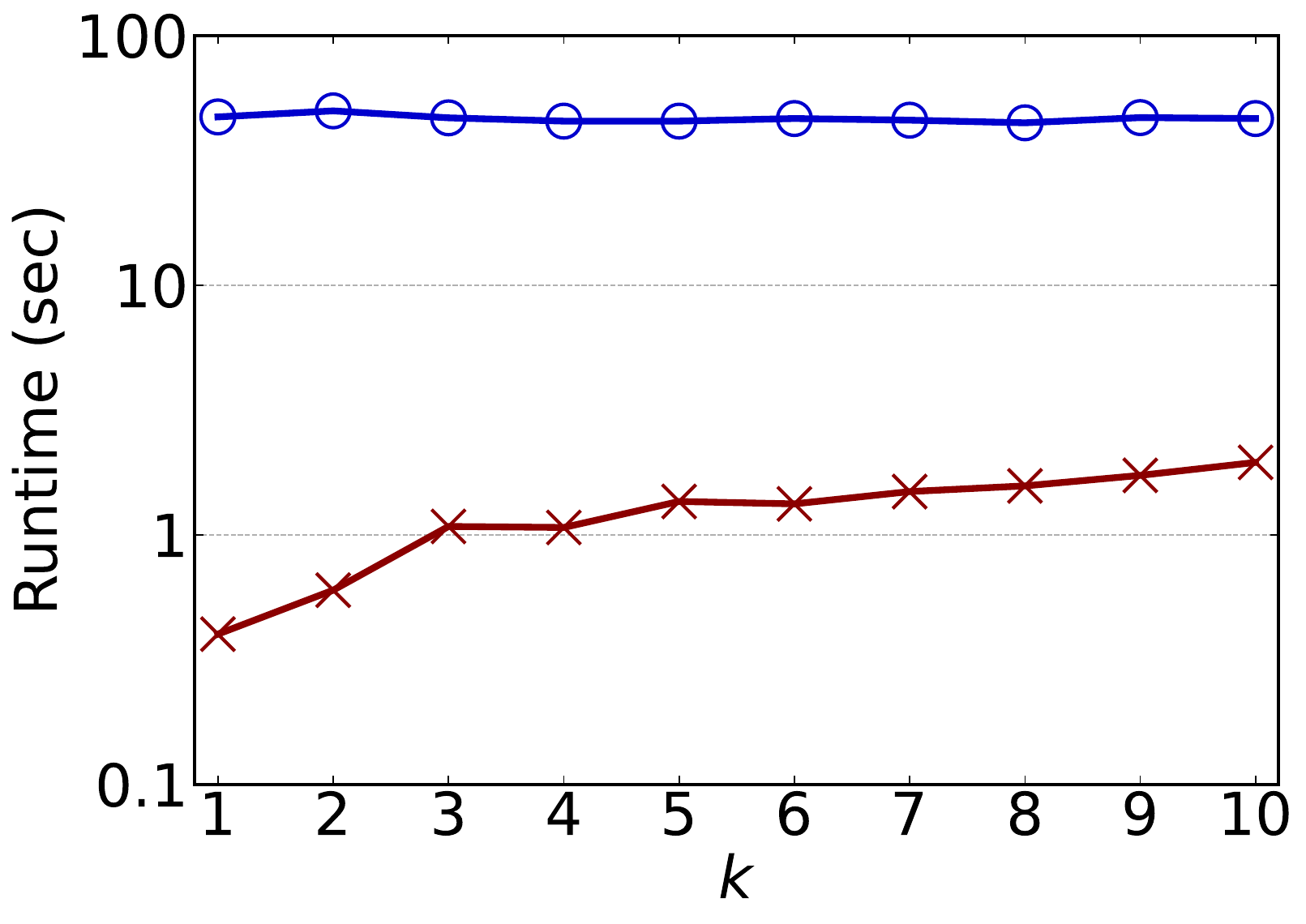} 
	\vspace{-6mm}
	\caption{$k$-NN NV (varying $k$)}
	\label{exp:nv_knn_evaluate}
\end{subfigure}
\hfill
\begin{subfigure}{0.45\linewidth}
	\centering
	\includegraphics[width=\columnwidth]{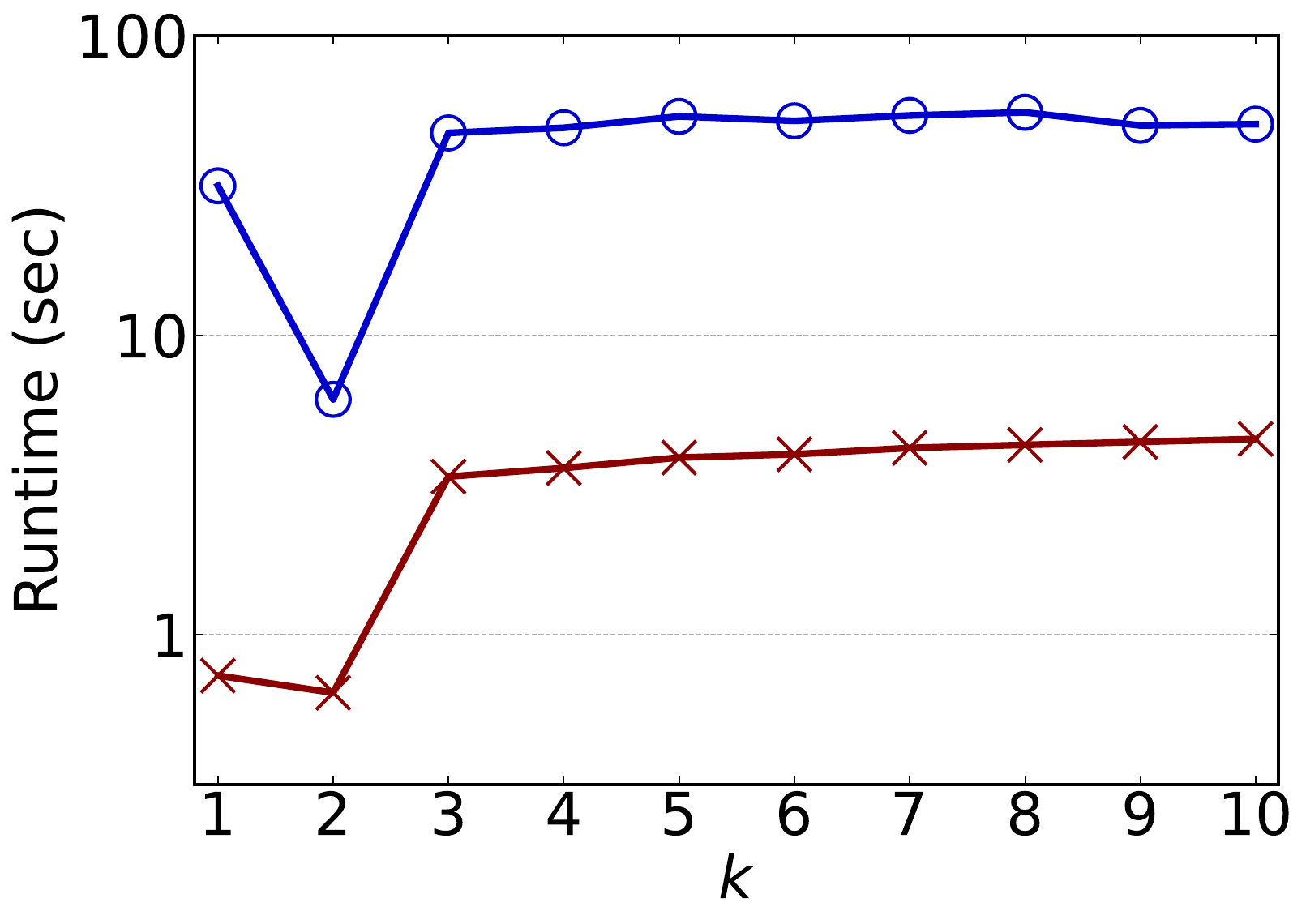}
	\vspace{-6mm}
	\caption{$k$-NN NN (varying $k$)}
	\label{exp:nn_knn_evaluate}
\end{subfigure}
\vspace{-3mm}
\caption{$k$-NN Object-Pair Pruning by CPU v.s.\ GPU}
\label{exp:knn_evaluate}
\vspace{-1mm}
\end{figure}

\setcounter{figure}{20}

\begin{figure*}[!t]
\centering
\includegraphics[width=0.8\linewidth]{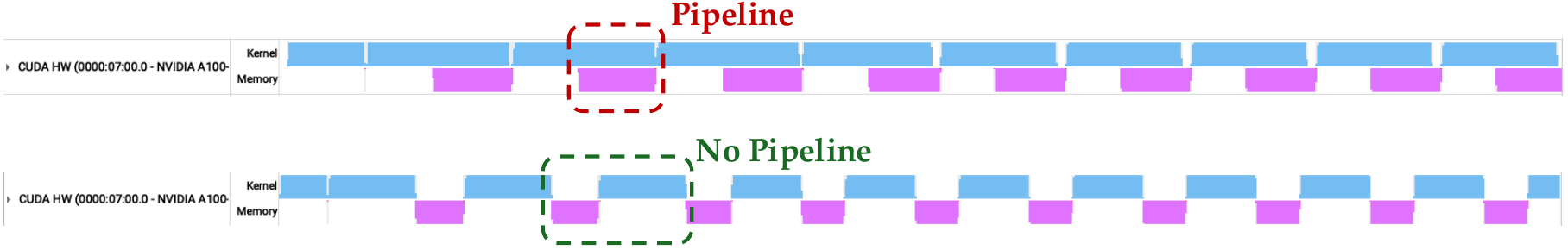} 
\vspace{-3mm}
\caption{Screenshots from Nvidia Nsight Systems. Top: Pipelining with CUDA Stream. Bottom: No Pipelining.}
\label{exp:pipeline_screenshot}
\vspace{-2mm}
\end{figure*}

Recall that for a $k$-NN query, object-pair pruning is repeatedly invoked after filtering and each refinement round. TDBase performs this step on CPU, whereas we design a GPU-based kernel (Algorithm~\ref{algo:knn}) to fully exploit parallelism.

Figure~\ref{exp:knn_evaluate} compares the total runtime of object-pair pruning (summed over all its invocations). Overall, the GPU-based approach achieves nearly two orders of magnitude speedup over the CPU baseline with OpenMP. On NV (Figure~\ref{exp:nv_knn_evaluate}), it is 23$\times$--99$\times$ faster, where CPU execution is dominated by expensive filtering rounds (up to 42\,s), while the GPU completes each round in 0.04\,s. On NN (Figure~\ref{exp:nn_knn_evaluate}), the speedup remains 9$\times$--42$\times$. In the final refinement round (LoD = 100\%), where CPU-based refinement dominates (up to 49\,s), the GPU still completes selection in about one second. 
These results show that GPU-based object-pair pruning is critical to performance and scales well with the number of candidates.

\setcounter{figure}{19}

\begin{figure}[!t]
\centering
\includegraphics[width=0.6\linewidth]{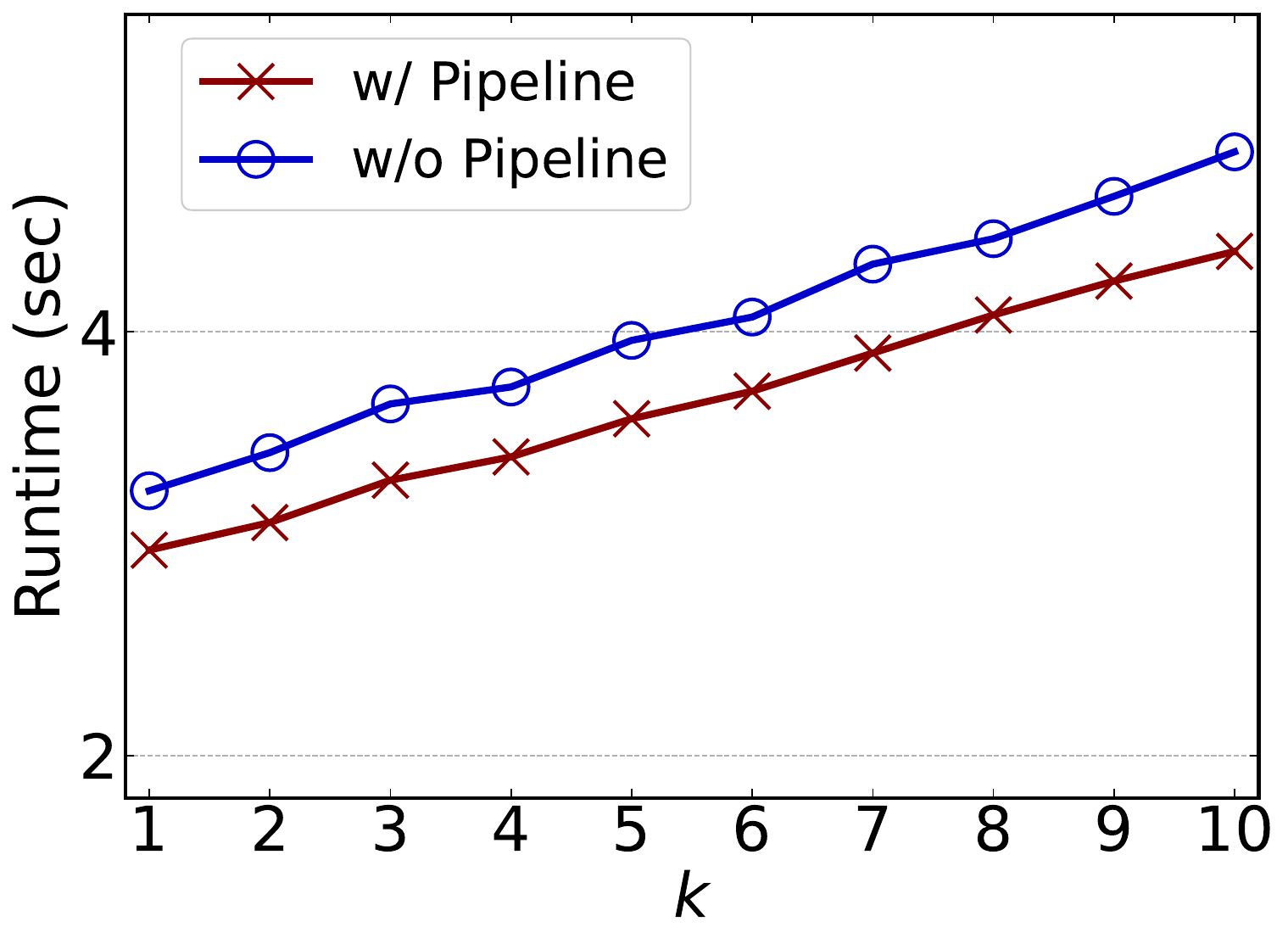}
\vspace{-3mm}
\caption{Time of Algorithm~\ref{algo:chunk_stream_pipe} With v.s.\ Without Pipeline}
\label{exp:nv_knn_pipeline}
\vspace{-3mm}
\end{figure} 

\setcounter{figure}{21}

\subsection{Other Ablation Studies}
We present additional ablation studies on secondary optimizations, complementing the major components evaluated above.

\vspace{1mm} 
\noindent \textbf{CUDA Stream Pipeline.}  
Our chunked streaming handles data beyond GPU memory capacity, but GPU may still remain idle during device-to-host data transfer. We introduce a pipelined implementation (Algorithm~\ref{algo:chunk_stream_pipe}) to overlap computation and data transfer.

Figure~\ref{exp:nv_knn_pipeline} compares chunked streaming with and without pipelining on $k$-NN queries over NV. We observe a consistent $\sim$10\% speedup, as pipelining utilizes idle GPU cycles by initiating computation for the next chunk while transferring results.

To verify this overlap, we profile execution using NVIDIA Nsight Systems. As Figure~\ref{exp:pipeline_screenshot} shows, kernel execution overlaps with memory transfer when pipelining is enabled, whereas no overlap is observed otherwise. This confirms that asynchronous execution effectively hides data transfer latency and improves GPU utilization.

\begin{figure}[!t]
\centering
\begin{subfigure}{0.48\linewidth}
	\centering
	\includegraphics[width=\linewidth]{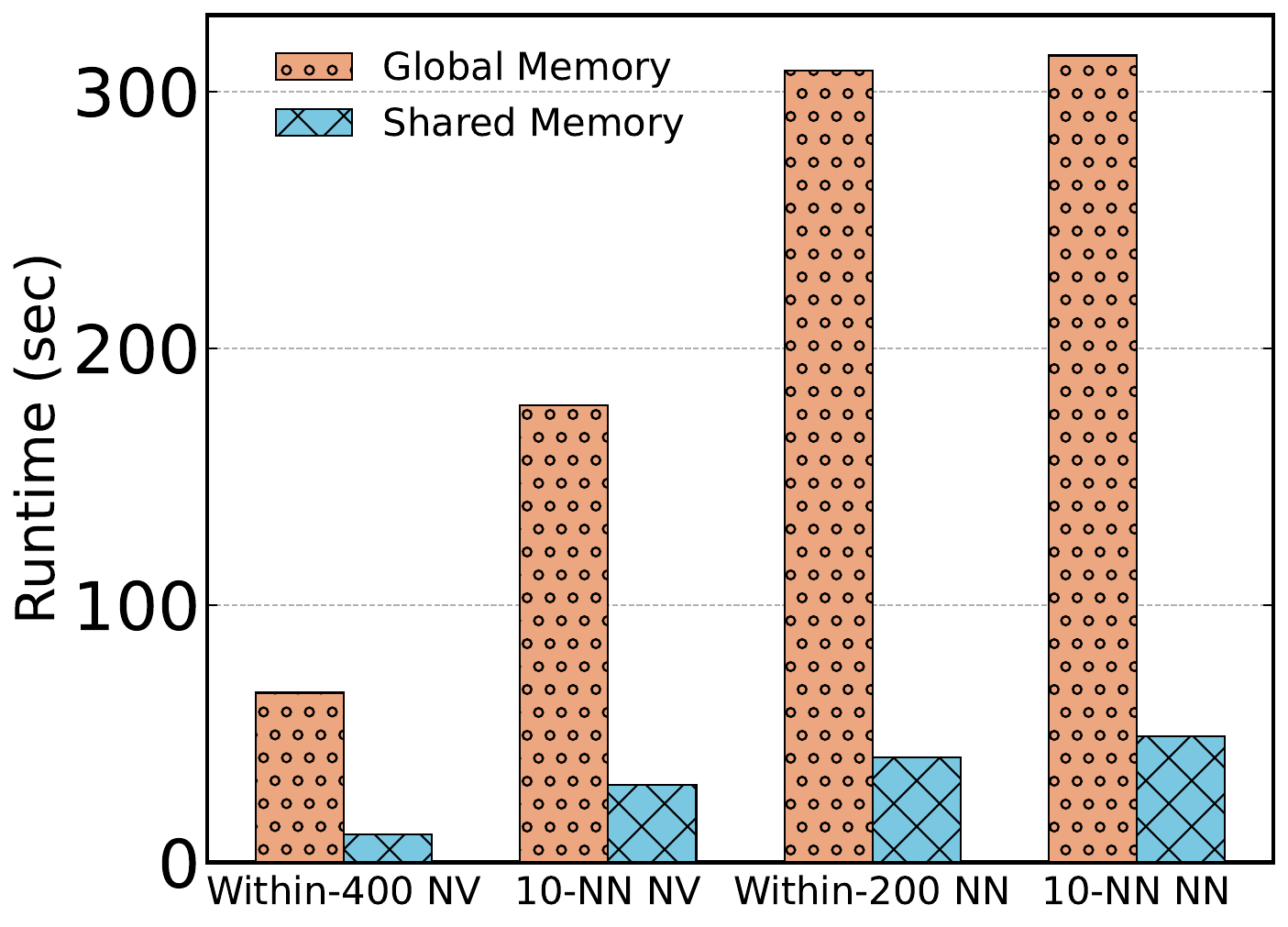} 
	\vspace{-3mm}
	\caption{Runtime of Refinement}
	\label{exp:all_shm_time}
\end{subfigure}
\hfill
\begin{subfigure}{0.48\linewidth}
	\centering
	\includegraphics[width=\columnwidth]{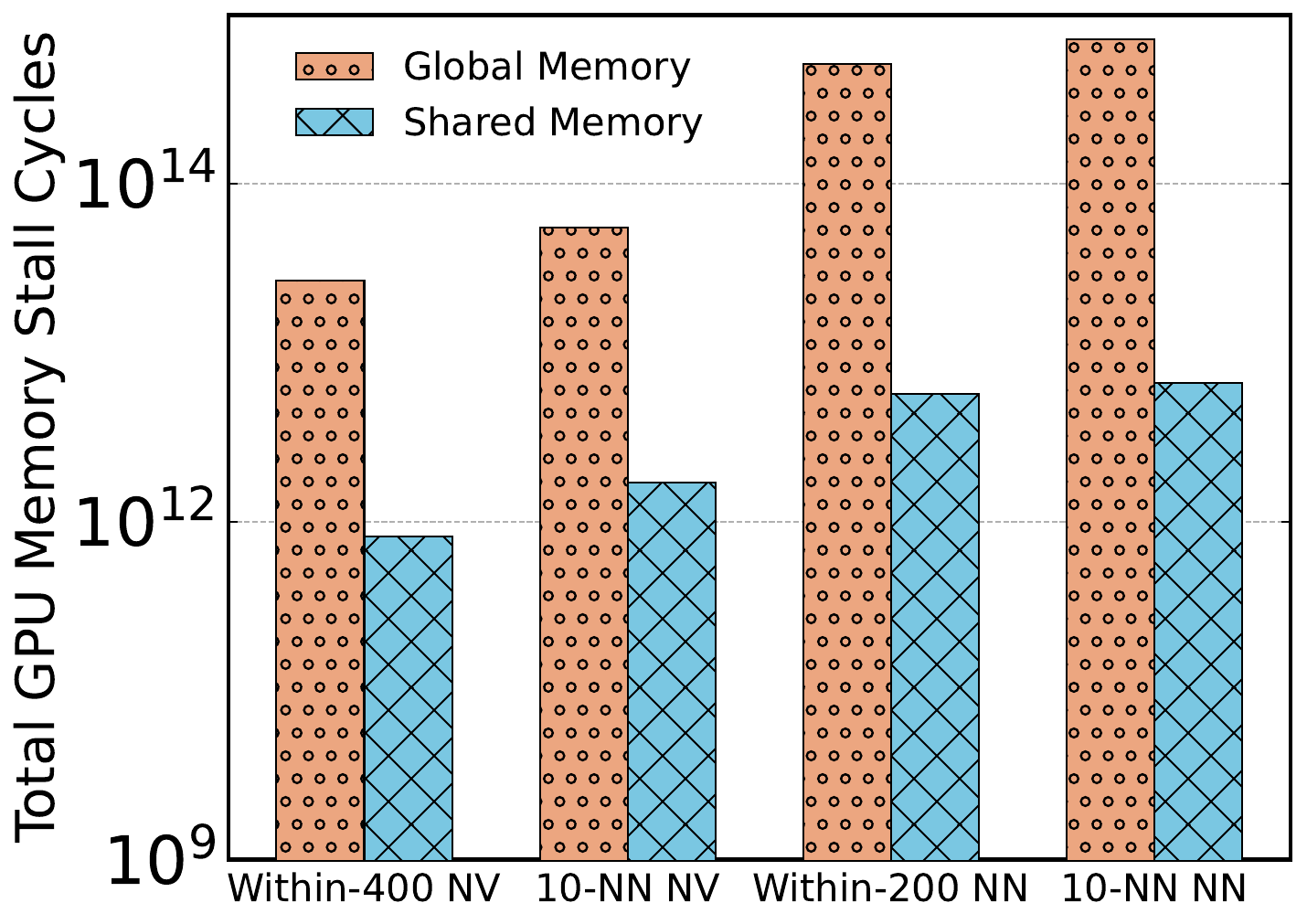}
	\vspace{-3mm}
	\caption{Memory Stall Cycles}
	\label{exp:all_shm_cycle}
\end{subfigure}
\vspace{-3mm}
\caption{Aggregation in GPU: Global v.s.\ Shared Memory}
\label{exp:all_shm}
\end{figure} 

\vspace{1mm} 
\noindent \textbf{Aggregation with Shared Memory.} 
Recall that block-wise aggregation (Figure~\ref{fig:prefix}) computes the aggregation across a thread block using shared memory, which is used in both filtering and refinement of 3DPipe. In particular, our refinement kernel, Algorithm~\ref{algo:refine}, uses it in Line~13. TDBase instead performs this minimum-aggregation in global memory using \texttt{atomicMin}. 

Figure~\ref{exp:all_shm} quantifies the impact of this shared-memory aggregation on refinement performance. As shown in Figure~\ref{exp:all_shm_time}, shared-memory aggregation achieves 6$\times$--7$\times$ speedup over global-memory aggregation. Meanwhile, Figure~\ref{exp:all_shm_cycle} reports GPU memory stall cycles (caused by cache/global memory accesses). Shared memory reduces stall cycles by 32$\times$--108$\times$, indicating significantly lower memory latency and improved warp scheduling efficiency.

\subsection{Scalability Analysis}

\begin{figure}[!t]
\centering
\includegraphics[width=0.7\linewidth]{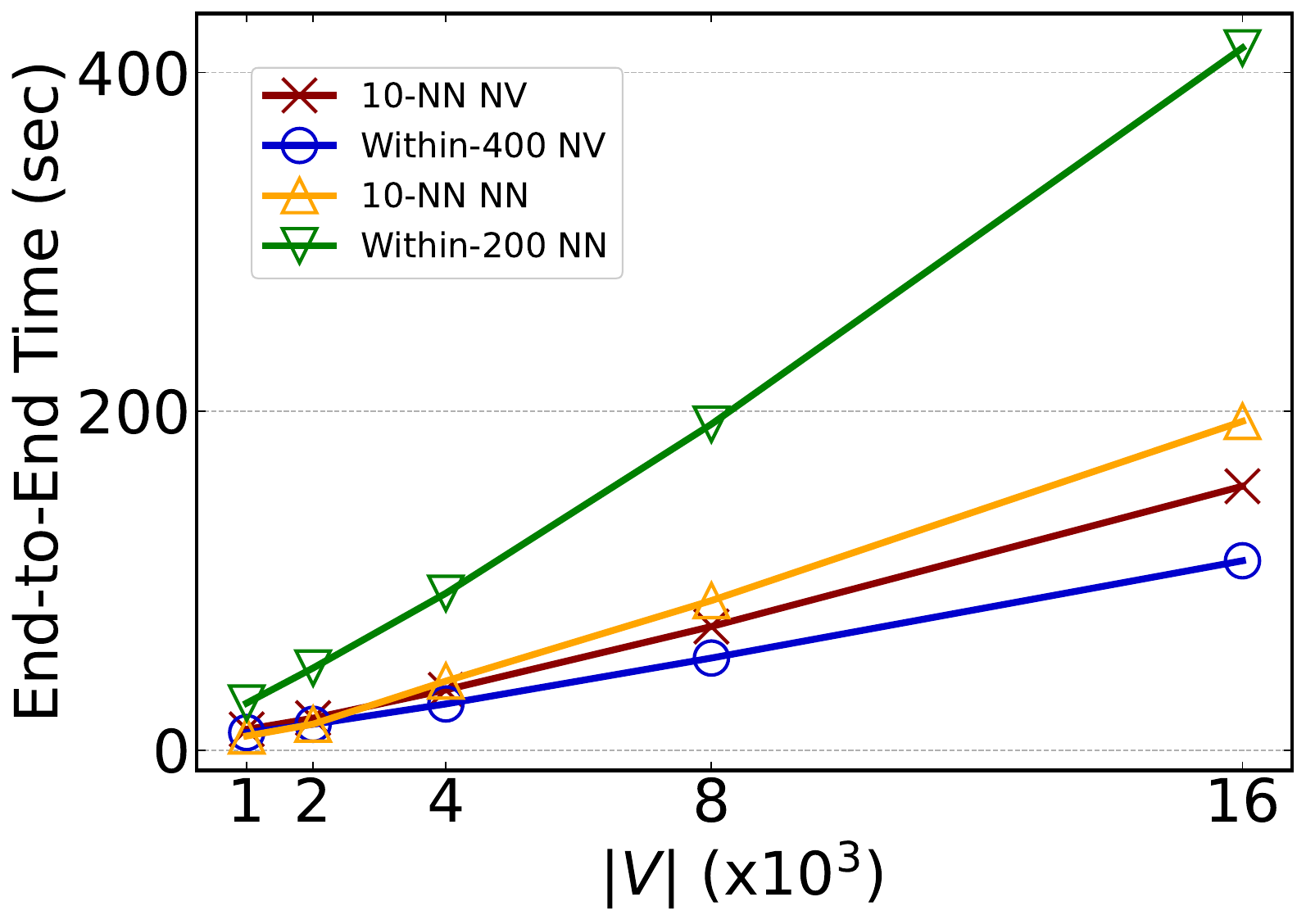} 
\vspace{-3mm}
\caption{End-to-End Time with Increasing Data Scale}
\label{exp:all_data_scale}
\end{figure} 

To evaluate scalability, we generate vessel and nuclei datasets with increasing sizes from 1K to 16K, doubling at each step. We use four representative queries with the large workloads.

Figure~\ref{exp:all_data_scale} shows that our approach scales nearly linearly with data size, while maintaining high efficiency across all settings. This demonstrates its ability to handle large-scale datasets effectively.


\begin{figure*}[!t]
\centering
\begin{subfigure}{0.48\linewidth}
	\centering
	\includegraphics[width=\linewidth]{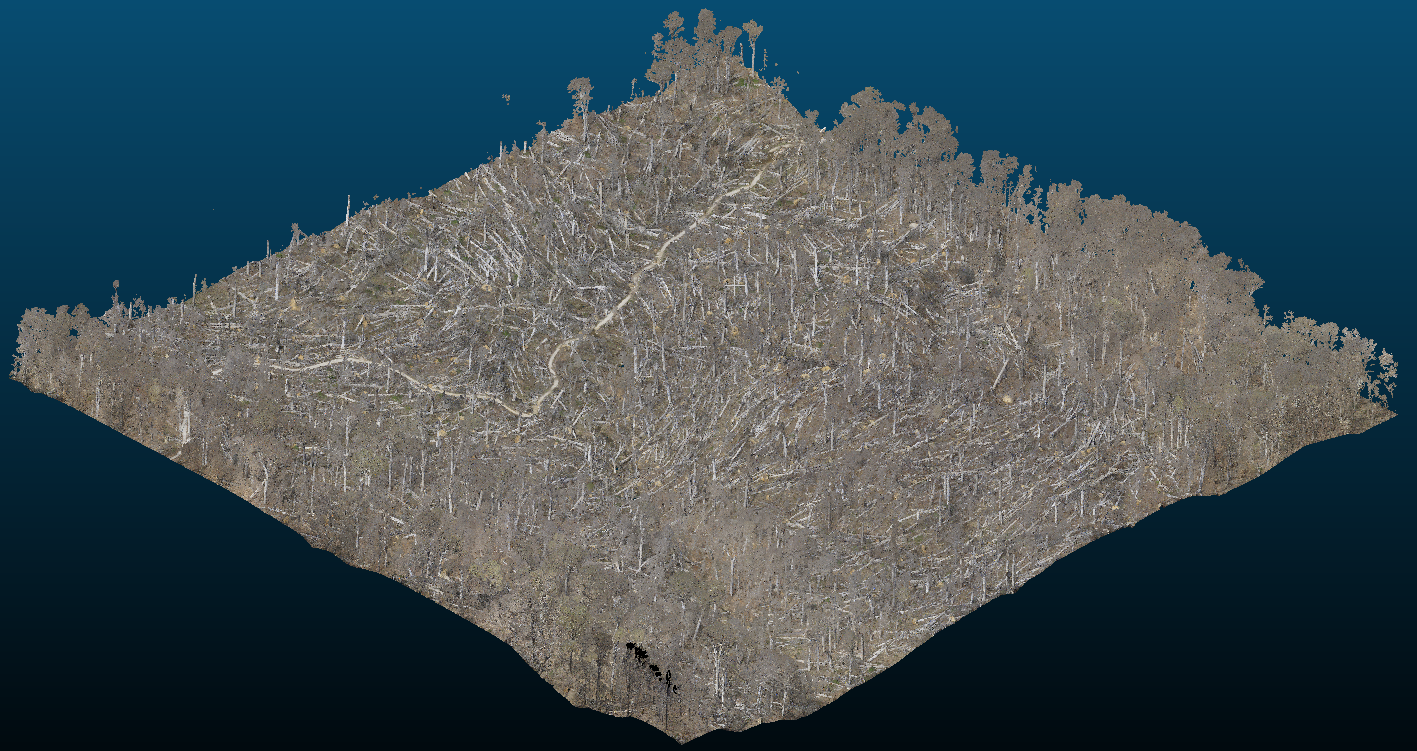} 
	\vspace{-4mm}
	\caption{The Point Cloud of Studied Region After Tornado}
	\label{tree:no_plot}
\end{subfigure}
\hfill
\begin{subfigure}{0.48\linewidth}
	\centering
	\includegraphics[width=\columnwidth]{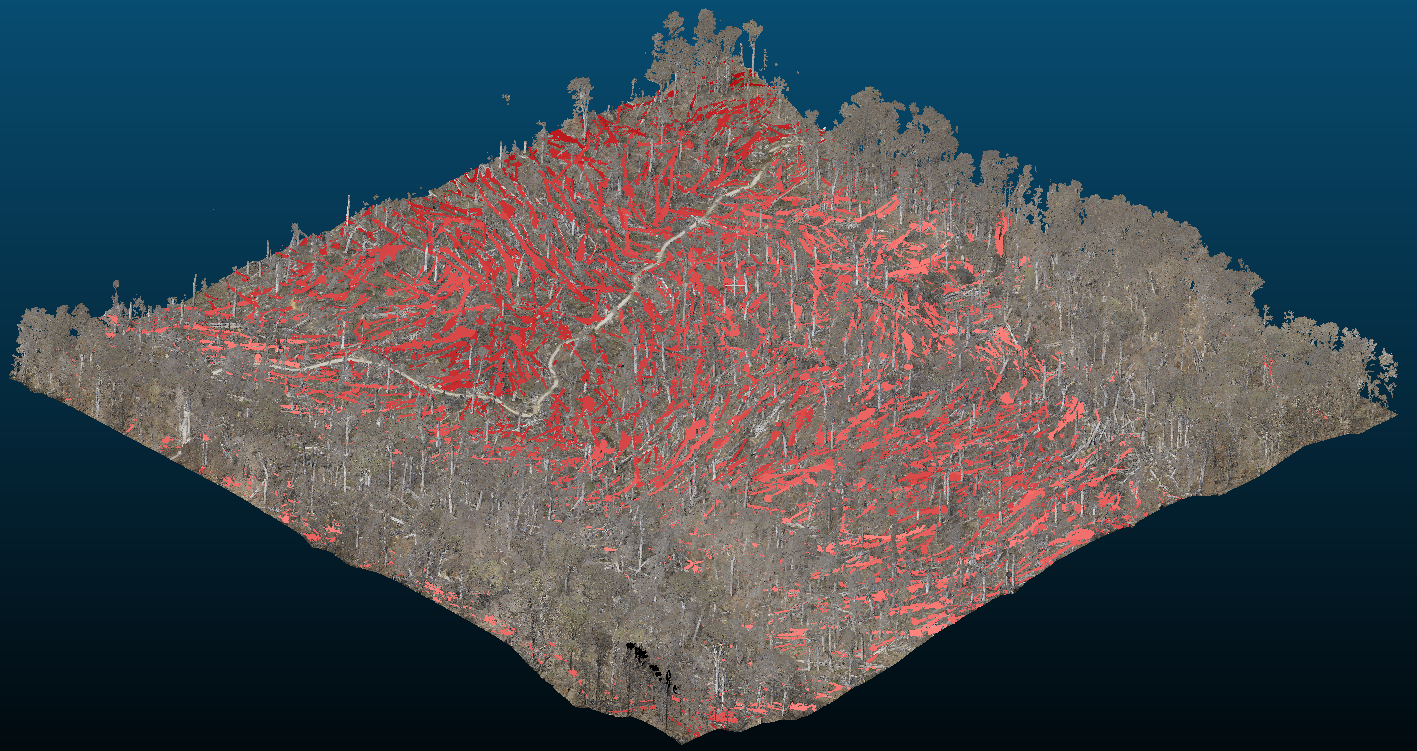}
	\vspace{-4mm}
	\caption{Downed Trees in theTornado Corridor Highlighted in Red}
	\label{tree:with_plot}
\end{subfigure}
\vspace{-3mm}
\caption{Tornado Path Detection using 3DPipe }
\label{tree}
\vspace{-4mm}
\end{figure*}

\subsection{A Real-World Case Study}\label{ssec:case_study}

We demonstrate the practical utility of 3DPipe by delineating the path of an EF-3 tornado and identifying the severely affected forest regions from large-scale 3D LiDAR data. Our study uses an aerial LiDAR survey collected after the tornado struck McCormick's Creek State Park in Spencer, Indiana, on March~31,~2023. 
The tornado caused widespread forest damage, leaving a lot of downed trees.
Figure~\ref{tree:no_plot} shows our studied area, covering $1398.48$~m $\times$ $984.38$~m.

We construct two object sets for spatial join: the standing-tree set $S$ and the downed-tree set $R$. Standing trees are detected directly from the point cloud using ForestFormer3D~\cite{forestformer}. Downed trees are extracted using a two-stage pipeline. We first apply cloth simulation filtering (CSF)~\cite{csf} to remove the tree canopy and generate a bird's-eye view, where Mask2Former~\cite{mask2former}, fine-tuned using manually annotated region proposals, generates candidate downed-tree regions. The regions are back-projected to the point cloud and refined by a PointNet++~\cite{pointnet} model, fine-tuned using manually annotated 3D tree instances created with CloudCompare, to obtain the final 3D downed-tree instances. This process produces 3,193 standing trees and 2,451 downed trees in our  studied area.

Since 3DPipe operates on watertight meshes, each segmented tree point cloud is converted into a watertight mesh using the alpha shape algorithm~\cite{alphashape}. For each tree, we increase the alpha value from 0.5 to 6.0 and use the first value that yields a watertight mesh.

We perform a within-$\tau$ spatial join between $R$ and $S$ with $\tau=60$~m. 3DPipe finishes the query in 4.5 seconds which is very efficient. After the join, we aggregate the join results by each downed tree to count its nearby standing trees.
Downed trees with fewer than 134 neighboring standing trees are discarded, while the remaining trees are visualized using a red color map, where darker colors indicate fewer nearby standing trees and therefore locations closer to the tornado center.
The distance threshold (60 m) and count threshold (134) are empirically tuned and chosen so that the downed trees in the tornado corridor are clearly delineated. 
As Figure~\ref{tree:with_plot} shows, the query clearly recovers the tornado corridor and highlights the regions with the highest concentration of tree damage, providing useful information for post-disaster assessment and management.

The efficiency of 3DPipe becomes even more important for larger-scale post-disaster assessment over entire parks or forests. For example, if $R$ consists of hiking trails and $S$ of downed trees, 3DPipe can accurately identify trees blocking trails to guide trail clearing.

\section{Related Work}
\label{sec:related_work}

\noindent \textbf{Spatial Join Processing.}
Spatial join has been studied extensively for 2D spatial data. Early work mainly focused on disk-based methods, such as hash-based~\cite{lo1996spatial} and tile-based spatial join~\cite{pbsm, sssj}, to cope with limited main memory. As memory capacity increased, in-memory methods~\cite{touch} and later parallel and distributed systems~\cite{brinkhoff1996parallel, zhou1998data, tspatial, hadoopgis, spatialhadoop, sparkgis, geospark} were developed to improve performance on large-scale datasets. In addition to CPU-based acceleration, modern hardware such as GPU and FPGA has also been explored for faster spatial join processing~\cite{zhang2017parallel, swiftspatial}. 

\vspace{1mm}
\noindent \textbf{Mesh Simplification and Multi-LoD Representations.}
To reduce the cost of geometric computation in 3D, prior work has explored progressive mesh simplification~\cite{valette2009progressive, khodakovsky2000progressive}, which constructs multiple levels of detail (LoDs) for the same polyhedron. Such multi-LoD representations are useful for spatial query processing because they enable coarse-to-fine evaluation: many object pairs can be pruned or resolved at low resolutions, avoiding unnecessary computation on the original high-resolution geometry.

\vspace{1mm}
\noindent \textbf{3D Spatial Data Management and Query Processing.}
Compared with the rich literature on 2D spatial join, 3D spatial join over polyhedral objects remains much less explored. Among existing systems, iSPEED~\cite{ispeed} adopts a MapReduce-based framework for distributed 3D spatial queries, with optimizations to reduce disk I/O. Real et al.~\cite{real2019large} accelerate basic 3D spatial operations on top of PostGIS~\cite{postgis}. However, general-purpose spatial databases such as PostGIS lack efficient support for large-scale 3D spatial join, especially when polyhedra contain many facets and complex geometry, which motivates specialized solutions. 
More closely related to our work, 3DPro~\cite{3dpro} introduces progressive refinement over multiple LoDs and leverages GPU to accelerate the refinement stage. TDBase~\cite{tdbase} further improves this framework by incorporating facet-level Hausdorff and proxy Hausdorff bounds, enabling tighter distance estimation and more effective pruning at low LoDs.

\section{Conclusions and Future Work}
\label{sec:conclusion}

We presented 3DPipe, a pipelined GPU framework for scalable 3D spatial join over polyhedral objects. Our design exploited GPU parallelism across both filtering and refinement, and introduced chunked streaming and CPU-GPU pipelining to overcome memory and utilization bottlenecks. Together with GPU-oriented optimizations such as shared-memory aggregation and parallel top-$k$ filtering, 3DPipe achieved efficient end-to-end execution that consistently outperformed the state-of-the-art solution TDBase, achieving up to 9.0$\times$ speedup while maintaining near-linear scalability.

Currently, all facet data are loaded in host memory, extracted and transferred to the GPU on demand during refinement. While effective, this design incurs non-trivial host memory overhead for large datasets. In future work, we plan to extend 3DPipe to an out-of-core setting by storing facet data on SSD and adopting a index nested-loop join strategy. Given that facet-level distance computation is the dominant cost and is highly GPU-parallelizable, we expect that pipelined execution can overlap SSD I/O with GPU computation and effectively hide I/O latency.

%
%

\bibliographystyle{ACM-Reference-Format}
\bibliography{ref}

\end{document}